\documentclass[11pt,a4paper]{article}
\usepackage{jcappub}
\usepackage[utf8]{inputenc}
\usepackage[T1]{fontenc}
\usepackage{booktabs}
\usepackage{amssymb}
\usepackage{graphicx}
\usepackage{xcolor}
\usepackage{mathtools}
\usepackage{amsthm}
\usepackage{comment}
\usepackage{amsbsy}
\usepackage{subcaption}
\usepackage{geometry}
\usepackage{color}
\usepackage{physics}
\usepackage{soul}
\usepackage{cancel}
\usepackage{float}
\usepackage{cleveref}
\usepackage[normalem]{ulem}

\usepackage{fontawesome5}
\definecolor{linkcolor}{rgb}{0.7752941176470588, 0.22078431372549023, 0.2262745098039215}
\definecolor{linkcolor}{HTML}{1393C1}
\definecolor{zima_blue}{HTML}{1393C1}
\hypersetup{
    colorlinks=true,
    linkcolor=zima_blue,
    citecolor=zima_blue,
    urlcolor=zima_blue,
    filecolor=zima_blue,
    anchorcolor=zima_blue,
    menucolor=zima_blue
}

\makeatletter
\renewcommand\section{\@startsection{section}{1}{\z@}%
    {-3.5ex \@plus -1ex \@minus -.2ex}%
    {2.3ex \@plus.2ex}%
    {\normalfont\Large\bfseries}}
\renewcommand\subsection{\@startsection{subsection}{2}{\z@}%
    {-3.25ex\@plus -1ex \@minus -.2ex}%
    {1.5ex \@plus .2ex}%
    {\normalfont\large\bfseries}}
\renewcommand\subsubsection{\@startsection{subsubsection}{3}{\z@}%
    {-3.25ex\@plus -1ex \@minus -.2ex}%
    {1.5ex \@plus .2ex}%
    {\normalfont\normalsize\bfseries}}
\makeatother

\title{Can the universe be matter-dominated after a supercooled first-order phase transition?}

\abstract{We show that the answer is generally no, at least not immediately. Bubble collisions leave behind a highly inhomogeneous scalar field with persistent relativistic gradients, producing an equation of state between matter and radiation. Using lattice simulations in one, two and three spatial dimensions, we find that the equation of state is controlled by the wall Lorentz factor at collision $\gamma_\star$: walls with larger $\gamma_\star$ populate higher-momentum modes and drive the fluid closer to radiation. Matter domination begins only after these modes redshift, at $a/a_\star \simeq \gamma_\star$, or after the field thermalises through self-scattering and number-changing processes. This delay has direct implications for gravitational waves, primordial black holes and dark matter production.}


\begin{document}

\pagestyle{myplain}\pagenumbering{arabic}
\flushbottom

\makeatletter

\noindent{\small\sc Prepared for submission to JCAP \hfill Preprint: TTP26-031, P3H-26-059}

\vspace{1.5em}

\begin{center}

{\LARGE \bf \setlength{\baselineskip}{1.\baselineskip}\@title \par}

\medskip\bigskip\vspace{0.3cm}

\renewcommand{\thefootnote}{\fnsymbol{footnote}}

{\large
Henda Mansour$^{a}$\footnote{\href{mailto:henda.mansour@kit.edu}{henda.mansour@kit.edu}},
Yann Gouttenoire$^{b,c}$\footnote{\href{mailto:yann.gouttenoire@iap.fr}{yann.gouttenoire@iap.fr}},
and Felix Kahlhoefer$^{a,d}$\footnote{\href{mailto:kahlhoefer@kit.edu}{kahlhoefer@kit.edu}}
}

\renewcommand{\thefootnote}{\arabic{footnote}}
\setcounter{footnote}{0}

\medskip

{\it \small $^a$Institute for Theoretical Particle Physics (TTP), KIT, D-76131 Karlsruhe, Germany}\\
{\it \small $^b$PRISMA$^+$ Cluster of Excellence \& MITP, JGU Mainz, Germany}\\
{\it \small $^c$Institut d'Astrophysique de Paris (IAP), CNRS, Sorbonne Universit\'e, FR-75014, France}\\
{\it \small $^d$Institute for Astroparticle Physics (IAP), KIT, 76344 Eggenstein-Leopoldshafen, Germany}

\end{center}

\bigskip

\centerline{\bf Abstract}
\begin{quote}
\@abstract
\end{quote}

\medskip 
\vspace*{5pt} \noindent \noindent{\bf GitHub:}  The lattice code {\tt CoolBubble} for computing bubble nucleation and expansion during supercooled first order phase transitions~\href{https://github.com/Henda-M/CoolBubble}{\faGithub\;Henda-M/CoolBubble}.

\medskip
\vspace*{5pt} 
\noindent{\bf Keywords:} \@keywords

\makeatother

\newpage
\noindent\makebox[\linewidth]{\rule{\textwidth}{1pt}}
\setcounter{tocdepth}{2}
\tableofcontents
\noindent\makebox[\linewidth]{\rule{\textwidth}{1pt}}
\newpage

\section{Introduction}
\label{sec:intro}

 While the Standard Model of particle physics predicts only crossover phase transitions in the early universe~\cite{Kajantie:1996mn,Aoki:2006we}, many Standard Model extensions and secluded dark sectors allow for first-order phase transitions (FOPTs). Such a FOPT takes place when the effective potential of a scalar field is altered by thermal corrections that make a new vacuum state energetically favourable but separate the two vacua by a barrier. Once the temperature of the surrounding plasma drops below a critical temperature, the transition proceeds through the tunnelling of the field \cite{Coleman:1977py,Callan:1977pt,Linde:1981zj}: bubbles of the true vacuum nucleate and expand, converting vacuum energy to kinetic and gradient energies of the walls \cite{Hawking:1982ga} and heating the radiation bath.

The strength of the phase transition can be evaluated as the ratio between the vacuum energy difference between the two phases $\Delta V$ and the radiation energy density $\rho_{\rm rad}(T_n)$ at the temperature $T_n$ when nucleation becomes efficient: 
\begin{equation}
\label{eq:latent_heat_fac}
    \alpha\equiv \frac{\Delta V}{\rho_{\rm rad}(T_n)}\equiv \left(\frac{T_{\rm eq}}{T_n}\right)^4.
\end{equation}
The second equality defines the temperature $T_{\rm eq}$ when the vacuum energy would start dominating the energy density of the universe, $\rho_{\rm rad}(T_{\rm eq})\equiv \Delta V$. We define a FOPT to be supercooled when $\alpha \geq 1$, see ref.~\cite{Gouttenoire:2022gwi} for a review. Vacuum energy then dominates before
nucleation, driving a stage of thermal inflation~\cite{Guth:1980zm} that
dilutes the plasma. The resulting weak friction allows the walls to become
ultra-relativistic~\cite{Bodeker:2017cim,
Hoche:2020ysm, Azatov:2020ufh, Gouttenoire:2021kjv, Azatov:2023xem,Ai:2025bjw}, with a Lorentz factor at collision
$\gamma_\star$ that can span many orders of magnitude~\cite{Gouttenoire:2021kjv}.
Supercooled FOPTs can trigger a variety of cosmological implications. They can dilute dark matter~\cite{Hambye:2018qjv,Gouttenoire:2022gwi}, affect its thermal production~\cite{Baker:2019ndr,Bringmann:2023iuz,Benso:2025vgm}, produce dark matter through bubble collisions~\cite{Watkins:1991zt, Kolb:1996jr,Konstandin:2011ds,Falkowski:2012fb,Mansour:2023fwj,Shakya:2023kjf,Giudice:2024tcp,Baldes:2023fsp,Cheng:2026npt,Ghoshal:2026pew,An:2026sdu} or bubble wall plasma interaction~\cite{Baldes:2020kam,Baldes:2021aph,Azatov:2021ifm,Baldes:2022oev,Gouttenoire:2023roe,Baldes:2024wuz,Azatov:2024crd,Ai:2024ikj,Cheng:2026npt}, generate the baryon asymmetry of the universe~\cite{Katz:2016adq,Azatov:2021irb,Baldes:2021vyz,Chun:2023ezg,Dichtl:2023xqd,Cataldi:2024pgt,Cataldi:2025nac}, or high frequency gravitational waves (GWs)~\cite{Ai:2025fqw,Qiu:2025tmn}. The slow nucleation rate characteristic of supercooled FOPTs, $\beta/H\sim \mathcal{O}(10)$ with $\beta\equiv d \log{\Gamma}/dt$ the time derivative of the nucleation rate, can source sizeable density perturbations though late-blooming~\cite{Sasaki:1982fi,Liu:2022lvz,Giombi:2023jqq,Elor:2023xbz,Lewicki:2024ghw,Buckley:2024nen,Cai:2024nln,Jinno:2024nwb,Sui:2025epg,Greene:2026gnw}, that can collapse into primordial black holes (PBHs)~\cite{Kodama:1982sf,Hsu:1990fg,Liu:2021svg,Hashino:2021qoq,Kawana:2022olo,Lewicki:2023ioy,Gouttenoire:2023naa,Baldes:2023rqv,Gouttenoire:2023bqy,Salvio:2023ynn,Gouttenoire:2023pxh,Flores:2024lng,Lewicki:2024ghw,Lewicki:2024sfw,Kanemura:2024pae,Cai:2024nln,Goncalves:2024vkj,Banerjee:2024fam,Arteaga:2024vde,Banerjee:2024cwv,Hashino:2025fse,Ghoshal:2025dmi,Zou:2025sow,Franciolini:2025ztf,Wang:2025hwc,Kierkla:2025vwp,Banerjee:2025qji,Wang:2026zvz,Ning:2026nfs,Ai:2026zrs,Banerjee:2026tgr,Jinno:2023vnr,Guo:2026cuv}. Non-supercooled phase transitions could also produce PBHs through incompleteness of percolation~\cite{Jinno:2023vnr,Guo:2026cuv,Ai:2024cka,Murai:2025hse,Hawking:1982ga,Moss:1994iq,Ashoorioon:2020hln,Jung:2021mku}, bubble collisions~\cite{Hawking:1982ga,Moss:1994iq,Ashoorioon:2020hln,Jung:2021mku}, or 
matter squeezing by bubble walls~\cite{Crawford:1982yz,Gross:2021qgx,Baker:2021sno,Kawana:2021tde,Huang:2022him}. Finally, supercooled FOPTs can also produce  primordial magnetic fields \cite{Hogan:1983zz,Quashnock:1988vs,Baym:1995fk,Baym:1995fk,Sigl:1996dm,Ahonen:1997wh,Caprini:2009yp,DeSimone:2011ek,Zhang:2019vsb,Ellis:2019tjf,Di:2020kbw,RoperPol:2022iel,Balaji:2024rvo,ArteagaTupia:2025awh} and loud GWs~\cite{Witten:1984rs, Kamionkowski:1993fg,Caprini:2015zlo,Caprini:2019egz,Matuszak:2026xsz} that can explain the stochastic GW signal observed by pulsar timing arrays~\cite{Gouttenoire:2023bqy,NANOGrav:2021flc,NANOGrav:2023hvm,EPTA:2023xxk,Nakai:2020oit,Ratzinger:2020koh,Neronov:2020qrl,Madge:2023dxc,Bringmann:2023opz,Bringmann:2026xcx, Banik:2024zwj,Balan:2025uke, Goncalves:2025uwh, Li:2026pjy}. 

All these predictions are sensitive to the equation of state (EoS) of the universe after the phase transition, which itself depends on how the vacuum energy is redistributed after
bubble collisions. Reheating is often treated as instantaneous, but if the
scalar decay rate is small, $\Gamma_\phi\ll H$, the scalar remains dominant
while reheating proceeds in stages. This epoch is commonly assumed to be an early matter-dominated (eMD) era ~\cite{Barenboim:2016mjm,Hambye:2018qjv,Baldes:2021aph,Ellis:2019oqb,Ellis:2020nnr,Ertas:2021xeh,Hook:2020phx,Brzeminski:2022haa,Gouttenoire:2021jhk,Gonstal:2025qky}. Such an era would change the entropy dilution of dark matter
~\cite{Benso:2025vgm,Wong:2023qon,Racco:2025ons}, enhance PBH formation
~\cite{Harada:2016mhb,Harada:2017fjm}, and reshape the GW spectrum. In particular, modes entering the horizon during eMD scale as $f^1$ in the infrared rather than the usual $f^3$ of radiation domination
~\cite{Ellis:2020nnr,Barenboim:2016mjm,Hook:2020phx}, while the overall GW amplitude is diluted~\cite{Gonstal:2025qky,Ghoshal:2026ros}.

Despite its interesting cosmological consequences, it remains unclear whether a period of eMD is physically achievable following a supercooled transition with slow reheating. A primary concern is that the scalar field after percolation is highly inhomogeneous and cannot be treated as coherent waves. This is different from inflationary reheating where in simple realizations the inflaton is homogeneous and oscillates coherently, such that the EoS is determined solely by the shape of the potential. The minimum can be approximated as a monomial potential $\sim\phi^n$ and the average EoS is derived using the virial theorem $\omega=\frac{n-2}{n+2}$ \cite{Turner:1983he}, such that in a quadratic potential a coherently oscillating classical scalar field behaves like matter ($\omega=0$). However, after bubble collisions, the field configuration is far from being homogeneous. The system is initially dominated by the gradients of the ultra-relativistic walls, which enter the EoS and lead to a radiation-like behaviour. While these gradients eventually redshift in an expanding background, it is an open question how quickly these inhomogeneities settle and whether the EoS ever truly mimics that of matter.

In this work, we investigate the EoS of the universe following a supercooled phase transition with delayed reheating using lattice simulations in 1+1, 2+1, and 3+1 dimensions, performed with {\tt CoolBubble}~\cite{CoolBubble}, a code we developed for this purpose. Considering initial field configurations of randomly nucleated bubbles with periodic boundary conditions, we find that in a static universe the EoS after percolation relaxes to a constant value that depends strongly on the Lorentz factor of the bubble walls. Specifically, more relativistic bubbles result in a more radiation-like fluid upon completion of the transition. We relate this behaviour to the power spectrum of the field, enabling the derivation of an analytical function that predicts the dependence of the EoS on the Lorentz factor of the bubble walls $\gamma_\star$. We then show that the subsequent evolution of the EoS in an expanding background depends on the efficiency of the thermalisation processes of the scalar field. If these are slow and the walls are ultra-relativistic ($\gamma_\star \gg 1$), a radiation-like era persists until $a_\mathrm{matter}/a_\star \simeq \gamma_\star$, where $a_\star$ corresponds to the scale factor at the completion of the phase transition. Efficient thermalisation can lead to a faster establishment of matter domination, depending on the presence or absence of number-changing cannibal processes~\cite{Carlson:1992fn, Pappadopulo:2016pkp}.

The remainder of this paper is structured as follows. In section~\ref{sec:EoS}, we establish the theoretical framework for studying the EoS of a classical scalar field in the context of FOPTs. In section~\ref{sec:lattice}, we describe the numerical lattice setup and present the main simulation results for the post-collision EoS. In section~\ref{sec:PS-evol}, we investigate the evolution of the scalar power spectrum and its implications for the EoS. Assuming a simple analytical expression for the power spectrum, we derive a fitting function in terms of the Lorentz factor $\gamma_\star$ and the scale factor $a$. Additionally, we discuss the impact of thermalisation on the late-time evolution of the EoS. In section~\ref{sec:discussion}, we highlight and examine implications of our results on reheating, relics dilution, GW production and primordial black holes. Finally, we summarise our findings and conclude in section~\ref{sec:conclusions}. Additional details are provided in the appendices~\ref{app:gamma_wall}, \ref{app:c_dependence}, \ref{app:analytical}, \ref{app:NumericalPS} and \ref{app:thermalisation}.

\section{Equation of state of a classical scalar field}
\label{sec:EoS}

\subsection{Basic formalism}
\label{subsec:EoS-in-flat-Universe}

\paragraph{Energy-momentum tensor and equation of state.}
We first consider the evolution of a real scalar field $\phi$ in a flat static universe with an equation of motion given by
\begin{equation}
\label{eq:eom_EL}
\ddot{\phi}-\vec{\nabla}^2{\phi} = -V_{,\phi}\, , 
\end{equation} 
where dotted variables are derivatives with respect to the cosmic time $t$ and $V_{,\phi}=d V /d\phi$, with $V$ being the effective potential of the scalar field.  
The stress-energy tensor reads
\begin{equation}
    T_{\mu\nu} = \partial_\mu \phi \, \partial_\nu \phi - g_{\mu\nu}\left(\frac{1}{2}g^{\alpha\beta}\,\partial_\alpha \phi  \, \partial_\beta \phi - V(\phi) \right) .
\end{equation}
The energy density $\rho_\text{tot}$ of the scalar field is given by 
\begin{align}
    T_{00}&=\frac{\dot{\phi}^2}{2}+\frac{(\vec{\nabla}\phi)^2}{2}+V(\phi) \equiv \rho_\mathrm{kin}+\rho_\mathrm{grad}+\rho_\mathrm{pot}\equiv \rho_\text{tot} \; ,
\end{align}
where $\rho_\mathrm{kin}$, $\rho_\mathrm{grad}$ and $\rho_\mathrm{pot}$ correspond respectively to the kinetic, gradient and potential energy densities of the scalar field. 
The pressure density is obtained by averaging the diagonal entries for the spatial dimensions, which are given by
\begin{align}
    T_{ii}&=\frac{\dot{\phi}^2}{2}+(\partial_i\phi)^2-\frac{(\vec{\nabla}\phi)^2}{2}-V(\phi) \; .
\end{align}
Although our main focus will be on scalar fields in 3+1 dimensions, we will also analyse simulations in 1+1 and 2+1 dimensions. Denoting the number of spatial dimensions by $d$, the pressure density is given by 
\begin{equation}
   p_\phi \equiv \frac{1}{d}\sum_{i=1}^{d}T_{ii} = \rho_\mathrm{kin}-\frac{d-2}{d}\rho_\mathrm{grad}-\rho_\mathrm{pot}
    =\begin{cases}
\rho_\mathrm{kin}-\rho_\mathrm{grad}/3-\rho_\mathrm{pot} \, , & (d=3)\\
\rho_\mathrm{kin}-\rho_\mathrm{pot} \, , & (d=2)  \\
\rho_\mathrm{kin} + \rho_\mathrm{grad} - \rho_\mathrm{pot} \; . & (d=1) 
    \end{cases}
\end{equation} 
The EoS of a $d+1$ dimensional scalar field is therefore given by 
\begin{equation}
\label{eq:EoS_General}
    \omega\equiv \frac{p_\phi}{\rho_\text{tot}} =\frac{\rho_\mathrm{kin} +(d-2) \rho_\mathrm{grad}/d - \rho_\mathrm{pot}}{\rho_\mathrm{kin} + \rho_\mathrm{grad} + \rho_\mathrm{pot} }\; .
\end{equation}

\paragraph{Virial theorem.}
We define the spacetime average of a quantity $X$ as
\begin{equation}
\langle X\rangle(t) \equiv
\lim_{T,V\to\infty}\frac{1}{TV}
\int_t^{t+T} dt'\int_V d^d r\,X(t',\vec r) \; ,
\end{equation}
where the averaging time satisfies $T\gg m_\phi^{-1}$ and the averaging volume is taken to be much larger than all relevant microscopic and macroscopic length scales. Integrating by parts in
eq.~\eqref{eq:eom_EL} and dropping boundary terms yields 
\begin{equation}
\label{eq:Virial_main}
\langle\rho_\mathrm{kin}\rangle \,\simeq\,
\langle\rho_\mathrm{grad}\rangle + \frac{1}{2}\langle\phi V_{,\phi}\rangle \; .
\end{equation}
This result is equivalent to the virial theorem in Statistical Mechanics.
For a power-law minimum $V(\phi)\sim\phi^{n}$ one has
\begin{equation}
\label{eq:Virial_quadratic}
\langle\rho_\mathrm{kin}\rangle \,\simeq\,
\langle\rho_\mathrm{grad}\rangle
+ \frac{n}{2}\langle\rho_\mathrm{pot}\rangle \; .
\end{equation}
The resulting EoS reads
\begin{equation}
\label{eq:EoS_Virial}
\omega \,=\,
\frac{2\langle\rho_\mathrm{grad}\rangle/d
+(1-n/2)\langle\rho_\mathrm{pot}\rangle}
{2\langle\rho_\mathrm{grad}\rangle
+(1+n/2)\langle\rho_\mathrm{pot}\rangle}.
\end{equation}
This expression has also been derived in refs.~\cite{Lozanov:2016hid,Lozanov:2017hjm}, and a similar derivation of the virial theorem can be found in ref.~\cite{Drees:2025iue}.
For a quartic potential with $n=4$, corresponding to a thermalised massless scalar field, and $d=3$ spatial dimensions, eq.~\eqref{eq:EoS_Virial} gives $\omega \simeq 1/3$, in agreement with results from the lattice study in ref.~\cite{Bettoni:2021zhq}. 

For a quadratic potential with $n=2$, corresponding to a non-interacting massive scalar field, one finds $\langle \rho_\mathrm{kin}\rangle  \simeq \rho_\text{tot}/2$, which is conserved in a static universe. In this case, the gradient terms contribute to lowering the EoS compared to the thermalised state:
\begin{equation}
\label{eq:EoS_Virial_quadratic}
\boxed{
    \omega ~=~ d^{-1} \frac{\langle \rho_\mathrm{grad}\rangle }{\langle \rho_\mathrm{grad}\rangle +\langle \rho_\mathrm{pot}\rangle } ~\simeq~ d^{-1} \frac{\langle \rho_\mathrm{grad}\rangle }{\langle \rho_\mathrm{kin}\rangle }\; .}
\end{equation}
For a homogenous field with $\langle \rho_\mathrm{grad} \rangle = 0$, one recovers the case of eMD ($\omega = 0$) especially familiar from many post-inflationary reheating scenarios \cite{Turner:1983he}. If, on the other hand, the mass term is small, such that the energy density is dominated by the gradient and kinetic energy, one can approximate $\langle  \rho_\mathrm{pot}\rangle  \simeq 0$ and $\langle  \rho_\mathrm{kin} \rangle  \simeq \langle  \rho_\mathrm{grad} \rangle $. In this case the EoS is that of a radiation-like fluid, i.e.\ $\omega \simeq 1/3$ in $3+1$ dimensions.

\paragraph{Expanding universe.}
Promoting eq.~\eqref{eq:eom_EL} to an expanding background with scale
factor $a$ and Hubble rate $H=\dot a/a$ yields
\begin{equation}
\label{eq:eom_EL-Hubble}
\ddot\phi - a^{-2}\vec\nabla^{2}\phi + 3H\dot\phi = -V_{,\phi} \; ,
\end{equation}
together with the Friedmann equation
$H^{2}=\rho_\mathrm{tot}/(3M_\mathrm{pl}^{2})$ if the scalar field
dominates the energy density. Including the Hubble friction term in the
virial derivation adds a contribution
$\tfrac{3}{2}\langle H\phi\dot\phi\rangle$ on the right-hand side of
eq.~\eqref{eq:Virial_main}, which is negligible for $H\ll m_\phi$. Crucially, the
gradient term redshifts in an expanding universe:
\begin{equation}
\rho_{\rm grad}=\frac{(\vec\nabla\phi)^{2}}{2}\xrightarrow[\rm ]{\rm ~Hubble~expansion~}\frac{(\vec\nabla\phi)^{2}}{2a^{2}} \; .
\end{equation} 
For a quadratic minimum of the potential, the gradient energy in the denominator of
eq.~\eqref{eq:EoS_Virial_quadratic} redshifts faster than the potential energy, so an
initially radiation-like inhomogeneous configuration with $\langle \rho_\text{grad} \rangle \gg \langle \rho_\text{pot} \rangle$ and $\omega \simeq d^{-1}$ evolves towards a
matter-like state with $\langle \rho_\text{grad} \rangle \ll \langle \rho_\text{pot} \rangle$ and $\omega \simeq 0$. 
The detailed $a$-dependence of the different contributions will be derived in section~\ref{subsec:derivation-omega-of-gamma}.

\subsection{First-order phase transitions}
\label{subsec:FOPT}

\begin{figure}[t]
\centering
\begin{subfigure}[t]{0.495\linewidth}
\centering
\includegraphics[width=\linewidth]{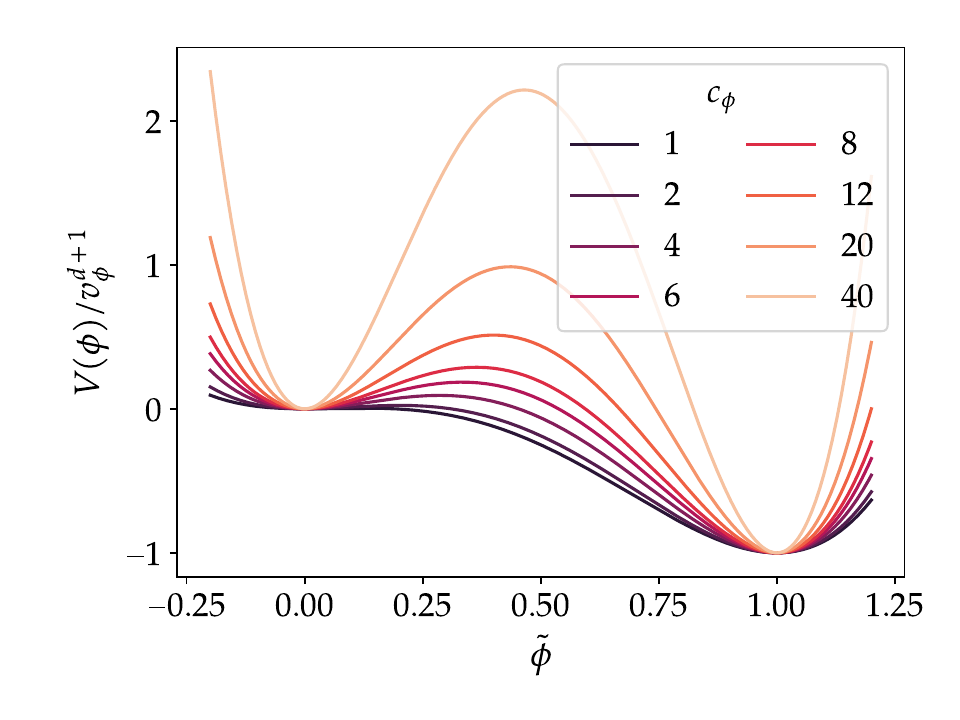}
\end{subfigure}
\begin{subfigure}[t]{0.495\linewidth}
\centering
\includegraphics[width=\linewidth]{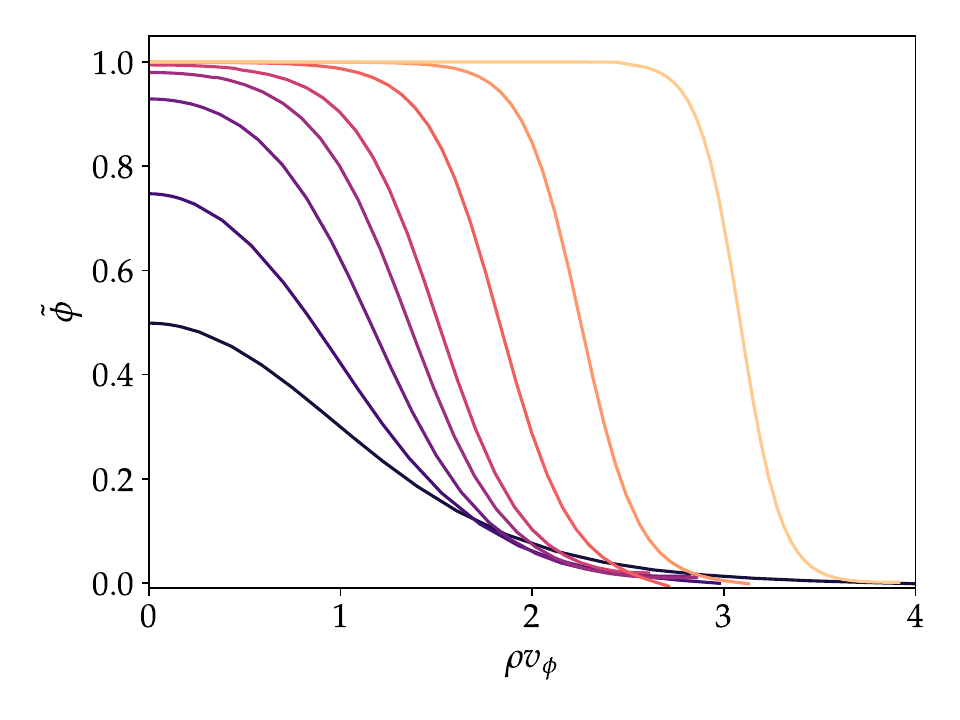}
\end{subfigure}
\caption{Left: scalar potential of eq.~\eqref{eq:toypotential} for
different values of $c_\phi$. Right: corresponding critical bubble
profiles obtained with \texttt{BubbleProfiler}~\cite{Athron:2019nbd}.
Higher potential barriers result in thinner walls and larger critical
bubbles.}
\label{fig:choiceofpotential}
\end{figure}

\paragraph{Set-up and toy potential.}
We focus on a scalar potential with two minima: a metastable minimum at
$\phi=0$ and a global minimum at $\phi=v_\phi$, separated by a potential
barrier. In $d=3$ spatial dimensions the field and $v_\phi$ have mass
dimension one. Such a potential typically arises from temperature
corrections in a thermal plasma~\cite{Athron:2022mmm}: the symmetric
minimum at $\phi=0$ is global for $T\gg v_\phi$ and becomes metastable
below a critical temperature $T_c$, so that the field can remain
trapped ({supercooling}) until thermal or quantum fluctuations
nucleate bubbles. We work directly in the supercooled regime,
where plasma and gradient energies are negligible compared to the
latent heat $\Delta V = V(0)-V(v_\phi)$~\cite{Caprini:2019egz}, so that
the universe starts from vacuum-energy domination, $\omega=-1$. For
concreteness, we use the temperature-independent
parametrisation~\cite{Jinno:2019bxw}
\begin{equation}
V(\phi) = c_\phi v_\phi^{2}\phi^{2}
-(2c_\phi+4)\,v_\phi\,\phi^{3}
+(c_\phi+3)\,\phi^{4} + \Delta V \; ,
\label{eq:toypotential}
\end{equation}
where $c_\phi$ tunes the barrier height while keeping the minima fixed
at $\phi=0$ and $\phi=v_\phi$ (figure~\ref{fig:choiceofpotential} left).
Modifying the parameter $c_\phi$ also changes the effective mass of the scalar field close to the two minima. In the true vacuum (close to the global minimum), it is given by $m_{\phi,\mathrm{true}}^2= 2(6+c_\phi)v_\phi^2$, while in the symmetric phase (close to $\phi = 0$) one finds $m_{\phi,\mathrm{false}}^2= 2c_\phi v_\phi^2$. In the following, we use the notation $m_\phi \equiv m_{\phi,\mathrm{true}}$.

\paragraph{Bounce solution and critical profile.}
Bubble nucleation is controlled by the Euclidean action of the bounce
solution~\cite{Coleman:1977py, Linde:1981zj}. For vanishing
temperature the bounce is $O(4)$-symmetric with action
\begin{equation}
S_E[\phi] = 2\pi^{2}\!\int_{0}^{\infty}\!d\rho\,\rho^{3}\!
\left[\frac{1}{2}\!\left(\frac{d\phi}{d\rho}\right)^{2}+V(\phi)\right],
\quad \rho=\sqrt{\tau^{2}+\vec x^{2}} \; .
\end{equation}
The smallest radius for which a bubble does not collapse is
the critical radius $R_c$, which we obtain using
\texttt{BubbleProfiler}~\cite{Athron:2019nbd}. The resulting profiles
are well fitted by
\begin{equation}
\label{eq:wall_profile}
\phi_\mathrm{nuc}(x) =
\frac{\phi_0}{2}\!\left[\tanh\!\left(\frac{R_c-x}{l_w}\right)+1\right] \; ,
\end{equation}
where $R_\mathrm{c}$ is the critical radius, $l_\mathrm{w}$ is the wall thickness and $\phi_0$ is the value the field tunnels to, which can be considerably smaller than $v_{\phi}$ for thick bubbles.
We fit these parameters as a function of $c_\phi$ and obtain the bubble profiles shown in the right panel of figure~\ref{fig:choiceofpotential}. As expected, higher potential barriers lead to thinner bubble walls, because larger gradients become more favourable. Unless otherwise stated we fix $c_\phi=4$. The
robustness of our results against this choice is discussed in
appendix~\ref{app:c_dependence}. Since the critical profile corresponds to a static solution, we choose an initial radius $R_n$ slightly larger than the critical radius in order to obtain expanding bubbles.

\paragraph{Wall Lorentz factor.}
A FOPT produces bubbles of true vacuum that
nucleate, expand, and eventually collide. The typical separation between
nucleation sites is characterised by the mean bubble separation $R_s$
or, equivalently, by the mean radius at collision
$R_\star = R_s/2 \simeq n_b^{-1/d}$, where $n_b$ is the bubble number
density~\cite{Cutting:2020nla}. Combined with the wall velocity $v_w$,
the scale $R_\star$ sets the duration of the transition, generally
parametrised by the quantity
$\beta\simeq \pi^{1/3}/R_\star$~\cite{Enqvist:1991xw,Hindmarsh:2019phv}.

The dynamics of an expanding bubble is set by the competition between
the driving pressure released across the wall,
\begin{equation}
\Delta V \,\equiv\, V(\phi=0) - V(\phi=v_\phi)\,,
\label{eq:DV}
\end{equation}
and the friction exerted by the surrounding plasma, whose dependence on
the wall Lorentz factor $\gamma_w$ controls the fate of the wall. At
leading order, particles crossing the wall and acquiring a mass inside
the bubble produce a $\gamma_w$-independent friction
pressure~\cite{Bodeker:2009qy}
\begin{equation}
\mathcal{P}_{\rm LO} \,\simeq\, \frac{g_\star}{24}\,\Delta m^2\,T_n^2\;,
\label{eq:PLO}
\end{equation}
where $\Delta m^2$ is the average mass gained by the plasma species
across the wall and $g_\star$ denotes the effective relativistic degrees of freedom of the thermal bath. Since $\mathcal{P}_{\rm LO}$ does not grow with the
wall velocity, the wall can overcome it and accelerate to
ultra-relativistic velocities only if
$\Delta V > \mathcal{P}_{\rm LO}$, the Bödeker-Moore condition, which
in terms of the the latent heat fraction $\alpha = \Delta V/\rho_{\rm rad}$ reads
$\alpha \gtrsim \alpha_{\rm LO} \equiv 0.1\,(\Delta m/T_n)^2$.

When this condition is satisfied, the wall starts to accelerate and its
Lorentz factor grows linearly with the bubble radius. In the absence of
any further friction, energy conservation between the released vacuum
energy and the wall kinetic energy implies that the wall reaches a
Lorentz factor~\cite{Ellis:2019oqb}%
\footnote{Consider a single spherical bubble of true vacuum with radius
$R$ in $d$ spatial dimensions, bounded by a thin wall of surface tension
$\sigma$ that expands into the false vacuum with Lorentz factor
$\gamma_w$. With $c_d \equiv \pi^{d/2}/\Gamma(d/2+1)$ the volume of the
unit $d$-ball, the wall area reads $S_{d-1}(R) = d\,c_d\,R^{d-1}$ and
the bubble volume reads $V_d(R) = c_d\,R^d$, so that the total energy
relative to the homogeneous false vacuum is
$E_{\rm tot}(R,\gamma_w) = d\,c_d\,\sigma\,\gamma_w\,R^{d-1} -
c_d\,\Delta V\,R^{d}$, with $\Delta V$ the latent heat. Defining the
nucleation radius by the zero-energy condition $E_{\rm tot}(R_n,1) = 0$,
$R_n = d\,\sigma/\Delta V$, and imposing energy conservation in the
runaway regime $E_{\rm tot}(R,\gamma_w) = E_{\rm tot}(R_n,1) = 0$, one
obtains $\gamma_w(R) = R/R_n$, valid for any $d \geq 1$ including $d=1$,
where $R_n = \sigma/\Delta V$ remains finite although the thin-wall
critical radius vanishes. The bubble radius at nucleation $R_n$, solution of
$E_{\rm tot}(R_n,1)=0$, is related to the critical radius $R_c$, solution of
$E'_{\rm tot}(R_c,1)=0$, by $R_c=(d-1)R_n/d$.}
\begin{equation}
\gamma_{\rm run} \,\simeq\, \frac{R_\star}{R_n}\,\simeq \,  4.2 \times 10^{10}~ \left(\frac{10^3~\rm TeV}{v_\phi}\right)\left(\frac{10 \,T_n}{v_\phi}\right)\left(\frac{v_\phi}{\Delta V^{\!\frac{1}{4}}}\right)^2\left(\frac{R_\star H}{0.1}\right)\left(\frac{1}{R_n T_n}\right)\,,
\label{eq:gammarun}
\end{equation}
where $R_\star$ is the bubble radius at collision and $R_n$ is the bubble radius at nucleation
given by the bounce solution. Eq.~\eqref{eq:gammarun} defines what
we will refer to as the {runaway regime}, in which plasma friction
is negligible and the wall keeps accelerating without bound until
collision.

A residual friction can however arise beyond leading order in scenarios with non-zero gauge coupling $g_D$. Charged
particles crossing the wall radiate soft gauge bosons in a process
analogous to transition radiation, producing a next-to-leading-order
friction pressure that grows linearly with the wall Lorentz factor and,
once leading logarithms are resummed for $T_n \ll v_\phi$, takes the form~\cite{Bodeker:2017cim,
Hoche:2020ysm, Azatov:2020ufh, Gouttenoire:2021kjv, Azatov:2023xem,Ai:2025bjw}
\begin{equation}
\mathcal{P}_{\rm LL}(\gamma_w) \,\simeq\, \gamma_w\,\alpha_D\,\Delta
m_V\,T_n^3\,\log(v_\phi/T_n)\,,
\label{eq:PLL}
\end{equation}
with $\alpha_D \equiv g_D^2/(4\pi)$ and $\Delta m_V$ the vector-boson
mass acquired in the broken phase. Because $\mathcal{P}_{\rm LL}$ grows
linearly with $\gamma_w$ while $\Delta V$ stays constant, the two
pressures eventually balance and define a terminal Lorentz
factor~$\gamma_{\rm LL}$ through\footnote{Additional effects, such as a
non-monotonic peak in the friction around the Jouguet velocity from
hydrodynamic obstruction~\cite{Konstandin:2010dm, Cline:2021iff,
Laurent:2022jrs, DeCurtis:2023hil, Ai:2024shx, Krajewski:2024gma} or
back-reaction from out-of-equilibrium particle shells trailing the
wall~\cite{Baldes:2024wuz}, can modify the form of the friction
pressure. Since the present work focuses on the runaway regime, none of
these effects directly affect our analysis.}
\begin{equation}
\mathcal{P}_{\rm LL}(\gamma_{\rm LL}) \,\simeq\, \Delta V \; ,
\label{eq:gammaLL_equation}
\end{equation}
that the wall would reach in the absence of collisions. Plugging eq.~\eqref{eq:PLL} into eq.~\eqref{eq:gammaLL_equation} leads to
\begin{equation}
\gamma_{\rm LL} ~\simeq~ \frac{1.2 \times 10^{7}}{\log(v_\phi/T_n)}~\left(\frac{0.01}{g_D}\right)^3\left(\frac{v_\phi}{T_n}\right)^3\left(\frac{g_D v_\phi}{\Delta m_V}\right)\left(\frac{\Delta V^{\!\frac{1}{4}}}{v_\phi}\right)^4 \; .
\label{eq:gammaLL}
\end{equation}
The actual Lorentz factor at collision is therefore set by the minimum of the
runaway and terminal values, 
\begin{equation}
    \gamma_\star\,\simeq\,\textrm{Min}\left[\gamma_{\rm run},~\gamma_{\rm LL}\right] \; ,
    \label{eq:gamma_star}
\end{equation}
and the two regimes correspond to the two
extremes $\gamma_{\rm run} \ll \gamma_{\rm LL}$ (runaway-dominated) and
$\gamma_{\rm LL} \ll \gamma_{\rm run}$ (friction-dominated). In the
friction-dominated case the wall reaches~$\gamma_{\rm LL}$ before
collision, the latent heat is mostly transferred to the plasma in the
form of relativistic shells and shock waves~\cite{Espinosa:2010hh}, and
only a fraction $\gamma_{\rm LL}/\gamma_{\rm run}$ of $\Delta V$ remains
stored in the wall gradient.
We focus instead on the runaway regime, $\gamma_{\rm run} \ll
\gamma_{\rm LL}$, in which plasma interactions are negligible and the
bubbles effectively expand into vacuum, a regime that arises naturally
in strongly supercooled transitions where the vacuum energy dominates
the energy budget~\cite{Lewicki:2022pdb} or for FOPTs associated to the spontaneous breaking of a global symmetry group for which the gauge coupling vanishes $g_{D}\to 0$. In this limit the walls keep
accelerating, the wall energy grows proportionally to $\gamma_w$ and the
wall width contracts as $l_w \propto (\gamma_w m_\phi)^{-1}$, leading to 
\begin{equation}
 \gamma_\star\,\simeq \, \frac{R_\star}{R_n} \qquad \text{(runaway regime)} \; .
    \label{eq:gammaR}
\end{equation}
We will explicitly
confirm this relation below and show that the ratio $R_\star/R_n$ is the
key parameter needed to characterise strongly supercooled phase
transitions in the runaway regime. In the following, starred variables correspond to values evaluated at collision time.

\paragraph{Distribution of energy after collision.}
After bubble collisions, the wall energy is converted into fluctuations of the scalar field about the true minimum. The subsequent EoS depends on how this energy is distributed among field modes, which is not fixed by energy conservation alone. Before presenting our result, we first discuss two naive limiting expectations. For slow walls, spatial gradients may be expected to remain subdominant, so that the field oscillates almost coherently around the true minimum and behaves as matter, with $\omega\simeq 0$~\cite{Ellis:2020nnr}. For ultra-relativistic walls, Lorentz contraction may instead deposit most of the energy into modes with momenta up to $k\sim\gamma_\star m_\phi$, so that, when the relativistic modes with $k\gg m_\phi$ dominate, the post-collision state is controlled by kinetic and gradient energy and approaches the radiation-like EoS $\omega\simeq 1/d$. If, in addition, self-interactions were sufficiently efficient, they would redistribute energy among modes and could rapidly drive the excited spectrum toward thermal equilibrium. We show below that neither limiting expectation is realised exactly. Instead, the collision populates both relativistic and non-relativistic modes, while self-interactions do not immediately thermalise the spectrum. The post-collision state therefore has an effective EoS interpolating between matter and radiation, $0<\omega<1/d$, with a value that depends on $\gamma_\star$. The functional form of $\omega(\gamma_\star)$ can be motivated analytically, but contains free parameters that must be calibrated by simulations. This motivates the lattice study presented next.

\section{Implementation on the lattice}
\label{sec:lattice}

\paragraph{Discretisation.}
We simulate the evolution and collision of vacuum bubbles on a
discretised lattice in $1{+}1$, $2{+}1$ and $3{+}1$ dimensions with
periodic boundary conditions
$\phi(\vec x)=\phi(\vec x+L\hat e_i)$, where $\hat e_i$ are the Cartesian
basis vectors and $L$ the lattice length, so that the simulation volume
is $L^{d}$. The lower dimensional simulations allow a wider dynamical
range and evade some resolution
constraints~\cite{Baeza-Ballesteros:2025tme}. We use $N$ grid points per
spatial dimension (so $N^{d}$ in total), grid spacing $\Delta x=L/N$,
and time step $\Delta t<\Delta x/\sqrt{d}$. Eq.~\eqref{eq:eom_EL} is
rewritten as two coupled first-order equations for $\phi$ and its
conjugate momentum $\pi\equiv\dot\phi$,
\begin{equation}
\partial_t\pi \,=\, \partial_i\partial^{i}\phi - V_{,\phi} \; ,
\qquad
\partial_t\phi \,=\, \pi \; ,
\label{eq:discrete-eom}
\end{equation}
with $\partial_i\partial^{i}$ the $d$-dimensional Laplacian. Following
ref.~\cite{Figueroa:2020rrl} we evaluate gradients with central finite
differences and integrate in time using a staggered leapfrog scheme. Our simulations are
performed with {\tt CoolBubble}~\cite{CoolBubble}, a lattice code developed for this work in \texttt{C++}, which evolves colliding vacuum
bubbles and computes the resulting equation of state. Although most of our simulations are performed in a static background, in section~\ref{subsec:returntoMD} we include cosmic expansion to study the late-time evolution of the scalar field. We then solve eq.~\eqref{eq:eom_EL-Hubble}, which takes the same form as eq.~\eqref{eq:discrete-eom} after redefining the conjugate momentum as $\pi \equiv a^{d}\dot\phi$. The scale factor $a(t)$ and its time derivative $\dot a(t)$ are obtained self-consistently from the Friedmann equations.

\paragraph{Units and initial conditions.}
Changing $d$ also changes the mass dimension of $\phi$. We always take
$v_\phi$ to have mass dimension one and rescale
$\tilde\phi\equiv\phi/v_\phi^{(d-1)/2}$, so that the potential reads
\begin{equation}
V(\tilde\phi) = v_\phi^{d+1}\!
\left[c_\phi\,\tilde\phi^{2} - (2c_\phi+4)\,\tilde\phi^{3}
+ (c_\phi+3)\,\tilde\phi^{4}\right] + v_\phi^{d+1} \; ,
\end{equation}
with global minimum at $\tilde\phi=1$ and latent heat
$\Delta V=v_\phi^{d+1}$. In $1{+}1$ dimensions in particular $\phi$ is
dimensionless. We set $v_\phi=1$ and, for our default $c_\phi=4$,
$m_\phi=\sqrt{20}$.\footnote{Direct comparison between different
$c_\phi$ requires adjusting the grid spacing because larger $c_\phi$
produces thinner walls and larger initial bubbles, see
appendix~\ref{app:c_dependence}.} As initial condition we set $\phi=0$ and
$\pi=0$ everywhere except inside a small number of bubbles
placed at random positions, each with the profile fitted from
$R_c(c_\phi)$, $\phi_0(c_\phi)$ and $l_w(c_\phi)$ of
section~\ref{subsec:FOPT}, but with a slightly supercritical radius $R_n=1.3\,R_c$ to ensure expansion.
The mean bubble separation, and hence $R_\star$, is varied by changing
$L$ at fixed $\Delta x$ (so $N$ scales accordingly). The largest
ratios $R_\star/R_n$ are obtained with a single bubble that collides
with its own periodic image.

\subsection{Simulation output}
\label{subsec:SimulationResults}

Before analysing our results in a quantitative way, let us first discuss the qualitative features seen in the simulations.

\begin{figure}[t]
\centering
\begin{minipage}{0.42\linewidth}
\centering
\begin{subfigure}[t]{\linewidth}
\includegraphics[width=\linewidth]{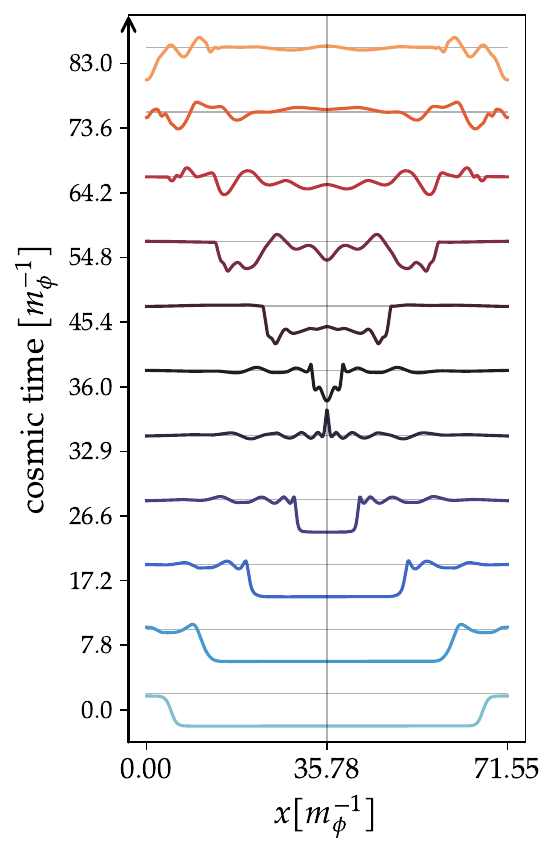}
\caption{Bubble profiles in space and time. Light blue: initial
configuration. Red/orange: late-time evolution.}
\end{subfigure}
\end{minipage}
\hfill
\begin{minipage}{0.52\linewidth}
\centering
\begin{subfigure}[t]{\linewidth}
\includegraphics[width=\linewidth]{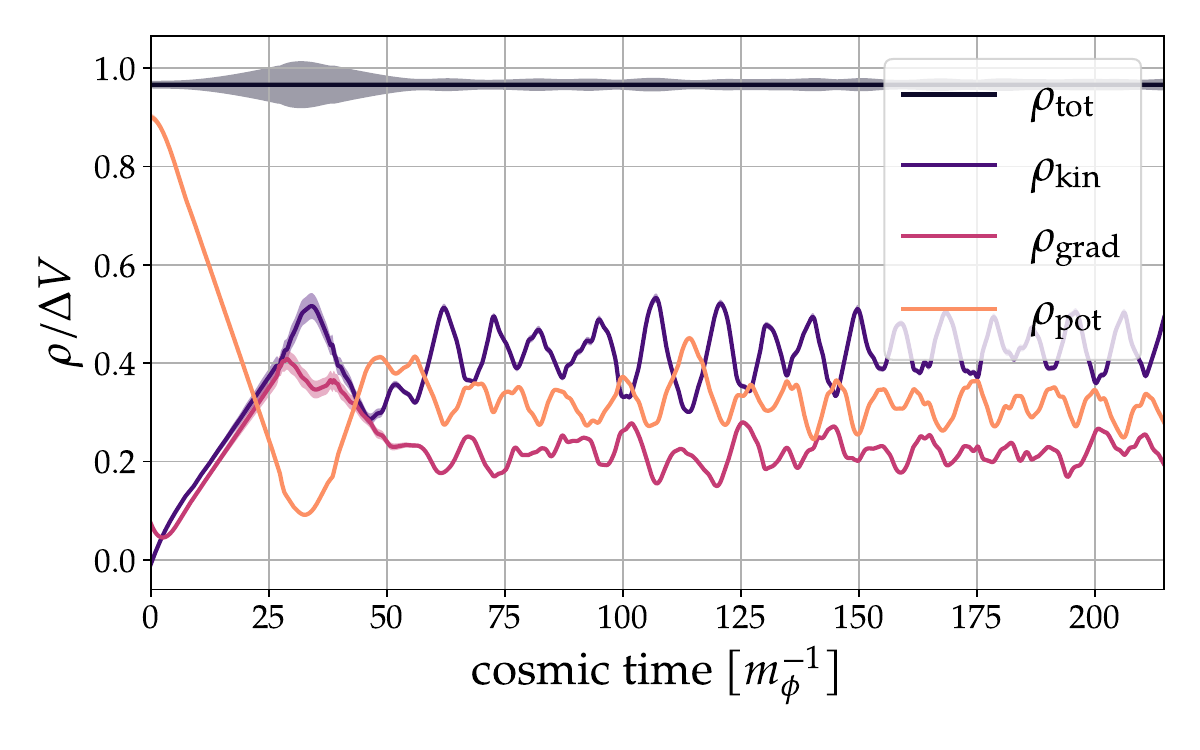}
\caption{Fractional energy densities.}
\end{subfigure}
\begin{subfigure}[t]{\linewidth}
\includegraphics[width=\linewidth]{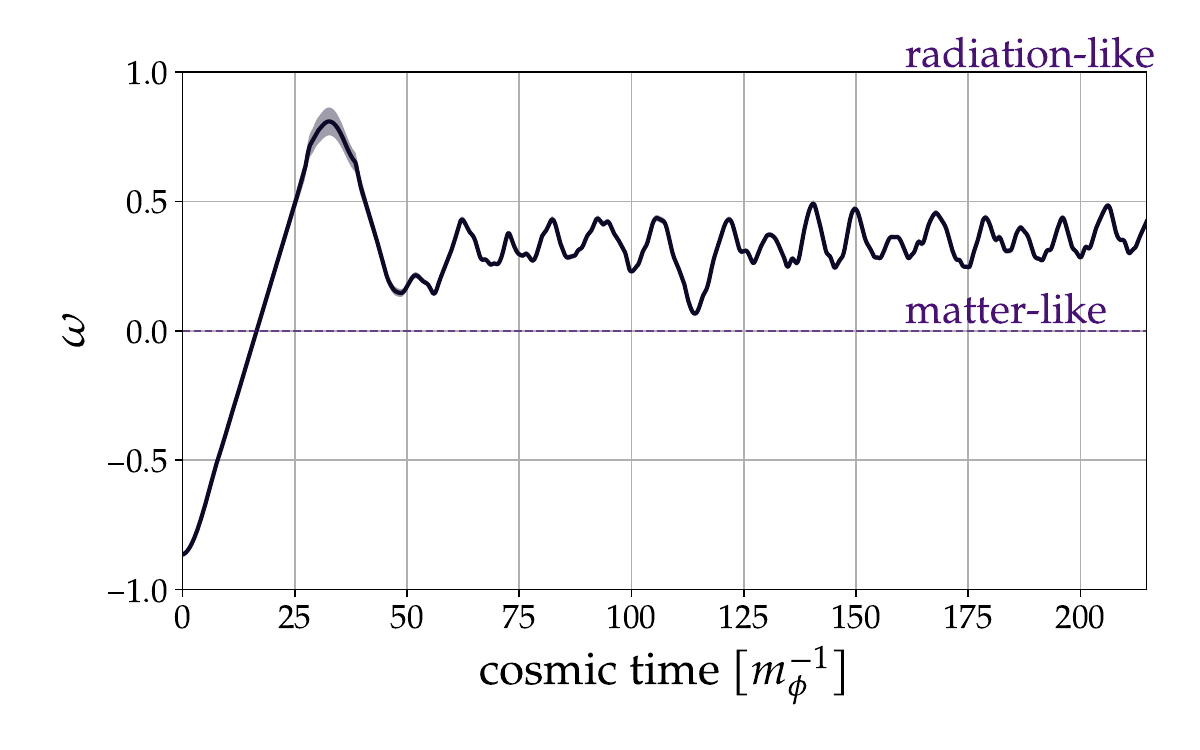}
\caption{EoS parameter. The radiation value in $d=1$ is
$\omega=1$.}
\end{subfigure}
\end{minipage}
\caption{Two-bubble collision in $1{+}1$ dimensions in a static
background with $L=16\simeq 71.55/m_\phi$, $\Delta t=0.001$ and
$N=8192$. After collision the walls dissipate and the field oscillates
around the true vacuum.}
\label{fig:1D_overview}
\end{figure}

\paragraph{Bubble collision in $1{+}1$ dimensions.}
The simulation is initialised with one bubble with a nucleation centre at $x=0$ and its periodic image at $x=L$, which then grows and collide at $x=L/2$. Since the bubbles expand with $v_w \simeq 1$, the collision time is $t_\star\simeq L/2$. The left panel of figure \ref{fig:1D_overview} shows an example of the field evolution for $L = 16 \simeq 71.55 / m_\phi$, $\Delta t = 0.001$ and $N = 8192$. 

The right panels show the corresponding evolution of the fractional energy densities (top) and the EoS parameter (bottom). 
At the beginning of the simulation, the energy is dominated by vacuum energy, corresponding to $\omega \simeq -1$. As the bubbles expand, the potential energy decreases and the gradient and kinetic energies grow  the collision time $t=t_\star$. Correspondingly, $\omega$ grows to positive values. After the bubble collision, the system oscillates at roughly constant average fractions of kinetic, gradient and potential energy. Hence, $\omega$ also oscillates around a constant value, which lies between the case of matter ($\omega = 0$) and radiation ($\omega = 1$). 

\begin{figure}
\centering
\begin{subfigure}{\linewidth}
\centering
\includegraphics[width=\linewidth]{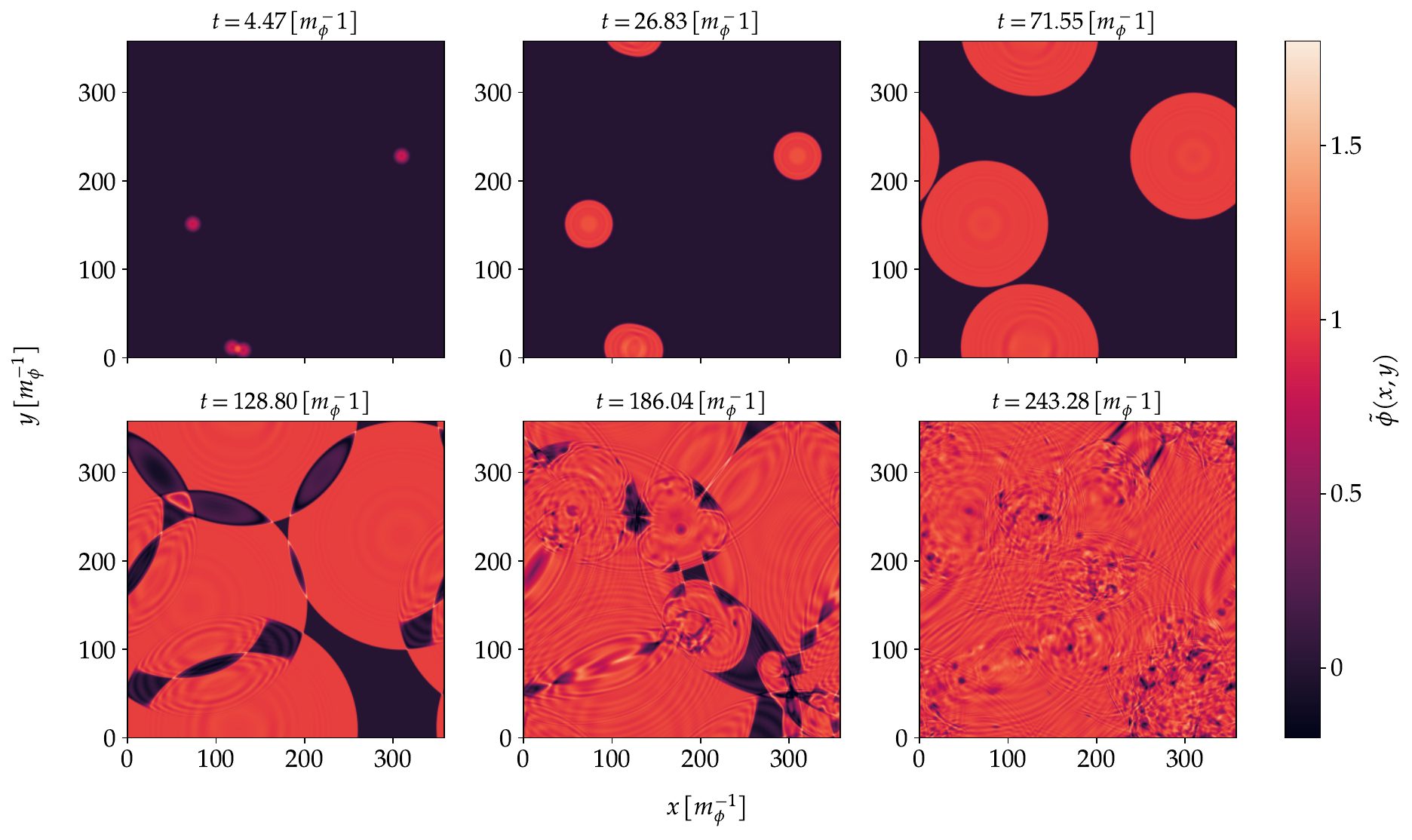}
\caption{Field configuration at different times, showing the
expansion, collision and subsequent evolution.}
\end{subfigure}
\begin{subfigure}{0.495\linewidth}
\includegraphics[width=\linewidth]{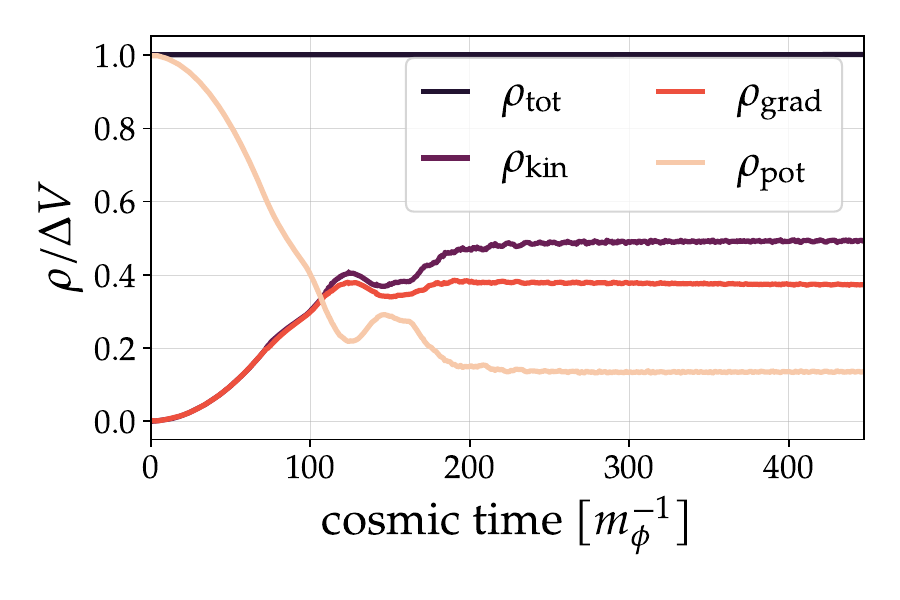}
\caption{Fractional energy densities.}
\end{subfigure}
\begin{subfigure}{0.495\linewidth}
\includegraphics[width=\linewidth]{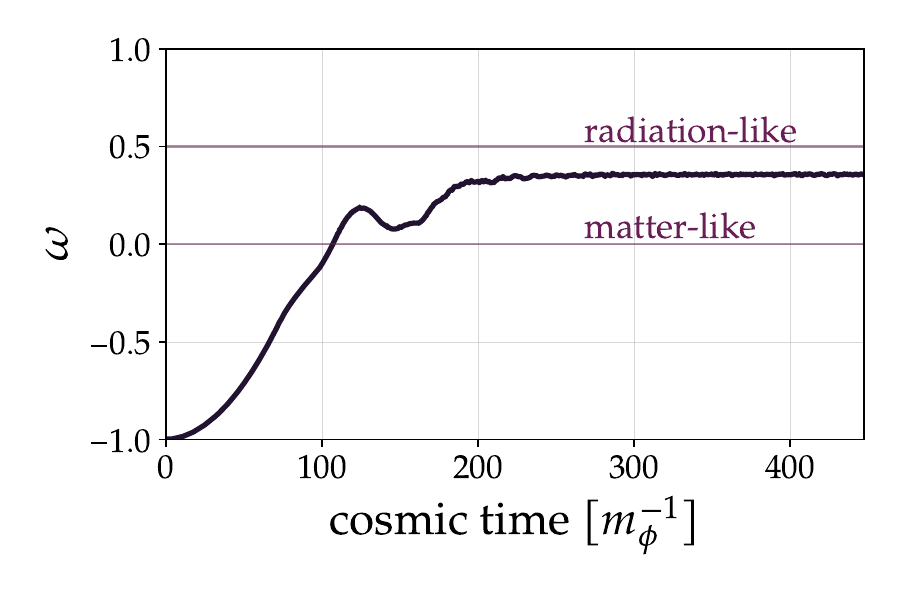}
\caption{EoS parameter.}
\end{subfigure}
\caption{Four randomly nucleated bubbles in $2{+}1$ dimensions in a
static background, with $L=80$, $\Delta t=0.002$, $N=20\,000$ and
$c_\phi=4$.}
\label{fig:2Dsimulation}
\end{figure}

\begin{figure}
\centering
\begin{subfigure}{0.49\linewidth}
\centering
\includegraphics[width=0.65\linewidth]{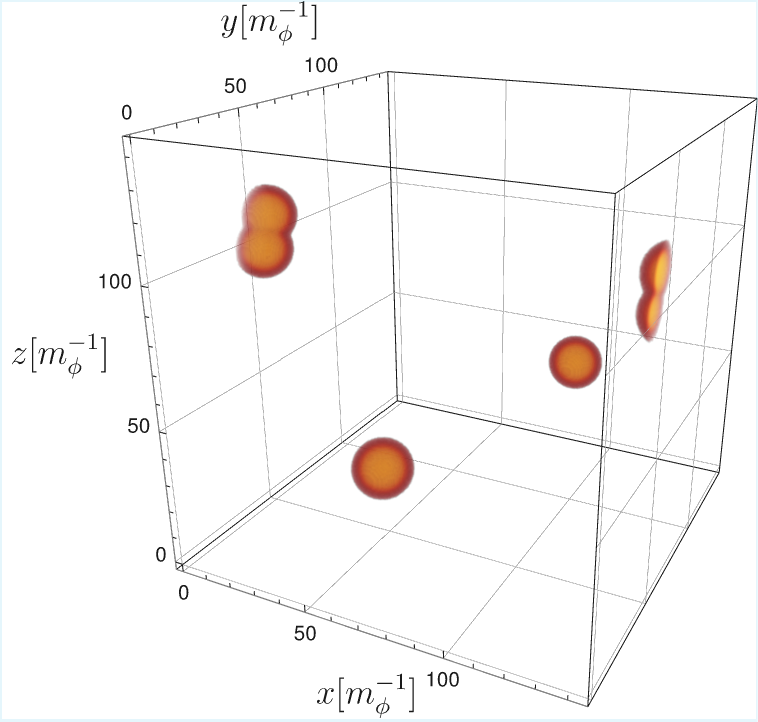}
\caption{Initial configuration ($t=0$).}
\end{subfigure}
\hfill
\begin{subfigure}{0.49\linewidth}
\centering
\includegraphics[width=0.65\linewidth]{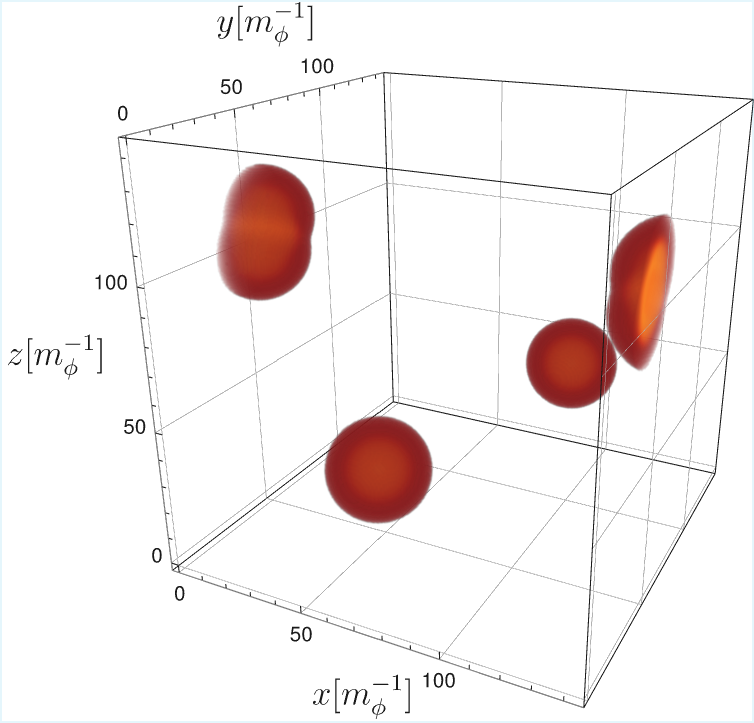}
\caption{Bubbles grow ($t=8.94\,m_\phi^{-1}$).}
\end{subfigure}
\caption{Early-stage evolution of the field configuration in three
spatial dimensions for four simultaneously nucleated bubbles with
periodic boundary conditions.}
\label{fig:3Dsimulation_3dplots}
\end{figure}

\begin{figure}[t]
\centering
\begin{subfigure}{\linewidth}
\centering
\includegraphics[width=\linewidth]{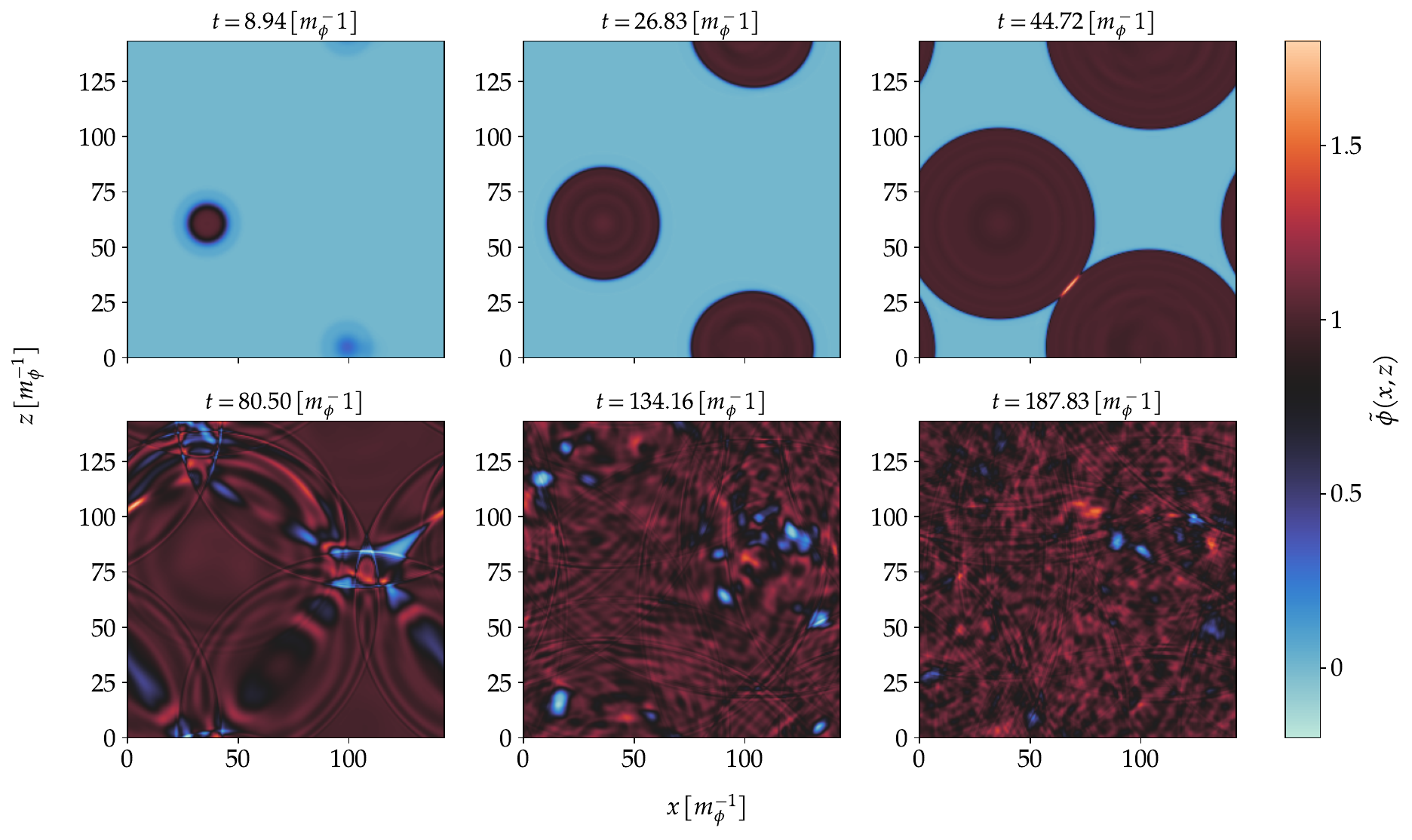}
\caption{Two-dimensional slice at $y=L/4$ at different times.}
\end{subfigure}
\begin{subfigure}{0.49\linewidth}
\includegraphics[width=\linewidth]{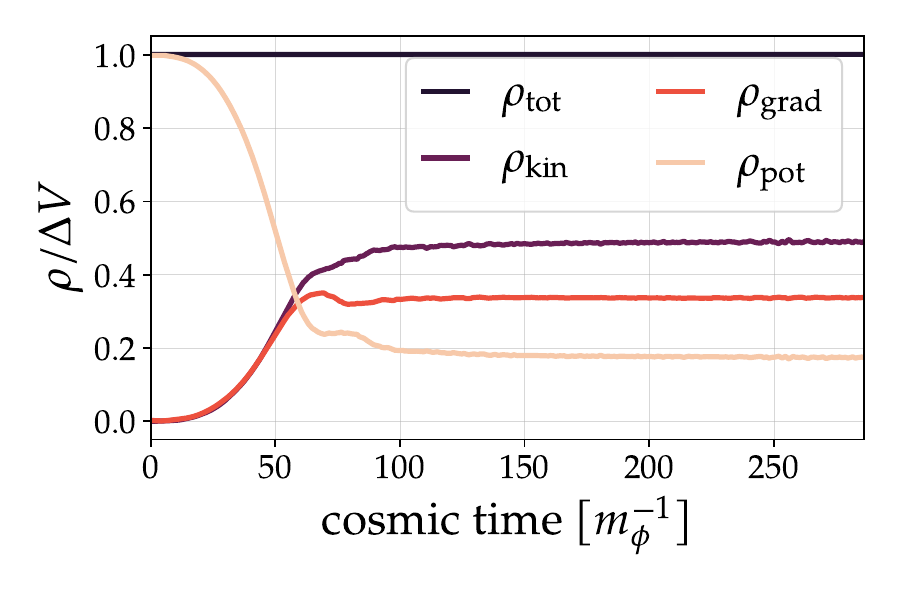}
\caption{Fractional energy densities.}
\end{subfigure}
\hfill
\begin{subfigure}{0.49\linewidth}
\includegraphics[width=\linewidth]{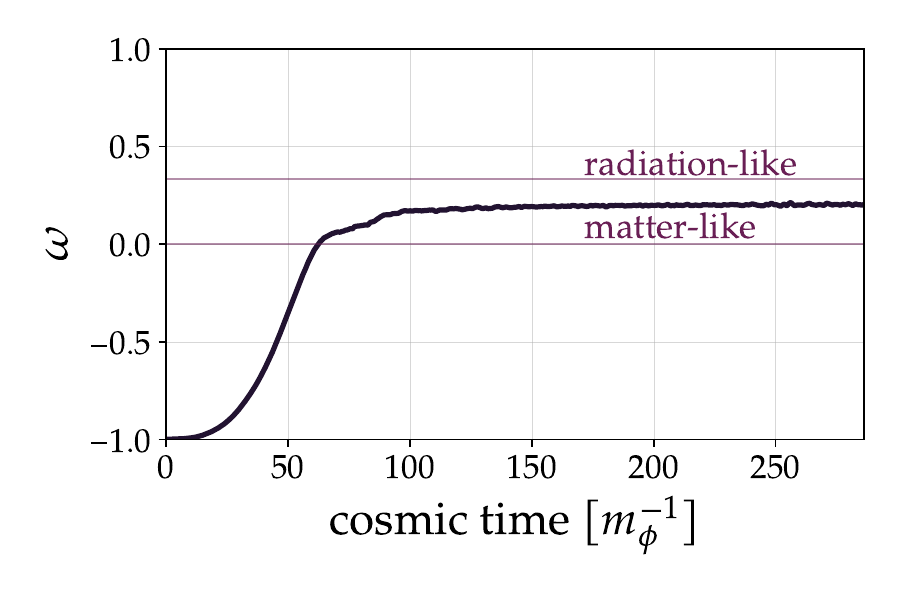}
\caption{EoS parameter.}
\end{subfigure}
\caption{Four randomly nucleated bubbles in $3{+}1$ dimensions in a
static background, with $L=32$, $\Delta t=0.004$, $N=1024$ and
$c_\phi=4$.}
\label{fig:3D_simu_overview}
\end{figure}

\paragraph{Randomly nucleated bubbles in $2{+}1$ and $3{+}1$ dimensions.}
In two and three spatial dimensions one can nucleate several bubbles
simultaneously at random positions but with identical profiles. If a bubble extends beyond the simulation volume, periodic boundary conditions are enforced by introducing image bubbles at the corresponding translated positions. To resolve the walls adequately we use at most five bubbles
per volume. Figure~\ref{fig:2Dsimulation} shows a $d=2$ simulation with
four bubbles, $L=80$, $\Delta t=0.002$, $N=20\,000$ and $c_\phi=4$.
In the first snapshots of the simulation, the bubbles expand with visible oscillations of the field inside the bubbles. The qualitative pre-collision dynamics resemble
the $d=1$ case, but shortly after the first collisions
{false-vacuum pockets} form, producing a transient increase in the potential energy
$\rho_\mathrm{pot}$ and a dip in the EoS $\omega$. The system then settles with
$\langle\omega\rangle\simeq 0.36$, and shows fewer oscillations in  $\rho$ and $\omega$ than in
$d=1$ as a result of the averaging of all quantities over two dimensions. 

For the $d = 3$ case, we have simulated four bubbles with $L = 32$, $\Delta t = 0.004$ , $N = 1024$ and $c_\phi = 4$. Figure~\ref{fig:3Dsimulation_3dplots} shows the field configuration in three-dimensional space at different points in time, while figure~\ref{fig:3D_simu_overview} shows two-dimensional slices at $y = L/4$, which can be more directly compared with the simulations in 2+1 dimensions. Indeed, the overall evolution is quite similar in both cases, and we obtain an approximately constant EoS parameter at late times also in 3+1 dimensions. However, while in the simulation shown in figure~\ref{fig:2Dsimulation}, the final fraction of gradient energy was around 40\%, it is closer to 30\% in the simulation shown in figure~\ref{fig:3D_simu_overview}. This is a direct consequence of the smaller value of $L$ in the $d = 3$ case, which corresponds to a smaller bubble separation $R_\star$ and therefore a smaller Lorentz factor at collision $\gamma_\star$, see eq.~\eqref{eq:gammaR}. Let us now study this relation in a more quantitative way.

\subsection{Equation of state for varying bubble separation}
\label{sec:EoSofgamma}

\begin{figure}[t]
\centering
\includegraphics[width=0.485\linewidth]{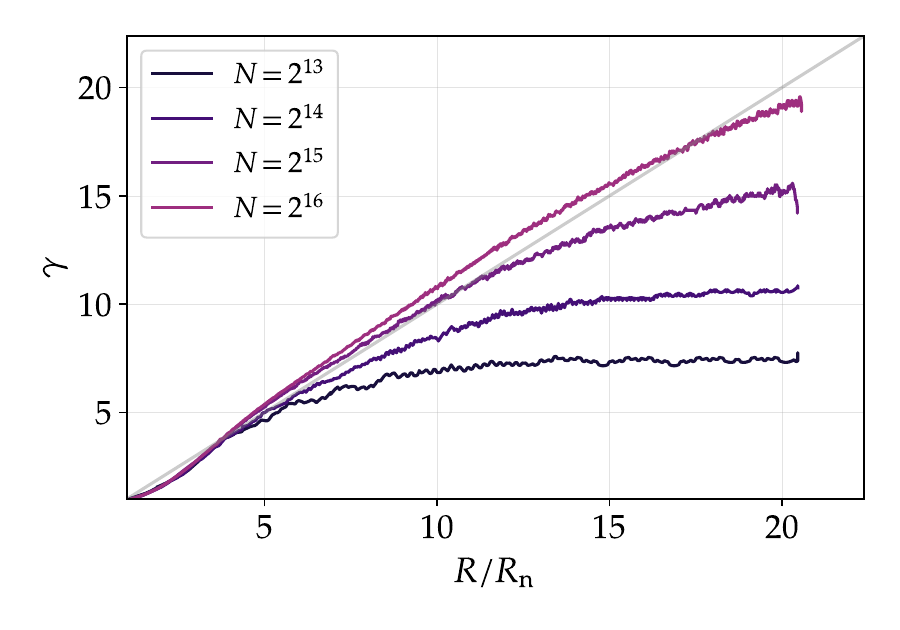}\hfill
\includegraphics[width=0.515\linewidth]{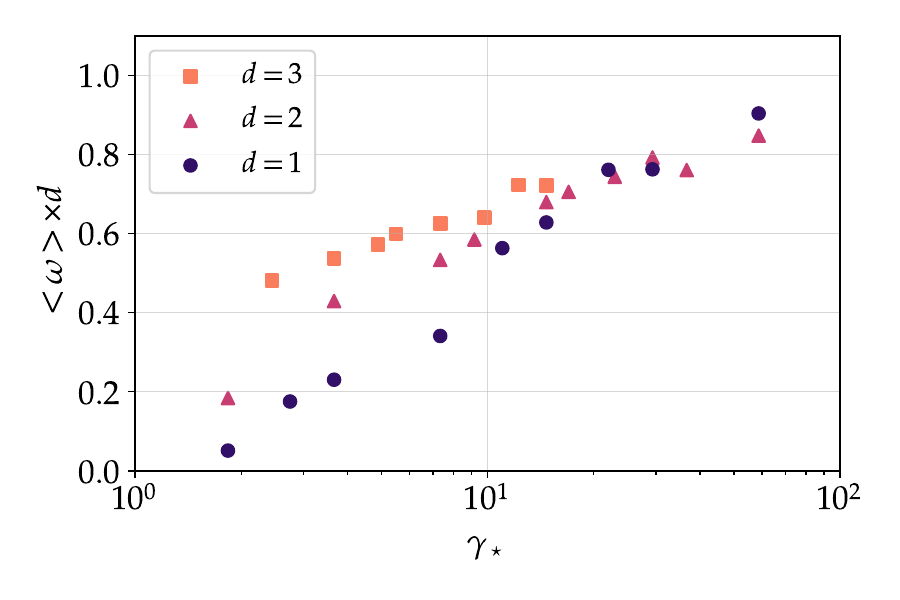}
\caption{Left: Lorentz factor inferred from the bubble-wall width as a
function of the bubble radius in $1{+}1$ dimensions, for different
grid resolutions at $L=64$ and $\Delta t=0.5\,\Delta x$. Right: averaged
EoS parameter after the collision as a function of the Lorentz factor
at collision, in $d=1,2,3$.}
\label{fig:Eosresults_and_Gamma}
\end{figure}

To test eq.~\eqref{eq:gammaR} directly we infer $\gamma$ from the
Lorentz contraction of the wall thickness: with $\gamma_\mathrm{ini}=1$
at nucleation and wall thickness $l_{w,\mathrm{ini}}$, the Lorentz
factor at later time is
\begin{equation}
\gamma(t) \,=\, \frac{l_{w,\mathrm{ini}}}{l_{w}(t)} \; .
\end{equation}
We determine $l_w(t)$ as the distance between two special points in the effective potential, see appendix~\ref{app:gamma_wall}. The left panel
of figure~\ref{fig:Eosresults_and_Gamma} shows the Lorentz factor $\gamma_\star$ in 1+1 dimension for $L=64$ and
several values of $N$, corresponding to different grid resolution. For the smaller values of $N$, the Lorentz factor saturates at some point, because the resolution is insufficient to resolve the wall width according to the procedure from appendix~\ref{app:gamma_wall}. Increasing the resolution, however, recovers the linear growth in eq.~\eqref{eq:gammaR}. 
The same conclusion can be reached from figure~\ref{fig:1D_overview}, which exhibits a linear growth of the kinetic and gradient energy fraction as a function of time before bubble collision. More generally, in $d$ dimensions the fraction of kinetic and gradient energy grow proportional to $t^d = R^d$, because the energy density of the bubble wall grows proportional to $R$ and the total surface area of the bubble grows proportional to $R^{d-1}$.

By varying the bubble separation, we can therefore vary the Lorentz factor at collision $\gamma_\star = R_\star / R_\text{ini}$. For this purpose, we consider a simplified set-up with a single bubble that expands until $R_\star=L/2$, at which point it collides with its own periodic image. The bubble separation can then be varied by changing the  lattice length $L$, keeping the spatial resolution $\Delta x$  fixed, as well as the initial profile of the bubbles and the shape of the effective potential ($c_\phi =4$). After the bubble collision, we can calculate the time-averaged EoS parameter $\langle \omega\rangle $. The time interval considered for this average is $t \in [L, 15L]$ for $d=1$, $t \in [L, 10L]$ for $d=2$  and $t \in [L, 5L]$ for $d=3$. Longer intervals are necessary for lower dimensional simulations, because there are fewer spatial directions to average over.

Our results are summarised in the right panel of figure~\ref{fig:Eosresults_and_Gamma}, which shows $\langle \omega \rangle \times d$ as a function of $\gamma_\star$.\footnote{In $d=2$ and $d=3$ the bubble collision takes a finite amount of time from the moment when the bubble walls first touch to the point when the true vacuum fills the entire simulation volume. During this time the bubble walls continue to accelerate, such that the Lorentz factor cannot be uniquely defined. We define $\gamma_\star$ as the Lorentz factor when the bubble walls first touch, which therefore gives a lower bound on the average Lorentz factor during the bubble collision.} We multiply the EoS parameter by $d$ such that the results from simulations in different dimensions become more comparable and the result is equal to unity for a radiation-like fluid in any dimension. Our results confirm the expectation that higher Lorentz factors at collision lead to a larger EoS parameter after collision. For $\gamma_\star < 10$ the inferred value of $\langle \omega \rangle \times d$ differs for the different dimensions, but for $\gamma_\star > 10$ the results become nearly independent of $d$. For $d=1$ and $d=2$, where we can make the simulation volume large enough to simulate $\gamma_\star \gg 10$, we find that $\langle \omega \rangle$ slowly approaches $1/d$. In the following section we will explain this asymptotic behaviour in terms of the power spectrum of the scalar field after the collision.

\section{Evolution of the scalar power spectrum}
\label{sec:PS-evol}

\subsection{Power spectrum formalism}
\label{subsec:PS}

In this section we derive an expression for $\omega$ in terms of the
Fourier modes of $\phi$, for which we will find analytical
approximations in terms of simple power laws. Unless stated otherwise
we consider $d=3$ and assume that the effective potential is
approximately quadratic close to the true minimum.

\paragraph{Fourier conventions and mode evolution.}
We treat $\phi(\vec x,t)$ as a statistically homogeneous and isotropic
(Gaussian) random field and adopt the Fourier convention
\begin{equation}
\phi(\vec x,t) \,=\, \int\!\frac{d^{3}k}{(2\pi)^{3}}\,
e^{i\mathbf{k}\cdot\mathbf{x}}\,\phi_{\mathbf k}(t) \, ,
\qquad
\phi_{\mathbf k}(t) \,=\, \int\!d^{3}x\,
e^{-i\mathbf{k}\cdot\mathbf{x}}\,\phi(\vec x,t) \; ,
\label{eq:phi_Fourier}
\end{equation}
so that reality of $\phi$ implies $\phi_{-\mathbf k}=\phi_{\mathbf k}^{\star}$.
The power spectrum $P_\phi(k)$ is defined by
\begin{equation}
\langle\phi_{\mathbf k}\,\phi_{\mathbf k'}^{\star}\rangle
\,=\, (2\pi)^{3}\,\delta^{(3)}(\mathbf k-\mathbf k')\,P_\phi(k) \; ,
\label{eq:power_spectrum_def}
\end{equation}
where $\left<\cdots\right>$ denotes the ensemble average. We work with the dimensionless
spectrum\footnote{$\Delta_\phi(k)$ has the same dimension as the
two-point correlator. It is made strictly dimensionless by the
rescaling $\phi\to\tilde\phi\equiv\phi/v_\phi$.}
$\Delta_\phi(k)\equiv k^{3}P_\phi(k)/(2\pi^{2})$,
in terms of which the real-space correlator is
\begin{equation}
\langle\phi(\vec 0)\,\phi(\vec x)\rangle
\,=\, \int_{0}^{\infty}\!\frac{dk}{k}\,j_{0}(kx)\,\Delta_\phi(k) \; ,
\label{eq:power_spectrum_Delta_def}
\end{equation}
with $j_{0}$ the spherical Bessel function of order zero. Setting
$\vec x=\vec 0$, we have
$\langle\phi^{2}\rangle=\int(dk/k)\,\Delta_\phi$.  In an expanding universe, the Fourier modes satisfy the equation of motion
\begin{equation}
    \ddot{\phi}_k + 3 H \dot{\phi}_k + \left(\frac{k^2}{a^2} + m_\phi^2\right) \phi_k = 0 \; ,
\end{equation}
which can be solved analytically using the WKB approximation for $H \ll m_\phi$:
\begin{align}
\label{eq:WKB}
    \phi_k(t) &\sim\frac{1}{a^{3/2} \sqrt{\omega_k}} \exp \left(\pm i \int \omega_k(t) dt\right)
\end{align}
with the dispersion relation $\omega_k^2 = {k^2}/{a^2} + m_\phi^2$. The slowly-varying prefactor can be neglected when taking time
derivatives so that $\langle|\dot\phi_{\mathbf k}|^{2}\rangle
\simeq \omega_{k}^{2}\,\langle|\phi_{\mathbf k}|^{2}\rangle$. The expression $\langle|\phi_{\mathbf k}|^{2}\rangle\sim a^{-3}/\omega_k$
asymptotes to
\begin{equation}
\langle|\phi_{\mathbf k}|^{2}\rangle \,\sim\, a^{-2}/k
\;\;\text{for}\;\; k/a\gg m_\phi \, ,
\qquad
\langle|\phi_{\mathbf k}|^{2}\rangle \,\sim\, a^{-3}/m_\phi
\;\;\text{for}\;\; k/a\ll m_\phi \; ,
\label{eq:redshift_modes}
\end{equation}
i.e.\ relativistic and non-relativistic modes redshift differently.
The transition between the two regimes is smooth on a scale
$\Delta\ln k \sim O(1)$. For simplicity, we will treat this transition as a step at $k/a=m_\phi$ in the following. This controlled approximation introduces $O(1)$ errors that can be absorbed into two fitting parameters as discussed below.

\paragraph{Energy densities and equation of state.}
Combining the WKB relation with statistical homogeneity, the average
kinetic energy density is
$\langle\rho_\mathrm{kin}\rangle=\tfrac{1}{2}\langle\dot\phi^{2}\rangle
=\tfrac{1}{2}\!\int d^{3}k/(2\pi)^{3}\,\omega_{k}^{2}\,P_\phi(k,a)$,
which after angular integration and substitution of $\Delta_\phi$ reads
\begin{equation}
\langle\rho_\mathrm{kin}\rangle
\,=\,\frac{1}{2}\!\int\!\frac{dk}{k}\,
\Big(m_\phi^{2}+\frac{k^{2}}{a^{2}}\Big)\,\Delta_\phi(k,a)\,.
\label{eq:EK_of_PS}
\end{equation}
The same manipulations applied to
$\langle(\nabla\phi)^{2}\rangle/(2a^{2})$ and
$\langle V\rangle=\tfrac{1}{2}m_\phi^{2}\langle\phi^{2}\rangle$ give
\begin{equation}
\langle\rho_\mathrm{grad}\rangle
\,=\,\frac{1}{2}\!\int\!\frac{dk}{k}\,\frac{k^{2}}{a^{2}}\,\Delta_\phi(k,a),
\qquad
\langle\rho_\mathrm{pot}\rangle
\,=\,\frac{1}{2}\!\int\!\frac{dk}{k}\,m_\phi^{2}\,\Delta_\phi(k,a),
\label{eq:EG_EV_of_PS}
\end{equation}
which by construction satisfy
$\langle\rho_\mathrm{kin}\rangle=\langle\rho_\mathrm{grad}\rangle
+\langle\rho_\mathrm{pot}\rangle$, in agreement with the $n=2$ virial
relation~\eqref{eq:Virial_quadratic}.  Substituting
eqs.~\eqref{eq:EK_of_PS}--\eqref{eq:EG_EV_of_PS} generalised to $d$ spatial
dimensions, into
eq.~\eqref{eq:EoS_Virial_quadratic}, yields the master formula
\begin{equation}
\boxed{\;
\omega \,=\, \frac{1}{d}\,
\frac{\displaystyle\int\!d\ln k\,(k/a)^{2}\,\Delta_\phi(k,a)}
{\displaystyle\int\!d\ln k\,\big[m_\phi^{2}+(k/a)^{2}\big]\,\Delta_\phi(k,a)}\,.\;}
\label{eq:omega_EoS_PS}
\end{equation}
Two limits are immediate: if
$\Delta_\phi$ is dominated by modes with $k/a\gg m_\phi$, numerator and
denominator coincide and $\omega\to 1/d$ (radiation-like). If instead the dominant contribution comes from
$k/a\ll m_\phi$, the numerator is parametrically suppressed and
$\omega\to 0$ (matter-like). The $a$-dependence of $\Delta_\phi$ and its impact on the evolution of the EoS is studied in section~\ref{subsec:derivation-omega-of-gamma}.

\subsection{Numerical results}
\label{subsec:numericalresults-PS}

\paragraph{Lattice power spectrum.}
We use the \texttt{fftw3} library~\cite{fftw3} to compute the discrete
Fourier transform of the field on the lattice. The finite lattice
length $L$ and grid spacing $\Delta x$ imply infrared and ultraviolet
cut-offs~\cite{Figueroa:2020rrl}
\begin{equation}
k_\mathrm{IR}=\frac{2\pi}{L} \, ,
\qquad
k_\mathrm{UV}=\frac{\pi}{\Delta x} \; ,
\end{equation}
so the discrete spectrum is meaningful only for
$k_\mathrm{IR}<k<k_\mathrm{UV}$. We normalise the discrete dimensionless
spectrum as~\cite{Baeza-Ballesteros:2025tme}
\begin{equation}
\Delta_\phi(k)
\,=\, \frac{k^{3}}{2\pi^{2}}\,\frac{1}{N^{3}}\,
\langle|\phi(k)|^{2}\rangle_{R(k)} \; ,
\end{equation}
where $\langle\cdot\rangle_{R(k)}$ denotes the shell average over modes
with $|\mathbf k|\in[k,k+\Delta k)$.

\begin{figure}
\centering
\includegraphics[width=0.85\linewidth]{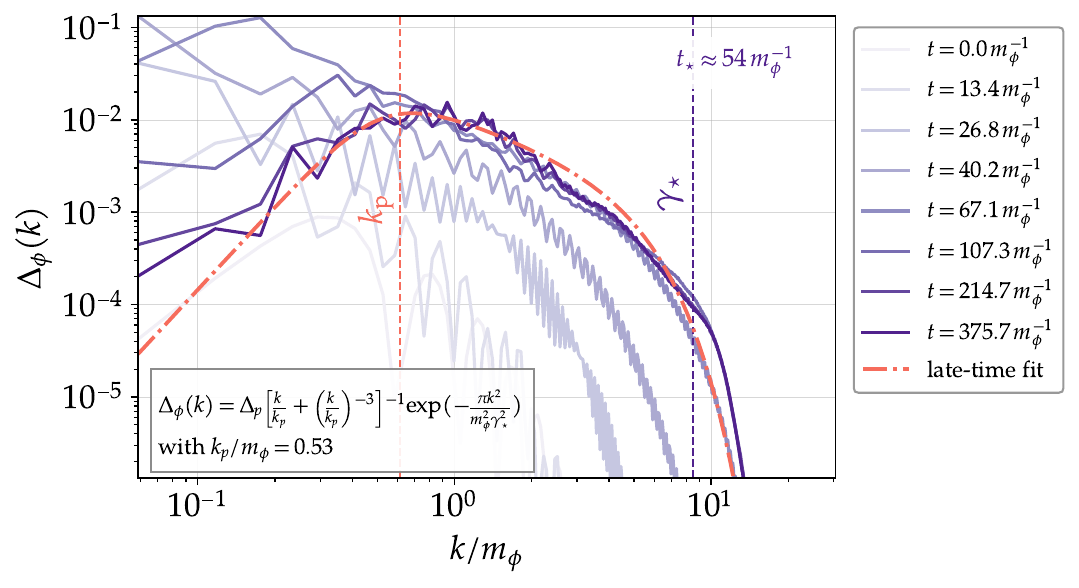}
\caption{Evolution of the scalar power spectrum for a single bubble configuration in $3{+}1$
dimensions. Darker curves correspond to later times of the evolution. The collision occurs at $t_\star=54\,m_\phi^{-1}$ and
the inferred Lorentz factor of the walls at collision $\gamma_\star$ is indicated. The analytical fit of eq.~\eqref{eq:analytical_smooth} and the corresponding best-fit value for $k_\mathrm{p}$ are shown in dot-dashed and dashed lines respectively.}
\label{fig:PS-Evol}
\end{figure}

\paragraph{Post collision power spectrum.}
Figure~\ref{fig:PS-Evol} illustrates the evolution of the power spectrum for a $3{+}1$ dimensional simulation of a single bubble colliding with its periodic images, as discussed in section~\ref{sec:EoSofgamma}. At early times, the field mainly populates low $k$ modes. Before the collision, the spectrum is determined by two characteristic length scales of the single expanding bubble: its radius $R$ and its wall width $l_w$. As detailed in appendix~\ref{app:analytical}, these scales dictate three distinct regimes. On large scales where $k \lesssim R^{-1}$, causality implies an infrared scaling of $\Delta_\phi \propto k^3$. At intermediate momenta where $R^{-1} \lesssim k \lesssim l_w^{-1}$, the spectrum inherits the scaling of an expanding bubble, $\Delta_\phi \propto k^{-1}$. Finally, modes in the UV where $k \gtrsim l_w^{-1}$ are strongly suppressed.
As the bubble expands, the power spectrum extends to increasingly higher values of $k$, reflecting the Lorentz contraction of the bubble wall width $l_w(t)= l_{w,0}/\gamma(t)$, which provides a UV cut-off that evolves with time. At the moment of collision, the bubble wall thickness is approximately $l_w \simeq (\gamma_\star m_\phi)^{-1}$, meaning the cut-off in momentum is located at
\begin{equation}
k_\star \simeq l_w^{-1} \simeq \gamma_\star m_\phi \; .
\end{equation}
After the collision, the UV cut-off and the relativistic branch ($k > k_p \simeq m_\phi$) stop evolving, remaining frozen at the scales established at the moment of collision. Meanwhile, the non-relativistic branch ($k < k_p$) rapidly settles towards its final causal $k^3$ stationary form, where $k_p$ acts as the turnover scale.
This asymptotic post-collision behaviour can be approximated by a broken power law:
\begin{equation}
\Delta_\phi(k) \simeq \Delta_p
\begin{cases}
(k/k_p)^{3} \, , & k \ll k_p \, , \\[2pt]
(k/k_p)^{-1}e^{- \pi k^2/k_\star^2} \, , & k_p \ll k \lesssim k_\star \; ,
\end{cases}
\label{eq:brokenPowerLaw}
\end{equation}
see appendix~\ref{app:analytical} for details.
In practice, the late-time numerical spectra shown in figure~\ref{fig:PS-Evol} are best reproduced by the smooth interpolation function
\begin{equation}
\Delta_\phi(k) = \Delta_p \left[ \left(\frac{k}{k_p}\right)^{-3} + \left(\frac{k}{k_p}\right) \right]^{-1} \exp\!\left[-\pi k^2/k_\star^2\right] \; ,
\label{eq:analytical_smooth}
\end{equation}
where the break momentum $k_p$ is treated as a free parameter, which we determine to be $k_p/m_\phi \simeq 0.5\text{--}0.7$ through fitting to our numerical results. The amplitude $\Delta_p$ can either be determined from a fit to the simulation results or from the requirement of energy conservation (see appendix~\ref{app:analytical} for details). Supplementary examples of spectra and fits are provided in appendix~\ref{app:NumericalPS}.

\begin{figure}[t]
\centering
\includegraphics[width=0.7\linewidth]{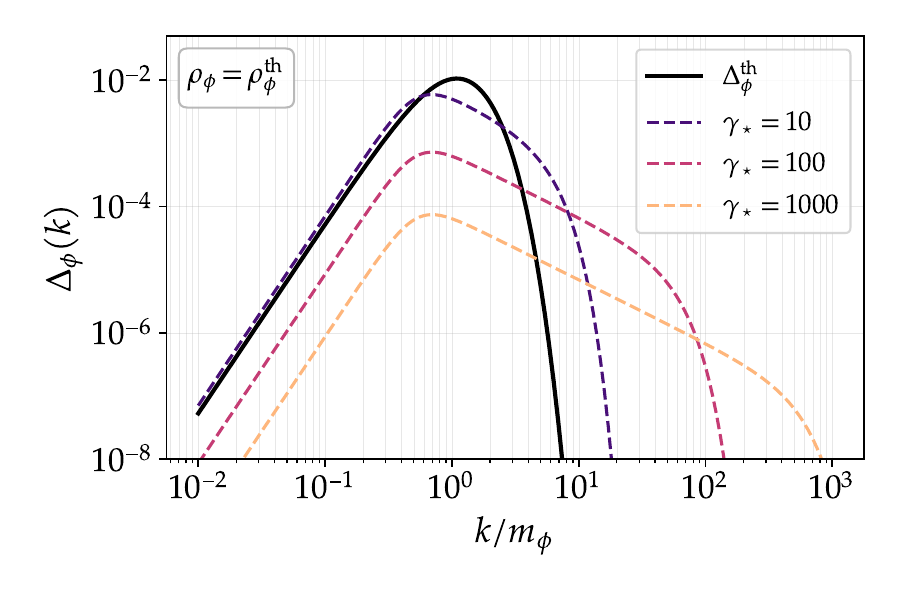}
\caption{Post-collision power spectra for different values of $\gamma_\star$  with peak $k_\mathrm{p}\simeq0.5\text{--}0.7\,m_\phi$ and cut-off $k_\star\simeq\gamma_\star m_\phi$ (dashed lines) against the thermal one as given in eq.~\eqref{eq:Delta_phi_th} of appendix~\ref{app:thermalisation} (solid black) at equal energy density $\rho_\phi=\rho_\phi^{\rm th}$. }
\label{fig:thermal_spectrum}
\end{figure}

\paragraph{Comparison with thermal equilibrium.}
Figure~\ref{fig:thermal_spectrum} compares this fit with the thermal spectrum $\Delta_\phi^{\rm th}$ at equal energy density, see eq.~\eqref{eq:Delta_phi_th} in appendix~\ref{app:thermalisation} for details. The thermal spectrum is peaked at $k\simeq m_\phi$ and Boltzmann-suppressed beyond, whereas the post-collision one extends up to the wall scale $k_\star\simeq\gamma_\star m_\phi$, holding ever more power in relativistic modes $k\gg m_\phi$ as $\gamma_\star$ grows. Thus, as anticipated in section~\ref{sec:EoS}, the collision does not thermalise the field. On the contrary, the more relativistic the walls, the further the scalar configuration lies from a thermal distribution. As we will discuss next, it is this surplus of hard modes that keeps the EoS radiation-like.

\subsection{Implications for the equation of state}
\label{subsec:derivation-omega-of-gamma}

\paragraph{Relativistic contribution.}
The broken-power-law ansatz~\eqref{eq:brokenPowerLaw} describes the
post-collision spectrum at $a = a_\star \equiv1$. The power spectrum at late time can be obtained by combining eq.~\eqref{eq:brokenPowerLaw} with the WKB redshifting of
eq.~\eqref{eq:redshift_modes}, approximating $k_p\simeq m_\phi$,
\begin{equation}
\Delta_\phi(k,a) \,\simeq\, \Delta_p\times
\begin{cases}
(k/k_p)^{3}(a_\star/a)^{3} \, , & k_p\gg k \, ,\\[2pt]
(k/k_p)^{-1}(a_\star/a)^{2}e^{- \pi k^2/(\gamma_\star m_\phi)^2} \, , & k_p \ll k \leq \gamma_\star m_\phi \; .
\end{cases}
\label{eq:brokenPowerLaw_a}
\end{equation}
To follow the EoS at later times, we insert the resulting
$\Delta_\phi(k,a)$ into eq.~\eqref{eq:omega_EoS_PS}.  Consider first the relativistic tail with $k/a\gg k_p\simeq m_\phi$. With
$\omega_k^2\simeq(k/a)^2$, the kinetic
energy in eq.~\eqref{eq:EK_of_PS} becomes equal to the gradient energy in eq.~\eqref{eq:EG_EV_of_PS}:
\begin{align}
\langle\rho_\mathrm{kin}\rangle_{\mathrm r}
\,\simeq\,\langle\rho_\mathrm{grad}\rangle_{\mathrm r}
\,\simeq\, \frac{\Delta_p\,k_p}{2\,a^{4}}\!\int_{am_\phi}^{\infty}\!dk\,
e^{-\pi k^2/(\gamma_\star m_\phi)^2}
\,=\, \frac{\Delta_p\,k_p\,\gamma_\star m_\phi}{4\,a^{4}}\, \text{erfc}\left(\frac{\sqrt{\pi} a}{\gamma_\star}\right) \; ,
\label{eq:rho_r_full}
\end{align}
where erfc is the complementary error function. This
equation displays the two sources of scale-factor dependence. The
prefactor $a^{-4}$ combines the physical momentum factor $(k/a)^2$ with
the WKB amplitude scaling $a^{-2}$. The argument of
$\operatorname{erfc}$ comes from the lower limit $k=am_\phi$: as $a$
grows, modes with fixed comoving momentum leave the relativistic
window. 

\paragraph{Nonrelativistic contribution.}
The non-relativistic part is dominated by modes near the spectral peak,
$k\sim k_p$. In this regime $\Delta_\phi$ in eq.~\eqref{eq:brokenPowerLaw_a} redshifts as $a^{-3}$. The
kinetic integrand in eq.~\eqref{eq:EK_of_PS} is dominated by the mass term,
$(k/a)^2+m_\phi^2\simeq m_\phi^2$, whereas the gradient energy in eq.~\eqref{eq:EG_EV_of_PS} carries
an extra factor $(k/a)^2$. Thus
\begin{equation}
\langle\rho_\mathrm{kin}\rangle_{\mathrm{nr}} \,\sim\,
\frac{m_\phi^{2}\,\Delta_p}{a^{3}} \, ,
\qquad
\langle\rho_\mathrm{grad}\rangle_{\mathrm{nr}} \,\sim\,
\frac{k_p^{2}\,\Delta_p}{a^{5}} \; .
\label{eq:rho_nr_scaling}
\end{equation}
The appropriate constants of proportionality depend on the details of the power
spectrum near~$k_p$, which is only approximately modeled by the broken-power-law ansatz, as well as on the transition from relativistic to non-relativistic scaling in eq.~\eqref{eq:redshift_modes} and in the dispersion relation. We therefore parametrise
the sum of the two regimes by introducing two positive dimensionless
parameters $b$ and $c$, such that
\begin{align}
\langle\rho_\mathrm{kin}\rangle \,&\simeq\,
\frac{m_\phi k_p \Delta_p}{a^{3}}\!
\left[\,c + \frac{\gamma_\star}{4a}\,\text{erfc}\left(\frac{\sqrt{\pi}a}{\gamma_\star}\right)\right] \, , \label{eq:EK_of_a}
 \\
\langle\rho_\mathrm{grad}\rangle \,&\simeq\,
\frac{m_\phi k_p \Delta_p}{a^{3}}\!
\left[\,\frac{b}{a^{2}} + \frac{\gamma_\star}{4a}\,\text{erfc}\left(\frac{\sqrt{\pi}a}{\gamma_\star}\right)\right] \; .
\label{eq:EG_of_a}
\end{align}
Because of the relation between kinetic and gradient energy in eq.~\eqref{eq:Virial_quadratic}, it follows that $c > b$. However, since $k_p$ is typically quite close to $m_\phi$, we expect both coefficients to be of similar magnitude, $b, c \sim \mathcal{O}(1)$. For $\gamma_\star\gg 1$ the
relativistic term in each expression dominates shortly after
collision, so kinetic and gradient energies are then comparable. Once $a\gtrsim\gamma_\star$ the relativistic term is
suppressed by a Gaussian tail, the gradient energy decays as $a^{-5}$ and the kinetic
energy as $a^{-3}$.

\paragraph{Time dependent equation of state.} 
Inserting eqs.~\eqref{eq:EK_of_a} and~\eqref{eq:EG_of_a} into the virial
relation in eq.~\eqref{eq:EoS_Virial_quadratic} with $n=2$ (generalised to $d$ spatial
dimensions) gives our main analytical result:
\begin{equation}
\boxed{\;
\omega(\gamma_\star,a) \,=\, \frac{1}{d}\,
\frac{\dfrac{b}{a^{2}}+\dfrac{\gamma_\star}{4a}\,\text{erfc}\left(\frac{\sqrt{\pi}a}{\gamma_\star}\right)}
{c+\dfrac{\gamma_\star}{4a}\,\text{erfc}\left(\frac{\sqrt{\pi}a}{\gamma_\star}\right)} \; .\;}
\label{eq:omega_of_a}
\end{equation}

\begin{figure}[t]
\centering
    \centering \includegraphics[width=0.65\linewidth]{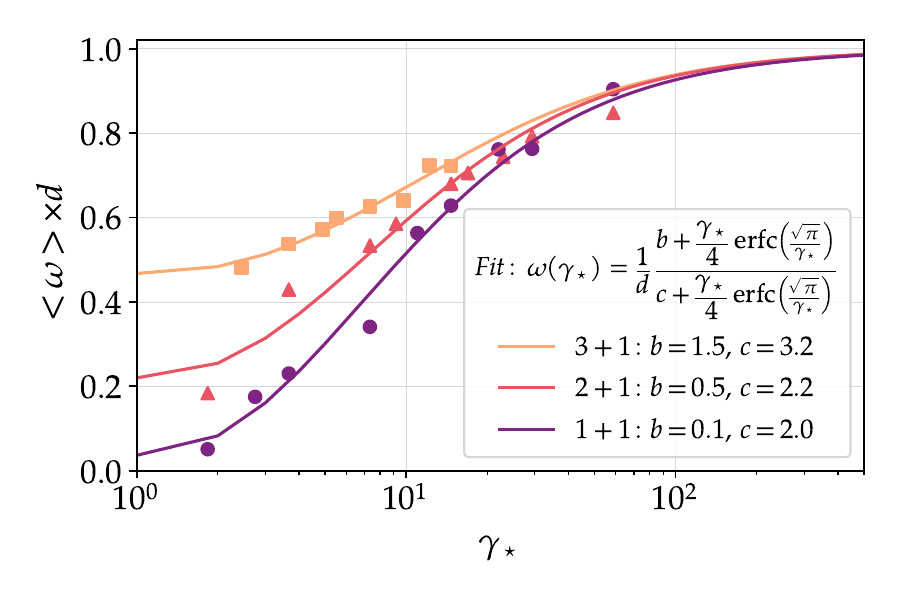}
\caption{Fits of eq.~\eqref{eq:omega_of_gamma} to the simulated EoS
parameter after bubble collision in $d=1,2,3$. }
\label{fig:EoS_fit}
\end{figure}

Two limits follow immediately:
(i)~at early times and large $\gamma_\star$, the relativistic terms
dominate both numerator and denominator and $\omega\to 1/d$
(radiation-like),
(ii)~for $a\gtrsim\gamma_\star$ the relativistic contribution becomes strongly suppressed $\propto \tfrac{\gamma_\star}{a} \exp(-\pi a^2/\gamma_\star^2)$, the numerator is
set by $b/a^{2}$ and the denominator by $c$, so
$\omega(a)\simeq b/(d c a^{2})\to 0$ (matter-like).
The transition occurs at $a\simeq\gamma_\star$. 
\paragraph{Static limit and comparison with simulations.}
In a static universe (or shortly after collision, where $a \simeq 1$), eq.~\eqref{eq:omega_of_a}
reduces to
\begin{equation}
    \omega(\gamma_\star) \,=\, \frac{1}{d}\,
\frac{b+\dfrac{\gamma_\star}{4}\,\text{erfc}\left(\frac{\sqrt{\pi}}{\gamma_\star}\right)}
{c+\dfrac{\gamma_\star}{4}\,\text{erfc}\left(\frac{\sqrt{\pi}}{\gamma_\star}\right)} \; .
\label{eq:omega_of_gamma}
\end{equation}
The best-fit curves to the simulation results of
section~\ref{sec:EoSofgamma} are shown in figure~\ref{fig:EoS_fit}. The
agreement confirms that eq.~\eqref{eq:omega_of_gamma} provides a good description of $\omega(\gamma_\star)$ and predicts that for
$\gamma_\star\gtrsim 100$ the post-collision EoS is essentially
radiation-like. 

\paragraph{Slow wall limit.}
For non-relativistic bubbles at collision ($\gamma_\star\to 1$), figure~\ref{fig:EoS_fit}
shows $\omega>0$ in $d=2,3$ rather than the naive
expectation of a  matter-like EoS $\omega\to 0$. Formally, this is because eq.~\eqref{eq:omega_of_gamma} asymptotes to $\omega\to b/(dc)$ in this limit,
 which would be zero only if $b=0$.
Physically, $b > 0$ encodes the gradient energy carried by
the bubble wall itself: even for bubbles that have barely expanded
($\gamma_\star\simeq 1$, $R_\star\simeq R_n$), the wall surface
contributes to $\rho_\mathrm{grad}$ while the interior is at $V=0$,
leading to $\omega_\mathrm{ini}>-1$. 
We have verified this interpretation explicitly by running $2{+}1$
simulations with bubbles already touching at nucleation, for which the
initial state is not purely vacuum-dominated but gives
$\omega_\mathrm{ini}\simeq -0.5$.

\subsection{Establishment of matter domination}
\label{subsec:returntoMD}

\begin{figure}[t]
\centering
\includegraphics[width=0.6\linewidth]{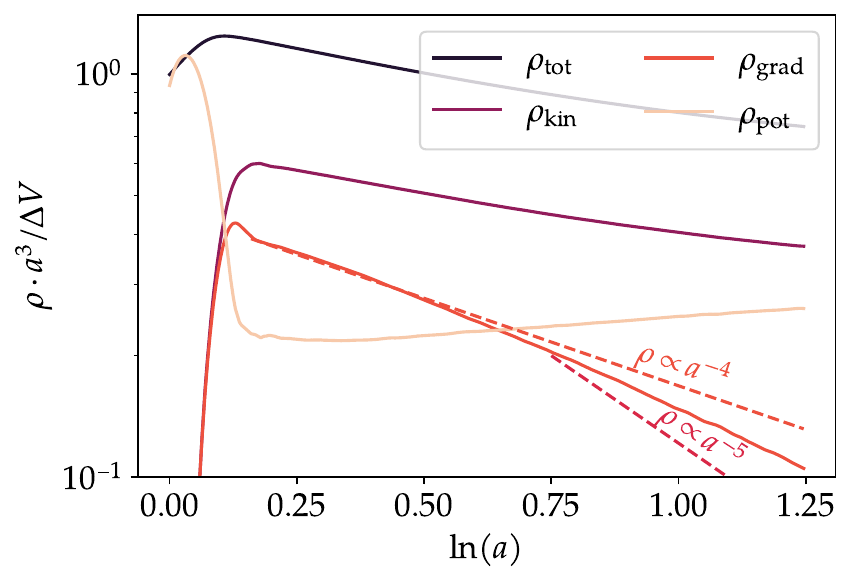}
\caption{Comoving energy densities versus the
scale factor $a$ of a single-bubble simulation in an expanding $3{+}1$D universe
with $M_\mathrm{pl}=200$, $L=32$, $N=1024$, $\Delta t=0.004$ and
$\gamma_\star\simeq 11$. }
\label{fig:3D-LateTime}
\end{figure}

\begin{figure}[t]
    \centering
    \includegraphics[width=\linewidth]{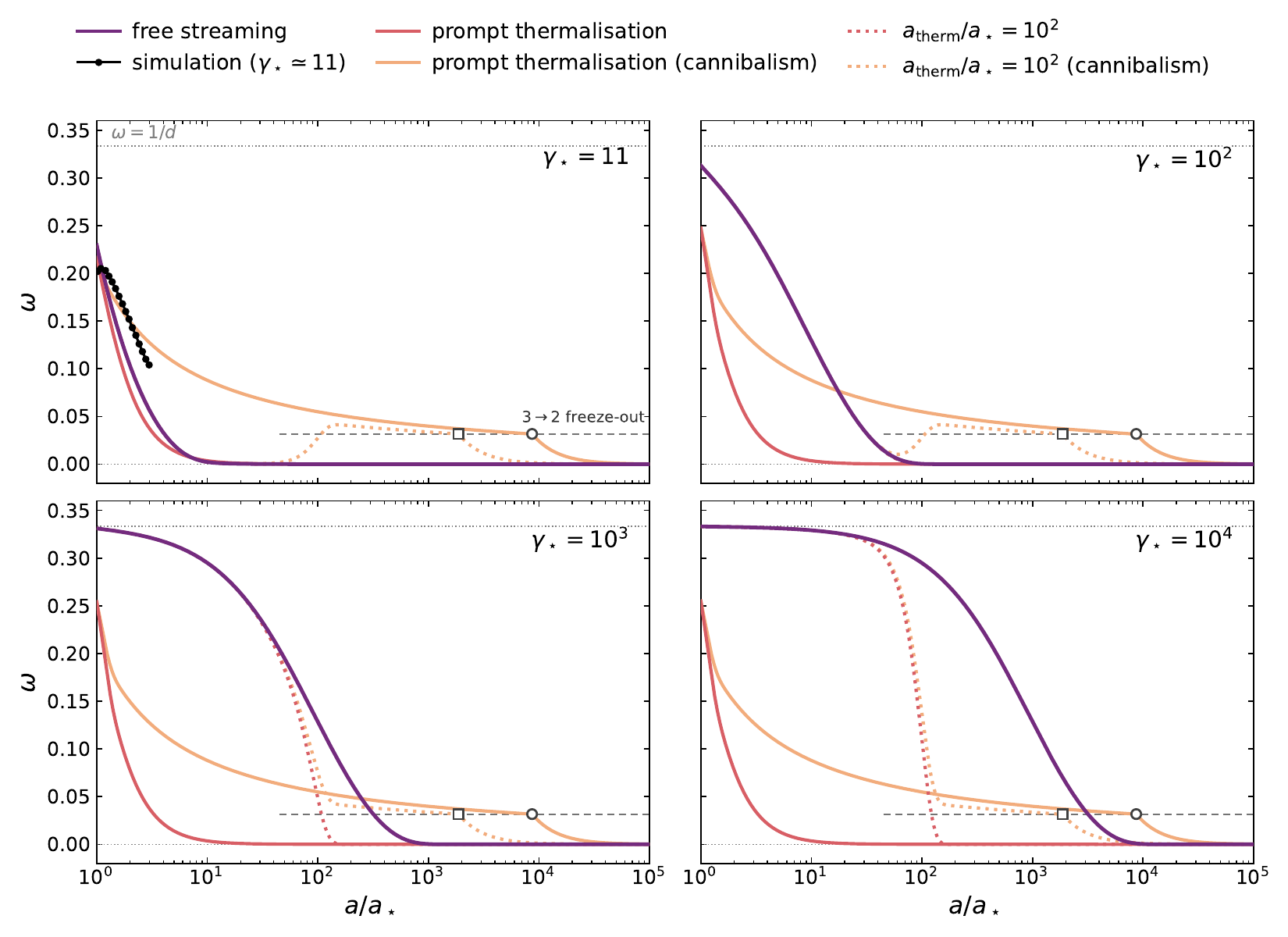}
    \caption{EoS $\omega$ versus scale factor $a/a_\star$ for the post-collision histories of the scalar, at $\gamma_\star=11,\,10^{2},\,10^{3},\,10^{4}$. Free streaming (purple, eq.~\eqref{eq:omega_of_a}) stays radiation-like, $\omega\simeq1/d$, until $a/a_\star\simeq\gamma_\star$ (dashed vertical) and only then turns pressureless. The thermalised cases, introduced below, instead set the onset by the thermal physics. Thermalisation with $x\propto a^2$ (red) becomes matter-like a few $e$-folds after the field thermalises, shown for instantaneous thermalisation (solid, $a_{\rm therm}/a_\star=1$) and for a later $a_{\rm therm}/a_\star=10^{2}$ (dotted). Cannibal $3\to2$ reactions (orange) instead hold the gas on a warm plateau ($x\propto\ln a$, $\omega\propto 1/\ln a$)  until a benchmark freezeout, here fixed at $x_{\rm fo}\simeq 30$ for which $\omega_{\rm fo}\equiv \omega^{\rm th}(x_{\rm fo})\simeq0.03$ (grey dashed line, circle at $a_{\rm fo}/a_\star\simeq10^{4}$), after which $\omega\propto a^{-2}$ cools to dust. Black points are the lattice EoS at $\gamma_\star\simeq11$. For supercooled walls, $\gamma_\star\sim10^{8\text{--}9}$, the free-streaming delay far exceeds either edge.}
    \label{fig:eos_cases}
\end{figure}

\paragraph{A delayed matter era.}
So far the late-time evolution $\omega(a)$ of eq.~\eqref{eq:omega_of_a}
rests on the parameters $b$ and $c$ extracted from a static fit. To test it
dynamically we run single-bubble simulations on an expanding background.
The required modifications were given in
section~\ref{subsec:EoS-in-flat-Universe} and their lattice implementation
in section~\ref{sec:lattice}. To produce an appreciable expansion within
the simulation volume we lower the reduced Planck mass to
$M_\mathrm{pl}=200$ in lattice units.\footnote{A lower bound on
$M_\mathrm{pl}$ comes from the requirement that the Hubble radius be larger
than the lattice length, so that the simulation volume is causally
connected.} Figure~\ref{fig:3D-LateTime} shows a run in $3{+}1$ dimensions
with $L=32$, $N=1024$, $\Delta t=0.004$ and $\gamma_\star\simeq 11$ and displays the comoving energy densities, normalised by
$\rho_\mathrm{tot,ini}/a^{3}$ with $\rho_\mathrm{tot,ini}\simeq\Delta V$.
After collision, the potential energy scales approximately as $a^{-3}$,
while the kinetic and gradient energies initially decay as $a^{-4}$. Once
the gradient and potential energies become comparable, the system leaves
relativistic scaling: the gradient energy then falls more steeply,
asymptotically as $a^{-5}$, while the kinetic energy approaches the
$a^{-3}$ behaviour of the potential. Assuming that the scalar modes redshift independently of each other, the EoS relaxes to that of
non-relativistic matter only once the universe has expanded by a factor $\gamma_\star$, i.e.\ matter domination begins at\begin{equation}
\label{eq:a_matter_kin}
\frac{a_{\rm matter}}{a_\star}\simeq\gamma_\star \qquad \text{(free streaming case)}
\end{equation}
with $a_\star$ the scale factor at collision. 
The upper-left panel of figure \ref{fig:eos_cases} compares the simulated
$\omega(a)$ with eq.~\eqref{eq:omega_of_a}, using the same $b,c$ as in the static fit. We find good agreement apart from a slight shift, which is likely due to the fact that the analytical description treats the bubble collision as instantaneous at $a = a_\star$, while in our simulation a finite amount of time passes between the first bubble collision and the completion of the phase transition. This agreement, with no additional freedom, dynamically
validates our formalism.

For the free-streaming case, this delay of the matter era is generic, as the purple lines in figure~\ref{fig:eos_cases} illustrate over a
range of $\gamma_\star$: the field stays radiation-like until
$a/a_\star\simeq\gamma_\star$ and only then turns pressureless. The origin of this behaviour
is purely kinematic: Since each mode redshifts
independently (eq.~\eqref{eq:redshift_modes}), the hardest modes populated at collision, $k\simeq\gamma_\star m_\phi$, stay relativistic
until their physical momentum has redshifted below $m_\phi$, which takes an expansion by $\gamma_\star$. The resulting delay can be enormous, since
supercooled transitions can reach
$\gamma_\star\sim 10^{10}$ (eq.~\eqref{eq:gammarun}, see also ~\cite{Ellis:2020nnr,Konstandin:2011ds}).

\paragraph{Thermalisation and cannibalism.}
The delay of matter domination rests on the assumption that the field is free-streaming and each mode redshifts independently. Self-interactions could instead relax the spectrum to its equilibrium form much more quickly and drive the EoS to the one of non-relativistic matter well before $a/a_\star \simeq \gamma_\star$. A thermalised scalar has the equilibrium EoS (derived in appendix~\ref{app:thermalisation})
\begin{equation}
\label{eq:omega_th_final}
\omega^{\rm th}(x)=\frac{K_2(x)}{3K_2(x)+xK_1(x)}\simeq\frac{1}{3+x} \; ,
\end{equation}
which is a function of $x\equiv m_\phi/T$ alone and runs from $\omega^{\rm th}\simeq1/3$ at $x\ll1$ to $\omega^{\rm th}\simeq1/x$ at $x\gg1$. Lacking couplings to any lighter states, the field thermalises into its own quanta ($g_\phi=1$) with the equilibrium temperature $T_\star$ given by energy conservation: $\rho_\phi(T_\star)=\Delta V$. If thermalisation happens immediately after the phase transition, the resulting temperature corresponds to a semi-relativistic gas\footnote{We have approximated $\rho_\phi(T_\star) \simeq g_\phi\pi^2T_\star^4/30$ where $T_\star$ denotes the temperature obtained under the assumption of instantaneous reheating. It is sometimes denoted $T_{\rm eq}$. In general, this temperature should be distinguished from both the critical temperature $T_c$ and the nucleation temperature $T_n$.}
\begin{equation}
\label{eq:Tstar_over_m}
x_\star\equiv\frac{m_\phi}{T_\star}=\left(\frac{\pi^2 g_\phi}{30\Delta V}\right)^{\!1/4} \!\!m_\phi \simeq 0.8g_\phi^{1/4}\frac{m_\phi}{\Delta V^{1/4}} \; .
\end{equation}
For the case of the potential in eq.~\eqref{eq:toypotential}, we get
\begin{equation}
\label{eq:Tstar_over_m_2}
x_\star=\frac{g_\phi^{1/4}\sqrt{2(c_\phi+6)}}{\big(30/\pi^{2}\big)^{1/4}}\simeq 3.4g_\phi^{1/4}\Big(\frac{c_\phi+6}{10}\Big)^{1/2} \; ,
\end{equation}
leading to $\omega^{\rm th}\simeq0.2$ for $c_\phi = 4$, the potential parameter choice introduced in eq.~\eqref{eq:toypotential} and used in our simulation setup. 

Thermalisation is however not instantaneous: the still-free-streaming fraction decays as $e^{-\tau}$ with the optical depth $\tau(a)=\int_{a_\star}^{a}(\Gamma_{\rm therm}/H)\,d\ln a'$. Since larger wall velocities lead to a spectrum further away from equilibrium, the thermalisation rate $\Gamma_\mathrm{therm}$ may depend implicitly on $\gamma_\star$. However, for simplicity, we remain agnostic about this relation in the following. The transition to a thermalised spectrum completes at $a_{\rm therm}$ where $\tau\simeq1$, so the macroscopic EoS crosses over smoothly from the free-streaming to the thermal form:
\begin{equation}
\label{eq:omega_corrected}
\omega(a)\simeq\omega^{\rm th}\!\big(x(a)\big)+\big[\omega(\gamma_\star,a)-\omega^{\rm th}\!\big(x(a)\big)\big]\,e^{-\tau(a)} \; .
\end{equation}
How quickly this transition happens depends on the slope of the rate at thermalisation,
\begin{equation}
\label{eq:switch_width}
\Delta\ln a\simeq\frac{1}{\delta}\,,\qquad \delta\equiv\left.\frac{d\ln(\Gamma_{\rm therm}/H)}{d\ln a}\right|_{a_{\rm therm}} \; ,
\end{equation}
which is of order a Hubble time for a typical number-changing rate ($\delta=1$).
If thermalisation does not happen instantaneously after percolation, we can approximately assume that the energy density of the scalar field, which is initially dominated by relativistic modes, redshifts as $a^{-4}$. Hence, when the field finally thermalises, its temperature will be given by $x_\text{therm}/ x_\star =  a_\text{therm} / a_\star$.

The onset of matter domination is then determined by how quickly the thermalised gas cools to the point where pressure is negligible. This transition depends on the efficiency of number-changing processes. If number-changing processes are inefficient and the comoving number density is conserved after the initial thermalisation period, the gas cools approximately as $T\propto a^{-2}$. Defining (somewhat arbitrarily) the onset of matter domination as $\omega^\text{th}(x_\text{matter}) = 10\% \times \tfrac{1}{3}$, we find $x_{\rm matter}\simeq27$ and
\begin{equation}
 \frac{a_{\rm matter}}{a_{\rm therm}} = \sqrt{\frac{x_{\rm matter}}{x_{\rm therm}}} \; .
\end{equation}
For example, for the case of instantaneous thermalisation ($a_\text{therm} = a_\star$, $x_\text{therm} = x_\star$), we find
\begin{equation}
\frac{a_\text{matter}}{a_\star} =\sqrt{\frac{x_{\rm matter}}{x_\star}} \simeq 4.2 \left(\frac{\Delta V}{g_\phi m_\phi^4}\right)^{1/8}\simeq\frac{2.8}{g_\phi^{1/8}}\left(\frac{10}{c_\phi+6}\right)^{1/4} \; .
 \end{equation}

If, on the other hand, 
number-changing processes sustain a vanishing chemical potential, the comoving entropy density is conserved instead of the comoving number density. The field then enters a ``cannibalism'' stage~\cite{Carlson:1992fn}, in which rest mass is converted into kinetic energy. This stage slows the cooling to logarithmic,
\begin{equation}
\label{eq:x_a_cannibal}
x(a)\simeq x_{\rm therm}+3\ln(a/a_{\rm therm}) \;. 
\end{equation}
In other words, the gas lingers on a warm plateau at $\omega\simeq1/x\sim0.1$ and turns pressureless only much later. If the number-changing processes remain efficient over a long period of time, one can have $a_\text{matter} / a_\text{therm} \simeq e^{(x_\text{matter} - x_\text{therm})/3}$, which can be as large as $10^4$ for $x_\text{therm}$ close to $x_\star$. 
More realistically, number-changing processes will freeze out at some temperature $x_\text{fo}$, beyond which the comoving number remains fixed and the EoS parameter decreases as $\omega\propto a^{-2}$. We can combine this case with the previous one by writing
\begin{equation}
    \frac{a_\text{matter}}{a_\text{therm}} = \sqrt{\frac{x_\text{matter}}{{x_\text{fo}}}} e^{(x_\text{fo} - x_\text{therm})/3}
\end{equation}
and hence
\begin{equation}
    \frac{a_\text{matter}}{a_\star} = \frac{x_\text{therm}}{x_\star} \sqrt{\frac{x_\text{matter}}{{x_\text{fo}}}} e^{(x_\text{fo} - x_\text{therm})/3} \; ,
\end{equation}
which is valid for $x_\star < x_\text{therm} < x_\text{fo} < x_\text{matter}$. Other cases can be obtained by taking the limit $x_\text{therm} \to x_\star$ (instant thermalisation), $x_\text{fo} \to x_\text{therm}$ (no cannibalism) and $x_\text{fo} \to x_\text{matter}$ (efficient cannibalism until the field becomes non-relativistic). For the case $x_\text{therm} > x_\text{matter}$ (very late thermalisation) instead one obtains simply $a_\text{matter} / a_\star = x_\text{therm} / x_\star = a_\text{therm} / a_\star$.

\paragraph{Delay of the matter era.}
We can summarise our results as follows:
\begin{equation}
\label{eq:a_matter_limits}
\boxed{\;
\frac{a_{\rm matter}}{a_\star}\simeq
\begin{cases}
\;\gamma_\star\,, & \text{free streaming}\ (a_{\rm therm} \to \infty)\,,\\[6pt]
\;a_{\rm therm}/a_\star\,, & \text{late thermalisation}\,,\\[6pt]
\;\sqrt{\dfrac{x_{\rm matter}}{x_\star}}\simeq 4.2 \Bigl(\tfrac{\Delta V}{g_\phi m_\phi^4}\Bigr)^{\!\frac{1}{8}}\!\!, & \text{prompt thermalisation, no cannibalism}\,,\\[6pt]
\;e^{(x_{\rm matter}-x_\star)/3}\sim10^{3\text{--}4}\,, & \text{prompt thermalisation, cannibalism}\,,
\end{cases}
\;}
\end{equation}
as illustrated in figure~\ref{fig:eos_cases}. Free streaming postpones matter domination by $a_{\rm matter}/a_\star\simeq \gamma_\star$, while instantaneous thermalisation reaches it after only a factor of $ \rm a~few$ (number-conserving) to $\sim10^{3\text{--}4}$ (cannibal), both fixed by the thermal physics and independent of $\gamma_\star$. At intermediate epochs the onset simply tracks $a_{\rm therm}$. Which regime is realised depends on the thermalisation rate, which we do not compute here.

\paragraph{Conditions for cannibalism.}
The cannibal regime requires four conditions. The first three
are satisfied in our isolated benchmark scalar theory, while the fourth is a dynamical requirement
that we do not evaluate here. First, a number-reducing vertex allowed by the symmetries. For a symmetry-breaking phase transition, the cubic term of the asymmetric potential supplies the $3\leftrightarrow 2$ reaction~\cite{Pappadopulo:2016pkp,Hochberg:2014dra}. In special cases with an exactly conserved charge, such as a complex field with a global $U(1)$ or a particle--antiparticle asymmetry, it may be possible to forbid odd-number vertices and leave only $4 \leftrightarrow 2$ processes as a way to change the particle number~\cite{Bernal:2018ins}. Second, a mass gap, so the quanta turn non-relativistic ($x\gtrsim1$) while the reaction is still active. A gapless spectrum would stay relativistic and act merely as dark radiation. Third, thermal isolation, so the liberated rest mass reheats the sector rather than a larger reservoir~\cite{Pappadopulo:2016pkp,Erickcek:2020wzd}. The scalar is its own bath ($g_\phi=1$, with no lighter states and no portal), so heat and entropy is retained and the temperature falls only logarithmically, while a coupling to a much larger bath would lock $T$ to that reservoir and erase the plateau. Fourth, the reaction must stay faster than Hubble, $\Gamma_{3\to2}/H\gtrsim1$, all the way from $x_\star$ to the freeze-out temperature $x_{\rm fo}$~\cite{Dondi:2019olm,Pappadopulo:2016pkp}. Beyond $x_{\rm fo}$ the rate drops below Hubble, the comoving number freezes, and the gas cools as dust. Number-changing into a separate lighter species, by contrast, drains energy out of the sector and does not lead to cannibalism.

\section{Discussion} 
\label{sec:discussion}

Having determined the EoS of the universe after a strong FOPT as well as its subsequent evolution, we now turn to a discussion of the different possible implications of our results for other phenomena in the early universe.

\subsection{Reheating after a supercooled phase transition}

\paragraph{Boltzmann equations with a modified equation of state.}

Until now we have only considered the self-interactions of the scalar field driving the transition (see section \ref{subsec:returntoMD}). In realistic models, however, $\phi$ is coupled to other particle species into which it decays after the FOPT, eventually re-establishing radiation domination. The findings of the previous sections allow studying the reheating phase more accurately: In the free streaming case, the fitting function proposed in eq.~\eqref{eq:omega_of_a} describes the evolution of the EoS $\omega(a,\gamma_\star)$ as a function of the scale factor and its dependence on $\gamma_\star$ encodes the impact of the scalar dynamics during the FOPT on reheating. Since the average Lorentz factor of the walls at collision $\gamma_\star$ can be inferred from the microphysics of a specific BSM model or estimated for a given $\beta/H$ value, our results provide a ready-to-use template. Concretely, if thermalisation is slow, eq.~\eqref{eq:omega_of_a} can be directly included in the Boltzmann equations for the energy densities of the scalar field and a radiation bath after the phase transition:
\begin{align}
     \frac{\mathrm{d} \rho_\phi}{\mathrm{d} a}  &= -\frac{3 [1+\omega(a,\gamma_\star)]}{a} \rho_{\phi} - \frac{\Gamma_\phi}{a H} \rho_{\phi} \;,  \label{eq:Boltzmann_Eq_phi} \\
     \frac{\mathrm{d} \rho_r}{\mathrm{d} a}  &= -\frac{4}{a} \rho_{r} +\frac{\Gamma_\phi}{a H} \rho_{\phi} \;,  \label{eq:Boltzmann_rho_r}
\end{align} 
where $\Gamma_\phi$ is a decay width, that can be computed perturbatively from the microphysics and $H$ the Hubble rate. If instead thermalisation effects are efficient, the EoS parameter from eq. \eqref{eq:omega_corrected} should replace the one from eq. \eqref{eq:omega_of_a}. 

\paragraph{Relation to scalar field fragmentation and non-perturbative particle production.}

Both after a supercooled phase transition and after inflation, the reheating of the universe is driven by the out-of-equilibrium dynamics of a scalar field. Inflationary preheating shows that the Boltzmann description used above can be a poor approximation, since it only captures perturbative decays (see ref.~\cite{Barman:2025lvk} for a recent review). Usually, the inflaton field decays much more efficiently through non-perturbative processes such as parametric resonance (periodic oscillations of the field lead to resonances and enhance particle production)~\cite{Kofman:1994rk,Kofman:1997yn,Khlebnikov:1996mc,Greene:1997fu,Micha:2002ey} and tachyonic instabilities (exponential growth of fluctuations when the effective mass squared becomes negative)~\cite{Felder:2000hj,Felder:2001kt}. This breaking of the homogenous inflaton condensate into inhomogeneous fluctuations is called \textit{fragmentation}. Comparable processes could take place during or after a FOPT~\cite{Zhang:2010qg,Braden:2014cra,Braden:2015vza,Bond:2015zfa}.

Our lattice results suggest that the evolution during a FOPT indeed showcases similarities with non-perturbative preheating.  Both in FOPTs and in inflaton fragmentation, the system evolves from an approximately homogeneous initial scalar configuration towards a highly inhomogeneous non-thermal state, with energy cascades from large scales (zero-mode) to smaller scales (high-k modes), as demonstrated by the evolution of the power spectrum in figure~\ref{fig:PS-Evol} and appendix~\ref{app:NumericalPS}. However, a fundamental distinction is to be noted: while inflaton fragmentation is primarily governed by the self-interactions of an oscillating scalar~\cite{Lozanov:2017hjm, Kim:2017duj,Kim:2021ipz,  Garcia:2023eol,Garcia:2023dyf}, the transition to the turbulent non-thermal state in supercooled FOPTs is driven by the expanding bubbles. As detailed in appendix \ref{app:c_dependence}, we have verified that self-interactions of the scalar field play at most a subdominant role in this context. The acceleration and collision of the bubbles populate wave-like excitations of the scalar field \cite{Hawking:1982ga,Watkins:1991zt, Kolb:1996jr}. The post collision field behaves as an ensemble of classical waves with occupation numbers $n_k \propto \omega_k P_\phi(k)$, constant for $k\lesssim k_p$ and falling as $n_k\propto k^{-3}$ along the $\Delta\phi\propto k^{-1}$ tail. These $\phi$-modes can decay into lighter particles if interactions such as $\mathcal{L}_{\text{int}} \sim g \phi \chi^2$ are present, and complete reheating \cite{Watkins:1991zt, Konstandin:2011ds}. 

A natural extension of this work would be to replace the idealised field profiles adopted in ref.~\cite{Mansour:2023fwj} with our fitted post-collision power spectrum, eq.~\eqref{eq:analytical_smooth}, sharpening the picture of particle production from bubble collisions \cite{Watkins:1991zt, Kolb:1996jr,Konstandin:2011ds,Falkowski:2012fb,Baldes:2023fsp,Mansour:2023fwj,Shakya:2023kjf,Giudice:2024tcp,Cataldi:2024pgt,Cataldi:2025nac,Cheng:2026npt,Inomata:2024rkt,Ghoshal:2026pew,An:2026sdu}.

\subsection{Dilution of relics} 

\paragraph{Entropy production}
When the long-lived-scalar finally decays, the released energy reheats the plasma and injects entropy, diluting any pre-existing relic, see e.g.\ refs.~\cite{McDonald:1989jd,Giudice:2000ex,Patwardhan:2015kga,Berlin:2016vnh,Berlin:2016gtr,Bramante:2017obj,Hamdan:2017psw,Allahverdi:2018aux,Cirelli:2018iax,Contino:2018crt,Chanda:2019xyl,Contino:2020god,Asadi:2022vkc,Gouttenoire:2023roe,Benso:2025vgm, Silva-Malpartida:2024emu, Racco:2025ons}. 
An early matter era has been widely invoked to dilute dark matter, e.g.\ refs.~\cite{McDonald:1989jd,Giudice:2000ex,Patwardhan:2015kga,Berlin:2016vnh,Berlin:2016gtr,Bramante:2017obj,Hamdan:2017psw,Allahverdi:2018aux,Cirelli:2018iax,Contino:2018crt,Chanda:2019xyl,Contino:2020god,Asadi:2022vkc}, thereby relaxing the unitarity bound of $100~\rm TeV$ on the mass of thermal dark matter~\cite{Griest:1989wd}. Our results are expected to affect all studies that rely on the long lifetime of the scalar field driving the phase transition to dilute such relics. In particular, the maximal dark matter mass reachable in these scenarios, discussed e.g.\ in refs.~\cite{Hambye:2018qjv,Baldes:2020kam,Baldes:2021aph,Gouttenoire:2023roe}, is expected to be modified. We first recall the standard estimate of the dilution before showing how the precise evolution of the EoS modifies it.

The comoving entropy of the radiation bath is given in terms of its energy density as
\begin{equation}
\label{eq:dil_entropy_definition}
S=C\rho_r^{3/4}a^3 \, ,
\qquad C=\frac{4}{3}\left(\frac{\pi^2g_\star}{30}\right)^{1/4} \; .
\end{equation}
Let $S_i$ be the comoving entropy just before percolation, at $a_\star$, and $S_f$ the entropy right after the decay of the scalar $\phi$ driving the phase transition, at $a_{\rm dec}$.
The dilution factor is then given by $D\equiv S_f/S_i$. The Boltzmann equation in
eq.~\eqref{eq:Boltzmann_rho_r} gives
\begin{equation}
\label{eq:dil_inject}
S_f-S_i=\int_{a_\star}^{a_f}\frac{\mathrm{d}S}{\mathrm{d}a}\,\mathrm{d}a
=\frac{3C}{4}\int_{a_\star}^{a_f}\frac{\Gamma_\phi}{H}\,
\frac{\rho_\phi}{\rho_r^{1/4}}\,a^2\,\mathrm{d}a \, ,
\qquad a_f\gg a_{\rm dec} \; .
\end{equation}
The redshifting term cancels in eq.~\eqref{eq:dil_inject}, only the decay
source changes the comoving entropy. 

\paragraph{Sudden decay approximation.}
In the sudden-decay approximation, where the scalar field decays instantly at $a = a_\text{dec}$, we can
write this source as 
\begin{equation}
\label{eq:Gamma_phi_inst}
Q(a) \equiv \frac{\Gamma_\phi}{aH}\rho_\phi
\simeq\rho_\phi(a_{\rm dec}^-)\delta(a-a_{\rm dec}) \; .
\end{equation}
Across the infinitesimal decay interval, only the singular part of
eq.~\eqref{eq:Boltzmann_rho_r} contributes. It gives
\begin{equation}
\label{eq:radiation_jump}
\rho_r^+-\rho_r^-
=\lim_{\epsilon\to0}\int_{a_{\rm dec}-\epsilon}^{a_{\rm dec}+\epsilon}
Q(a)\,\mathrm{d}a
=\rho_\phi(a_{\rm dec}^-) \; ,
\end{equation}
where $\rho_r^\pm$ are the radiation densities immediately before and after
the decay.
The finite redshifting term vanishes across this interval, so
$Q(a)\,\mathrm{d}a$ can be replaced by $\mathrm{d}\rho_r$ along the jump.
Eq.~\eqref{eq:dil_inject} therefore becomes
\begin{equation}
\label{eq:dil_inject_sudden}
S_f-S_i=\frac{3C}{4}a_{\rm dec}^3
\int_{\rho_r^-}^{\rho_r^+}\rho_r^{-1/4}\,\mathrm{d}\rho_r
=Ca_{\rm dec}^3\left[(\rho_r^+)^{3/4}-(\rho_r^-)^{3/4}\right] \; .
\end{equation}
In the sudden-decay approximation, the decay term vanishes for
$a<a_{\rm dec}$, and eq.~\eqref{eq:Boltzmann_Eq_phi} gives
\begin{equation}
\label{eq:g_evolution}
\rho_\phi(a)=\Delta V\left(\frac{a_\star}{a}\right)^3g(a,a_\star) \, ,\qquad
g(a_2,a_1)\equiv\exp\!\left[-3\int_{a_1}^{a_2}\omega(a')
\frac{\mathrm{d}a'}{a'}\right] \, ,\qquad
g_{\rm dec}\equiv g(a_{\rm dec},a_\star) \; .
\end{equation}
The standard treatment assumes $\omega=0$ immediately after the transition,
so $g_{\rm dec}=1$.

\paragraph{Dilution factor.}
Using eqs.~\eqref{eq:radiation_jump}, \eqref{eq:dil_inject_sudden}, \eqref{eq:g_evolution} and $\rho_r\propto a^{-4}$, we can write\footnote{A shorter, equivalent derivation starts directly from
energy conservation,
$\rho_r^+=\rho_r^-+\rho_\phi(a_{\rm dec}^-)$, and uses
$S\propto\rho_r^{3/4}a^3$ at fixed $a_{\rm dec}$. This immediately gives
$D=(\rho_r^+/\rho_r^-)^{3/4}$.}
\begin{equation}
\label{eq:dil_entropy_ratio}
D\equiv \frac{S_f}{S_i}
=\left(\frac{\rho_r^+}{\rho_r^-}\right)^{3/4}
=\left[1+\frac{\rho_\phi(a_{\rm dec}^-)}{\rho_r^-}\right]^{3/4}=\left[1+\alpha g_{\rm dec}
\frac{a_{\rm dec}}{a_\star}\right]^{3/4} \; ,
\end{equation}
where $\alpha\equiv \rho_\phi(a_{\star})/\rho_r(a_{\star})$ is the latent heat parameter, introduced in eq.~\eqref{eq:latent_heat_fac}.
To separate the latent heat from the subsequent evolution, let $S_\star$ be
the reference entropy obtained by thermalising all the energy at $a_\star$.
Eq.~\eqref{eq:dil_entropy_ratio} then gives
\begin{equation}
\label{eq:dil_key}
D=\underbrace{(1+\alpha)^{3/4}}_{S_\star/S_i}
\underbrace{\left[
\frac{1+\alpha g_{\rm dec}a_{\rm dec}/a_\star}{1+\alpha}
\right]^{3/4}}_{S_f/S_\star} \; .
\end{equation}
The
first factor isolates the entropy injected by latent heat conversion, while the second accounts for the entropy injected by $\omega\neq 1/3$ after percolation through $g_{\rm dec}$. Note that $D\to1$ for
$\alpha\to0$.

To express $D$ in terms of the decay rate, assume that the scalar dominates
at decay. The Friedmann equation at $a_\star$, the condition
$H(a_{\rm dec})\simeq\Gamma_\phi$, and eq.~\eqref{eq:g_evolution} give
\begin{equation}
\label{eq:adec_decay_alpha}
H_\star^2\simeq\frac{\Delta V(1+\alpha^{-1})}{3M_{\rm pl}^2} \, ,
\qquad
\left(\frac{a_{\rm dec}}{a_\star}\right)^3
\simeq\frac{\alpha}{1+\alpha}\,g_{\rm dec}
\left(\frac{H_\star}{\Gamma_\phi}\right)^2 \; .
\end{equation}
Substituting eq.~\eqref{eq:adec_decay_alpha} into
eq.~\eqref{eq:dil_key} yields
\begin{equation}
\label{eq:dil_decay_alpha}
D\simeq\left[1+\frac{\alpha^{4/3}}{(1+\alpha)^{1/3}}
g_{\rm dec}^{4/3}
\left(\frac{H_\star}{\Gamma_\phi}\right)^{2/3}\right]^{3/4}\quad \xrightarrow[D\gg 1]\qquad ~~~~\alpha^{3/4}g_{\rm dec}
\sqrt{\frac{H_\star}{\Gamma_\phi}} \; .
\end{equation}
Eq.~\eqref{eq:dil_decay_alpha} assumes
scalar domination at decay. For earlier decay one should use
eq.~\eqref{eq:dil_key} together with the Boltzmann system
\eqref{eq:Boltzmann_Eq_phi}--\eqref{eq:Boltzmann_rho_r}.

\paragraph{Thermalisation histories.}
If
thermalisation occurs at $a_{\rm therm}$, eq.~\eqref{eq:g_evolution}
separates the evolution before and after thermalisation (assuming $a_\star<a_{\rm therm}<a_{\rm dec}$)
\begin{equation}
\label{eq:gdec_split}
g_{\rm dec}=g(a_{\rm therm},a_\star)\times g(a_{\rm dec},a_{\rm therm}) \; .
\end{equation}
The first factor in eq.~\eqref{eq:gdec_split} describes free streaming up to
$a_{\rm therm}$, which for $\gamma_\star \gg a_{\rm therm}/a_\star$, the scalar being radiation-like before
thermalisation, reads
\begin{equation}
\label{eq:gdec_split_fac_1}
    g(a_{\rm therm},a_\star)=\frac{a_\star}{a_{\rm therm}}\simeq \frac{x_{\rm therm}}{x_\star}\quad \longrightarrow\quad \begin{cases}
\gamma_\star^{-1},
& \text{free streaming}~(a_{\rm therm}>\gamma_\star a_{\star}) \, ,\\[10pt]
1,
& \text{prompt thermalisation}~(a_{\rm therm}\simeq a_{\star}),\\[10pt]
\dfrac{a_\star}{a_{\rm therm}} \, ,& \text{delayed thermalisation} \; ,
\end{cases}
\end{equation}
where the limits follow from eq.~\eqref{eq:a_matter_limits}.
The second factor in eq.~\eqref{eq:gdec_split} describes the subsequent thermal evolution.
For $a>a_{\rm therm}$, we use the expression of the EoS of a thermal plasma in eq.~\eqref{eq:omega_th_final}:
\begin{equation}
 \omega^{\rm th}(x)\simeq 
  \frac{1}{3+x} \; .  
\end{equation}
For number-conserving thermalisation, assuming $x_{\rm therm}>x_{\rm matter}$ the scalar field cools as a relativistic particle
$x(a)\simeq x_{\rm therm}(a/a_\star)^2$, so
\begin{equation}
\label{eq:number_conserving_integral}
-3\int_{a_{\rm therm}}^{a_{\rm dec}}\omega^{\rm th}(a)\frac{\mathrm{d}a}{a}
\simeq-3\int_{a_{\rm therm}}^{a_{\rm dec}}
\frac{\mathrm{d}a/a}{3+x_{\rm therm}(a/a_\star)^2}
=\frac{1}{2}\ln\!\left(\frac{x_{\rm therm}+3(a_{\rm therm}/a_{\rm dec})^2}{x_{\rm therm}+3}\right) \; .
\end{equation}
Exponentiating eq.~\eqref{eq:number_conserving_integral} and plugging into eq.~\eqref{eq:gdec_split} gives
\begin{equation}
\label{eq:gdec_number-conserving}
    g_{\rm dec}=\frac{x_{\rm therm}}{x_{\star}}\sqrt{\frac{x_{\rm therm}+3(a_{\rm therm}/a_{\rm dec})^2}{x_{\rm therm}+3}} \, ,\qquad \text{number-conserving thermalisation} \; .
\end{equation}
For efficient cannibalism, we have (assuming $a_\star<a_{\rm therm}<a_{\rm fo}<a_{\rm dec}$)
\begin{equation}
\label{eq:gdec_split_can}
g_{\rm dec}=g(a_{\rm therm},a_\star)\times g(a_{\rm fo},a_{\rm therm})\times g(a_{\rm dec},a_{\rm fo}) \; .
\end{equation}
where the first factor is given by eq.~\eqref{eq:gdec_split_fac_1}. Using
eq.~\eqref{eq:x_a_cannibal} gives
$\mathrm{d}\ln a\simeq\mathrm{d}x/3$. Integrating until
$x=x_{\rm fo}$ gives the second factor
\begin{equation}
\label{eq:cannibal_integral}
\ln{g(a_{\rm fo},a_{\rm therm})}=-3\int_{a_{\rm therm}}^{a_{\rm fo}}
\omega^{\rm th}(a)\frac{\mathrm{d}a}{a}
\simeq-\int_{x_{\rm eq}}^{x_{\rm fo}}
\frac{\mathrm{d}x}{3+x}
=\ln\!\left(\frac{3+x_{\rm therm}}{3+x_{\rm fo}}\right) \; .
\end{equation}
The later
number-conserving stage is responsible for the third factor in eq.~\eqref{eq:gdec_split_can}, which is equivalent to eq.~\eqref{eq:number_conserving_integral}, and reads
\begin{equation}
    g(a_{\rm dec},a_{\rm fo}) = \sqrt{\frac{x_{\rm fo}+3(a_{\rm fo}/a_{\rm dec})^2}{x_{\rm fo}+3}} \; ,
\end{equation}
which for $a_{\rm dec}\gg a_{\rm fo}$ and $x_{\rm fo}\gg 1$, can be neglected. We conclude 
\begin{equation}
\label{eq:gdec_can}
   g_{\rm dec}\simeq \frac{x_{\star}}{x_{\rm therm}}\dfrac{3+x_{\rm therm}}{3+x_{\rm fo}}\sqrt{\frac{x_{\rm fo}+3(a_{\rm fo}/a_{\rm dec})^2}{x_{\rm fo}+3}} \, ,\qquad \text{cannibal thermalisation} \; .
\end{equation}
\paragraph{Limiting histories.}
Combining eqs.~\eqref{eq:gdec_split_fac_1}, \eqref{eq:gdec_number-conserving} and \eqref{eq:gdec_can}, the four
histories in eq.~\eqref{eq:a_matter_limits} give
\begin{equation}
\label{eq:dilution_limits}
{\small
g_{\rm dec}=1\quad\xrightarrow{\;\text{this work}\;}\quad
g_{\rm dec}\simeq
\begin{cases}
\dfrac{a_\star}{a_{\rm matter}}\simeq\gamma_\star^{-1} \, ,
& \text{free streaming}~(a_{\rm therm}/a_\star\gg \gamma_\star ) \, ,\\[10pt]
\dfrac{a_\star}{a_{\rm matter}}
\simeq\dfrac{a_\star}{a_{\rm therm}} \, ,
& \text{delayed thermalisation}
~(\gamma_\star \gg a_{\rm therm}/a_\star) \, ,\\[10pt]
\sqrt{\dfrac{x_{\star}}{3+x_{\star}}}\simeq0.7 \, ,
& \text{number-conserving, prompt therm.}~(a_{\rm therm}\simeq a_{\star}) \, ,\\[10pt]
\dfrac{3+x_{\star}}{3+x_{\rm fo}}\simeq0.2 \, ,
& \text{cannibal, prompt thermalisation}~(a_{\rm therm}\simeq a_{\star}) \; .
\end{cases}
}
\end{equation}
The last two lines assume prompt thermalisation,
$a_{\rm therm}=a_\star$ and a long lived scalar $a_{\rm dec}\gg a_{\rm therm}$, $a_{\rm fo}$, with $x_{\star}=3.4$, cf.\ eq.~\eqref{eq:Tstar_over_m_2} and $x_{\rm fo}$, a free parameter that we set to $x_{\rm fo}\simeq 30$ as in figure~\ref{fig:eos_cases}.

\subsection{Gravitational waves}

\paragraph{Peak redshift and suppression.}
We denote by $\Omega_{\rm GW,0}^{\rm RD}(f)$ the conventional present day
spectrum obtained by thermalising the total energy at $a_\star$ and assuming
radiation domination thereafter.  Since this reference already includes the
entropy present after prompt thermalisation, subsequent entropy production is
measured by
\begin{equation}
\label{eq:D_late_GW}
D_{\rm late}\equiv\frac{S_f}{S_\star}
=\frac{D}{(1+\alpha)^{3/4}}
=\left[
\frac{1+\alpha g_{\rm dec}a_{\rm dec}/a_\star}{1+\alpha}
\right]^{3/4} \; ,
\end{equation}
where $D=S_f/S_i$ is the relic dilution factor in
eq.~\eqref{eq:dil_key}, and $S_\star$ is the comoving entropy of the RD
reference history at $a_\star$.  Thus $D_{\rm late}$ removes from $D$ the
factor $(1+\alpha)^{3/4}$ already included in the reference spectrum.  If the
scalar continues to redshift like radiation until it decays, then
$g_{\rm dec}=a_\star/a_{\rm dec}$ and $D_{\rm late}=1$, as required.  The
present day spectrum is
\begin{equation}
\label{eq:GW_transfer_general}
\Omega_{\rm GW,0}(f)=D_{\rm late}^{-4/3}
\Omega_{\rm GW,0}^{\rm RD}\!\left(D_{\rm late}^{1/3}f\right)
S_\omega(f) \; .
\end{equation}
with $S_\omega=1$ at the subhorizon peak.  Hence
$f_{\rm peak}=D_{\rm late}^{-1/3}f_{\rm peak}^{\rm RD}$ and
$\Omega_{\rm peak}=D_{\rm late}^{-4/3}\Omega_{\rm peak}^{\rm RD}$
\cite{Ertas:2021xeh,Gouttenoire:2023pxh}.

\paragraph{Transfer of infrared modes.}
For superhorizon infrared modes, $S_\omega$ depends on the expansion history
at horizon reentry.  To evaluate it, define
\begin{align}
\omega_{\rm tot}(a)&\equiv\frac{p_{\rm tot}(a)}{\rho_{\rm tot}(a)}
=
\frac{\omega(a)\rho_\phi(a)+\rho_r(a)/3}
{\rho_\phi(a)+\rho_r(a)} \, ,
\nonumber\\
\mathcal E(a)&\equiv
\frac{\rho_{\rm tot}(a)a^4}{\rho_{\rm tot,\star}a_\star^4}
=\exp\!\left[\int_{a_\star}^{a}
\bigl(1-3\omega_{\rm tot}(a')\bigr)\frac{\mathrm da'}{a'}\right] \; .
\label{eq:E_GW}
\end{align}
Let $a_f$ denote the onset of the final adiabatic radiation era.  A mode
$k=2\pi a_0f$ is superhorizon at production if $k<a_\star H_\star$.  Its
reentry time $a_k$ is defined by $k=a_kH(a_k)$.  Neglecting tensor anisotropic
stress, the transfer relative to the RD history is
\begin{equation}
\label{eq:S_omega}
S_\omega(f)\simeq
\begin{cases}
1 \, , & k\geq a_\star H_\star  \, ,\\[2pt]
\mathcal E(a_k) \, , & k<a_\star H_\star \, ,\quad a_\star<a_k<a_f \, ,\\[2pt]
\mathcal E(a_f)=D_{\rm late}^{4/3} \, , & k<a_\star H_\star,\quad a_k\geq a_f \; .
\end{cases}
\end{equation}
The last equality follows from $S=sa^3\propto T^3a^3\propto(\rho_r a^4)^{3/4}$ and $\rho_r(a_f)=\rho_{\rm tot}(a_f)$, which lead to
\begin{equation}
\mathcal E(a_f)=\frac{\rho_{\rm tot}(a_f)a_f^4}
{\rho_{\rm tot,\star}a_\star^4}=\left(\frac{S_f}{S_\star}\right)^{4/3}=D_{\rm late}^{4/3} \; .
\end{equation}
For modes re-entering while $\omega_{\rm tot}$ is constant, or varies slowly,
the causal infrared spectrum has the local slope~\cite{Hook:2020phx}
\begin{equation}
\label{eq:GW_slope_general}
\frac{\mathrm d\ln\Omega_{\rm GW}}{\mathrm d\ln f}
\simeq\frac{1+15\omega_{\rm tot}(a_k)}
{1+3\omega_{\rm tot}(a_k)} \; .
\end{equation}
Eq.~\eqref{eq:GW_slope_general} gives $f^3$ during radiation
domination~\cite{Durrer:2003ja,Caprini:2009fx,Cai:2019cdl} and $f^1$
during matter domination.  A gradual EoS yields a smooth interpolation.  This
result applies only to causal modes that were superhorizon at production.

For an abrupt matter era beginning at $a_{\rm dom}$ and ending at
$a_{\rm dec}$, the transfer reduces to the familiar limit
\cite{Barenboim:2016mjm,Ellis:2020nnr,Hook:2020phx,Gouttenoire:2023pxh}
\begin{equation}
\label{eq:S_M}
S_M(f)=
\begin{cases}
1 \, , & f\geq f_{\rm dom} \, ,\\[2pt]
(f/f_{\rm dom})^{-2} \, , & f_{\rm dec}<f<f_{\rm dom} \, ,\\[2pt]
(f_{\rm dec}/f_{\rm dom})^{-2} \, , & f\leq f_{\rm dec} \;,
\end{cases}
\end{equation}
where
\begin{equation}
\label{eq:f_dom_f_dec}
f_i=\left(\frac{a_i}{a_0}\right)\frac{H_i}{2\pi} \, ,
\qquad i\in\{\mathrm{dom},\mathrm{dec}\} \; .
\end{equation}
For the histories dominated by the scalar considered here, the matter era
begins when the scalar becomes nonrelativistic, so
$a_{\rm dom}\simeq a_{\rm matter}$.
Between $f_{\rm dec}$ and $f_{\rm dom}$,
$S_M\propto f^{-2}$ changes the causal $f^3$ tail to $f^1$.  Below $f_{\rm dec}$ the $f^3$ slope is restored with a different normalisation.
Together with eq.~\eqref{eq:GW_transfer_general}, the associated entropy
production suppresses and redshifts the peak, while the transfer enhances the
infrared relative to the RD extrapolation~\cite{Hook:2020phx}.

\paragraph{Thermalisation histories}
In the free streaming history, the scalar has $\omega\simeq1/3$ until
$a_{\rm matter}\simeq\gamma_\star a_\star$, when
$H_{\rm matter}\simeq H_\star/\gamma_\star^2$.  If
$\Gamma_\phi\gtrsim H_\star/\gamma_\star^2$, it decays before becoming
nonrelativistic, giving $D_{\rm late}\simeq1$ and no $f^1$ interval.  For
$\alpha\gg1$ and $\Gamma_\phi\ll H_\star/\gamma_\star^2$, it survives into the
nonrelativistic regime and produces a matter era.  Sudden decay then gives
\begin{equation}
D_{\rm late}\simeq\frac{1}{\gamma_\star}
\sqrt{\frac{H_\star}{\Gamma_\phi}} \, ,\qquad
f_{\rm dom}\simeq\frac{f_\star}{\gamma_\star} \, ,\qquad
\frac{f_{\rm dec}}{f_{\rm dom}}\simeq D_{\rm late}^{-2/3}
=\left(\frac{\gamma_\star^2\Gamma_\phi}{H_\star}\right)^{1/3} \; ,
\label{eq:free_streaming_GW_scales}
\end{equation}
Here $f_\star\equiv(a_\star/a_0)H_\star/(2\pi)$ is the present day frequency
of the mode $k_\star=a_\star H_\star$ associated with the Hubble scale at
production, not the source peak.  If $k_{\rm peak}$ is the comoving peak wave number, then
$f_{\rm peak}/f_\star=k_{\rm peak}/(a_\star H_\star)
=\mathcal O(\beta/H_\star)$.  Both frequencies receive the same
$D_{\rm late}^{-1/3}$ shift relative to the RD history.  The $f^1$ interval is
$f_{\rm dec}<f<f_{\rm dom}$.  For thermalised or cannibal histories,
$\mathcal E(a)$ must instead be computed from the corresponding
$\omega_{\rm tot}(a)$.  The gradual EoS evolution smooths the sharp breaks of
eq.~\eqref{eq:S_M}.  Eq.~\eqref{eq:S_omega} assumes that the source
has ceased and tensor anisotropic stress is negligible.  Collisionless free
streaming may steepen $f^3$ towards $f^4$~\cite{Hook:2020phx}, although its
applicability to the classical wave state is uncertain.

\subsection{Primordial black holes}

\paragraph{PBHs from phase transitions.}
The completion time of a FOPT in a given finite volume --- which we may define as the moment when the averaged kinetic and potential energies become equal, $\rho_{\rm kin}\simeq \rho_{\rm pot}$, as shown in figure~\ref{fig:3D_simu_overview} --- is in fact a stochastic variable. It arises from the percolation of a finite number of bubbles that nucleate at random times and positions. Hence, a FOPT sources density perturbations~\cite{Sasaki:1982fi,Liu:2022lvz,Giombi:2023jqq,Elor:2023xbz,Lewicki:2024ghw,Buckley:2024nen,Cai:2024nln,Jinno:2024nwb,Sui:2025epg,Greene:2026gnw}. For a supercooled FOPT, late percolation coincides with an extended phase of vacuum domination within this finite volume, while the rest of the background universe is already diluting like radiation, thereby generating an overdensity. If the overdensity of such \textit{late-blooming}, Hubble-size patches becomes of order $\mathcal{O}(1)$, they are expected to collapse into Primordial Black Holes~\cite{Kodama:1982sf,Hsu:1990fg,Liu:2021svg,Hashino:2021qoq,Kawana:2022olo,Lewicki:2023ioy,Gouttenoire:2023naa,Baldes:2023rqv,Gouttenoire:2023bqy,Salvio:2023ynn,Gouttenoire:2023pxh,Flores:2024lng,Lewicki:2024ghw,Lewicki:2024sfw,Kanemura:2024pae,Cai:2024nln,Goncalves:2024vkj,Banerjee:2024fam,Arteaga:2024vde,Banerjee:2024cwv,Hashino:2025fse,Ghoshal:2025dmi,Zou:2025sow,Franciolini:2025ztf,Wang:2025hwc,Kierkla:2025vwp,Banerjee:2025qji,Wang:2026zvz,Ning:2026nfs,Ai:2026zrs,Banerjee:2026tgr}. Late-blooming can also be induced through catalysis~\cite{Jinno:2023vnr,Guo:2026cuv}. Other proposed PBH formation mechanisms operating during FOPTs include bubble collapse in incomplete phase transitions~\cite{Ai:2024cka,Murai:2025hse}, bubble collisions~\cite{Hawking:1982ga,Moss:1994iq,Ashoorioon:2020hln,Jung:2021mku}, and 
matter squeezing by bubble walls~\cite{Crawford:1982yz,Gross:2021qgx,Baker:2021sno,Kawana:2021tde,Huang:2022him}.

Recently, it has been claimed that PBH formation from late-blooming may be far less efficient than initially thought~\cite{Flores:2024lng, Franciolini:2025ztf, Wang:2026zvz}: the density-contrast threshold for collapse is $\delta_c\simeq 0.5$ in the comoving gauge during radiation domination ($\omega = 1/3$), but is about ten times larger in the spatially-flat gauge~\cite{Franciolini:2025ztf}. Refs.~\cite{Ai:2026zrs,Banerjee:2026tgr} proposed that a matter era ($\omega =0$) following a FOPT --- which could arise from the long lifetime of the scalar field driving the transition --- could lower the PBH formation threshold to $\delta_c\sim 0$~\cite{Harada:2016mhb}, thereby making PBH formation much more efficient, not only from late-blooming but, in practice, from any of the above-mentioned mechanisms operating right after the FOPT.

\paragraph{This work.}
Our results show, however, that the EoS parameter at percolation can remain sizeable because of large field gradients, as shown in figure~\ref{fig:EoS_fit}. In the free-streaming case, matter domination is reached only after an expansion by a factor $a/a_\star\simeq\gamma_\star$, which can be very large (see eqs.~\eqref{eq:gammarun} and~\eqref{eq:gammaLL}). In this regime, the long lifetime of the scalar field is therefore unlikely to substantially enhance PBH formation following a FOPT. Scalar self-interactions may lead to an earlier onset of matter domination by thermalising the scalar field, after which the EoS decreases as $\omega\simeq 1/(3+x)$. Even in this case, however, the transition to matter domination is gradual rather than instantaneous, as illustrated by the red line in figure~\ref{fig:eos_cases}. We leave a quantitative calculation of the resulting enhancement of PBH formation to future work.

\section{Conclusions}
\label{sec:conclusions}

Immediately after a first-order phase transition, the universe is in a
highly inhomogeneous state. Even after the true-vacuum bubbles have
expanded and collided, and the walls have disappeared, large gradients
in the scalar field persist and drive incoherent oscillations around
the true vacuum. Provided the field does not decay rapidly and
dominates the energy budget, the resulting EoS lies between matter and
radiation, with larger gradients implying a more radiation-like
behaviour. As the universe expands, the gradients redshift, the
oscillations become more coherent and the EoS relaxes to that of
non-relativistic matter.

In this work, we have quantified this evolution using lattice simulations of a scalar field undergoing a phase transition in vacuum. Our simulations confirm that the wall Lorentz factor grows linearly with the bubble radius and that substantial gradients survive the collisions. Using the virial theorem, we relate the gradient energy to the EoS parameter $\omega$. By varying the initial bubble separation, and hence the wall Lorentz factor $\gamma_\star$ at collision, we determine $\omega(\gamma_\star)$ in $1{+}1$, $2{+}1$, and $3{+}1$ dimensions (figure~\ref{fig:EoS_fit}).
This result is well described by a fitting function obtained from expressing $\omega$ in terms of the dimensionless power spectrum $\Delta_\phi$, modelled as a broken power law with a UV cut-off above the inverse Lorentz contracted wall width $k_\star\simeq\gamma_\star m_\phi$. Assuming that scalar modes free-stream and redshift independently of each other, we derive an analytic prediction for $\omega(\gamma_\star,a)$ (eq.~\eqref{eq:omega_of_a}) which we  qualitatively confirm using simulation results on an expanding background.

The main implication is that the Universe is generally \emph{not} matter-dominated immediately after a supercooled first-order phase transition, contrary to a common assumption. Instead, the post-percolation scalar configuration is highly non-thermal and, for $\gamma_\star\gtrsim100$, has an EoS within $10\%$ of the radiation value $\omega=1/3$ (figure~\ref{fig:EoS_fit}). The rate at which its EoS approaches matter-like behaviour due to the redshifting of momenta depends on the subsequent thermalisation processes (results summarised in eq.~\eqref{eq:a_matter_limits} and figure~\ref{fig:eos_cases}). 

For the large Lorentz factors reached in strongly supercooled transitions, $\gamma_\star\gg 100$, the delay is most dramatic if the scalar modes redshift independently of each other (free streaming). In this case, eMD is achieved only once the universe has expanded by a factor $a/a_\star\simeq\gamma_\star$. For a matter era to set it, the scalar field must survive until this epoch, which requires $\Gamma_\phi\ll H_\star/\gamma_\star^2$.
The extreme opposite is the case of prompt thermalisation.  In the absence of number changing processes, it gives the fastest onset of matter domination at $a_{\rm matter}/a_\star$ of only a few (eq.~\eqref{eq:a_matter_limits}), a number fixed by the self-thermalisation temperature $T_\star$ defined in eq.~\eqref{eq:Tstar_over_m} rather than by $\gamma_\star$. For a finite thermalisation timescale, the onset occurs at some intermediate $a_\mathrm{therm}/a_\star$. Since the post-collision field configuration deviates more strongly from a thermal distribution for higher $\gamma_\star$ values, $a_\mathrm{therm}$ may itself grow with $\gamma_\star$, and we leave a quantitative study of this relation for future work. 
Finally, in an interacting scalar theory, number changing (cannibal) interactions are expected to be present and delay eMD even for efficient thermalisation. By converting rest mass into kinetic energy, these hold both the temperature of the bath and the EoS at an intermediate plateau, and delay matter domination to $a_\mathrm{matter}/a_\star \simeq 10^{3-4}$. The thermalisation epoch $a_{\rm therm}$ cannot be captured by the classical lattice setup and is treated as a free parameter for simplicity.

These results directly affect several cosmological observables, as discussed in section~\ref{sec:discussion}. Delaying matter domination reduces the entropy injected into the plasma relative to the instantaneous-eMD approximation (eq.~\eqref{eq:dilution_limits}). This weakens the dilution of pre-existing relics and modifies the maximal dark matter mass attainable in dilution scenarios~\cite{Griest:1989wd,Hambye:2018qjv,Baldes:2020kam,Baldes:2021aph,Gouttenoire:2023roe}. The reduction is modest after prompt, number-conserving thermalisation because the pressure is lost rapidly, but it can be much stronger in the free-streaming case or when cannibal reactions keep the scalar bath warm. The modified expansion history also changes the relation between temperature and scale factor during reheating, affecting non-thermal dark-matter production mechanisms such as the phase-in mechanism~\cite{Benso:2025vgm}.

An early matter era modifies the gravitational-wave spectrum in two distinct ways. First, the late entropy release shifts the peak frequency by $D_{\rm late}^{-1/3}$ and suppresses its amplitude by $D_{\rm late}^{-4/3}$. Second, the causal $f^3$ infrared tail develops an intermediate $f^1$ scaling for modes entering the horizon during matter domination. Delaying the onset of matter domination shortens the matter era and shifts the range of affected modes. The position and shape of the infrared feature probe the EoS at horizon entry, whereas the peak modification probes the integrated entropy release. A measurement over a broad frequency range could therefore constrain $a_{\rm therm}$ within a given cooling model, distinguish prompt thermalisation, free streaming, and a prolonged cannibal phase, and thereby probe the scalar self-interactions.

The delayed onset of matter domination also affects PBH formation. Recent studies have assumed that scalar domination immediately after percolation produces pressureless matter, thereby lowering the collapse threshold and potentially enhancing PBH production from late-blooming perturbations and other FOPT mechanisms~\cite{Ai:2026zrs,Banerjee:2026tgr}. Our results suggest that this enhancement may be significantly reduced: field gradients keep the EoS sizeable, and in the free-streaming regime the scalar remains radiation-like until $a/a_\star\simeq\gamma_\star$, which can be very large. A long-lived scalar therefore does not necessarily lead to an extended matter era during which PBH formation is enhanced. Thermalisation can trigger matter domination earlier, but the EoS then decreases gradually rather than abruptly, so pressure and the sound speed may remain relevant during collapse. A reliable PBH abundance calculation should therefore evolve perturbations through the time-dependent EoS and sound speed, rather than assume an instantaneous transition to dust.

Future work should embed this framework in realistic supercooled models at non-zero temperature and determine the wall Lorentz factor at percolation, $\gamma_\star$, together with the relevant thermalisation and number-changing (cannibal) rates. This would allow us to predict the post-transition evolution of the EoS, $\omega(a)$, and derive more precise predictions for the expansion history and its implications for dark matter, gravitational waves, and primordial black holes.

\acknowledgments

For discussions and feedback at different stages of this project we are grateful to Wenyuan Ai, Kim Berghaus, Daniel Figueroa, Mark Hindmarsh, Marek Lewicki, Bogumiła Świeżewska, Daniel Schmitt, Jorinde van de Vis, David Weir and Ke-Pan Xie. We thank Ryusuke Jinno for providing an initial version of the code. HM is funded by the Deutsche Forschungsgemeinschaft (DFG) through
Grant No. 396021762 --- TRR 257. YG acknowledges support by the Cluster of Excellence ``PRISMA$^{++}$'' funded by the German Research Foundation (DFG) within the German Excellence Strategy (Project No. 390831469).
\appendix

\section{Determination of $\gamma_w$}
\label{app:gamma_wall}

\begin{figure}[t]
\centering
\includegraphics[width=0.41\linewidth,clip,trim=0 -15 0 15]{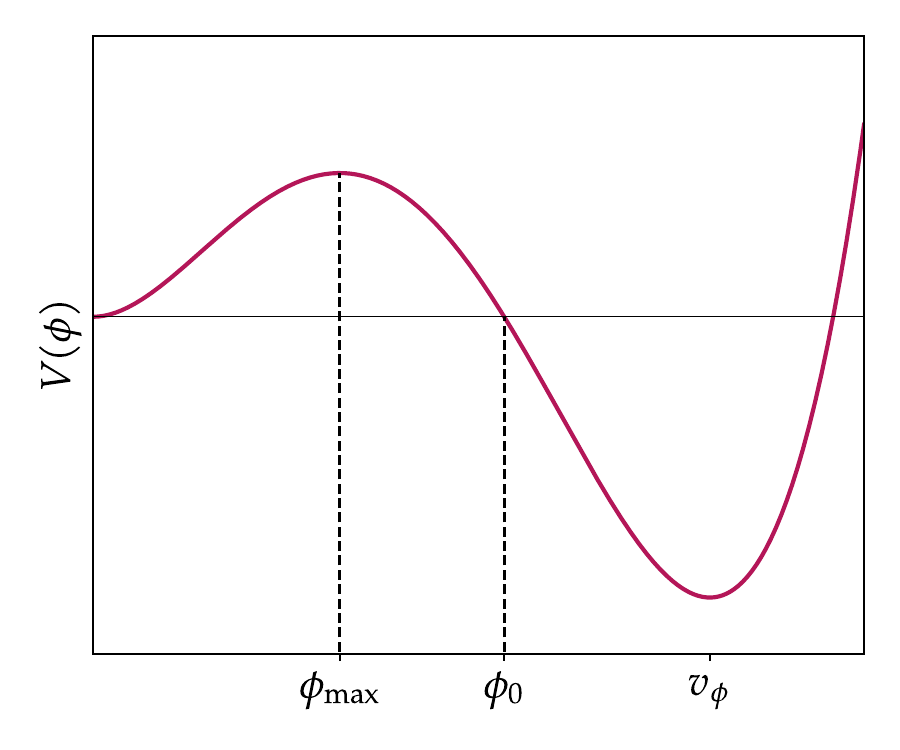}\hfill
\includegraphics[width=0.59\linewidth,clip,trim=0 15 0 0]{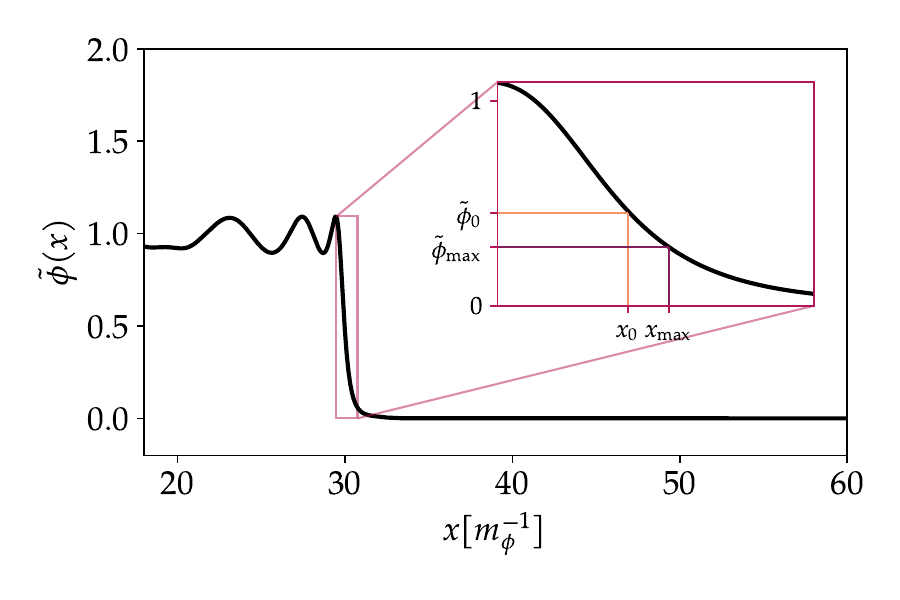}
\caption{Determination of the Lorentz factor from the bubble wall
thickness in $1{+}1$ dimensions. Left: definition of
$\phi_\mathrm{max}$ and $\phi_0$ from the scalar potential. Right:
definition of $x_0$ and $x_\mathrm{max}$ used to track the wall width
during expansion.}
\label{fig:gamma_measurement}
\end{figure}

To validate the linear relation between the wall Lorentz factor
$\gamma$ and the bubble radius we track the wall thickness in $1{+}1$
dimensions. Since the field reaches the false vacuum ($\phi=0$) only
asymptotically outside the bubble and oscillates around the true
vacuum ($\phi=1$) inside, the wall boundaries are not unambiguously
defined by these two values. We instead use two reference field
values that are numerically easier to track: $\phi_\mathrm{max}$, at
the maximum of the barrier, and $\phi_0$, the exit value of the bounce
solution (figure~\ref{fig:gamma_measurement} left). The wall length is
defined as $\Delta\ell=x_\mathrm{max}-x_0$, with $x_\mathrm{max},x_0$
the spatial coordinates corresponding to $\phi_\mathrm{max},\phi_0$
(figure~\ref{fig:gamma_measurement} right). Taking $\gamma=1$ initially
and assuming subsequent rigid Lorentz contraction, we obtain
$\gamma(t)=\Delta\ell(0)/\Delta\ell(t)$. Note that the part of the wall profile lying between $\phi_0$ and $\phi_{\rm max}$ is only a small fraction of the full wall, typically thinner by a factor of $5$--$10$. 
We leave to future work the construction of a method based on more widely separated characteristic points, for instance $\phi_{\rm max}$ and the first point at which the field crosses $\phi=v_\phi$. 

\section{Dependence on the shape of the scalar potential}
\label{app:c_dependence}

\begin{figure}[b]
\centering
\includegraphics[width=0.7\linewidth]{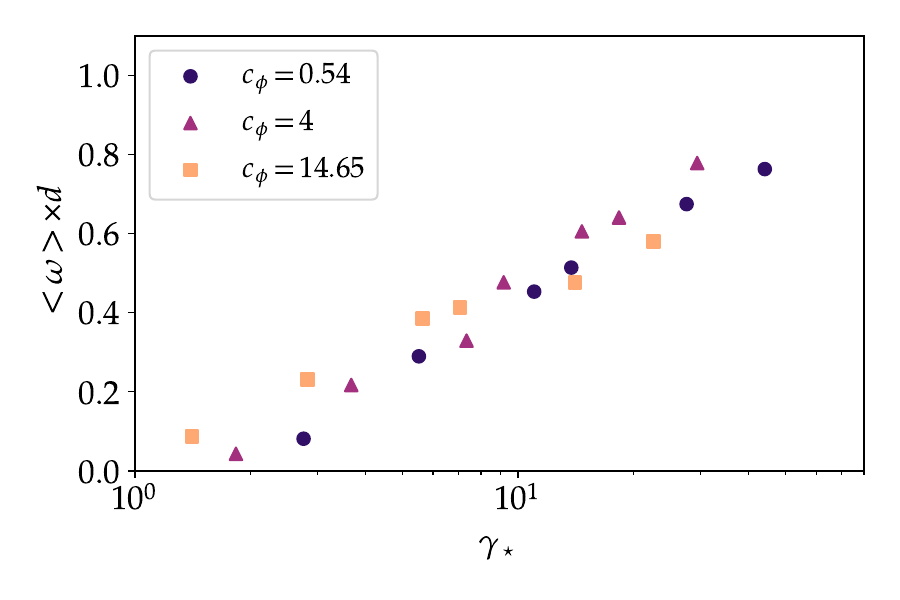}
\caption{Post-collision EoS in $1{+}1$ simulations for varying
potential couplings. Larger $c_\phi$ produces steeper barriers.}
\label{fig:test_potentialdependence}
\end{figure}

Throughout the main text we fixed the shape of the scalar
potential to eq.~\eqref{eq:toypotential} with $c_\phi=4$. One might
worry that the results depend sensitively on this choice, because
larger $c_\phi$ increases the mass $m_\phi$ relative to the potential
difference. However, changing $c_\phi$ also changes the critical
profile (and hence $R_n$). To isolate the dependence on potential
shape we therefore adjust the initial bubble separation $R_s$ so that
$R_s/R_n$, and hence $\gamma_\star$ at collision, is held fixed.
Figure~\ref{fig:test_potentialdependence} compares our benchmark
$c_\phi=4$ with a shallower ($c_\phi=0.54$) and a steeper
($c_\phi=14.65$) case in $d=1$. The post-collision EoS depends
dominantly on the initial bubble separation, with only small residual
differences across the three potentials. These should be interpreted
with care: larger $c_\phi$ values quickly become numerically
expensive because the initial bubble radius grows while the walls
become thinner, and steeper barriers also enhance wall
recollisions~\cite{Jinno:2019bxw, Konstandin:2011ds}, which delay the
relaxation of $\omega$.

\section{Scalar power spectrum before collision}
\label{app:analytical}

\paragraph{Power spectrum definitions.}
We begin by recalling the definition of the power spectrum
\begin{equation}
P_\phi(k)=\lim_{V\to\infty}\langle |\tilde\phi(\mathbf{k})|^2\rangle_{\rm ens} \, ,
\qquad
\tilde\phi(\mathbf{k})=\int d^dx\,e^{-i\mathbf{k}\cdot\mathbf{x}}\phi(\mathbf{x})\; ,
\end{equation}
where \(V\) is a finite volume and \(d\) the number of spatial dimensions. It is convenient to work with the dimensionless power spectrum
\begin{equation}
\Delta_\phi(k)=\frac{k^d}{(2\pi)^d}S_{d-1}\,P_\phi(k) \, ,
\qquad
S_{d-1}=\frac{2\pi^{d/2}}{\Gamma(d/2)} \; .
\end{equation}
An explicit expression for the power spectrum can be obtained whenever the Fourier transform $\tilde \phi(\textbf{k})$ can be evaluated in closed form. In the case of FOPTs, the full post-collision field configuration is highly non-linear, making a direct analytical treatment difficult, if not impossible. Hence, we focus on the early-time evolution and approximate the field by a single expanding bubble. We demonstrate in the following that this simplified setup already offers results compatible with our lattice results and motivates the broken power law template in eq.~\eqref{eq:brokenPowerLaw} and the smoothened version in eq.~\eqref{eq:analytical_smooth}. In particular, the intermediate $1/k$ behaviour of relativistic modes is set by the expanding bubble wall, while the UV-suppression of the tail follows from considering a smooth wall. 

\paragraph{Thin-wall limit.}
We first approximate a single stationary bubble of radius \(R\) by the step profile
\begin{equation}
\label{eq:Heaviside_bubble}
\phi(\mathbf{x})=v_\phi\,\Theta(R-|\mathbf{x}|) \, ,
\end{equation}
where \(v_\phi\) is the true-vacuum value and $\Theta$ is the Heaviside function. By spherical symmetry,
\begin{equation}
\tilde\phi(\mathbf{k})=v_\phi\,U_d(k) \, ,
\qquad
U_d(k)
=(2\pi)^{d/2}\frac{R^{d/2}}{k^{d/2}}\,J_{d/2}(kR)
\end{equation}
with \(k\equiv |\mathbf{k}|\), $d$ the number of spatial dimensions and \(J_\nu\) the Bessel function of the first kind. The corresponding power spectrum is
\begin{equation}
P_{\rm bubble}^{\rm step}(k)
=\frac{v_\phi^2}{V}(2\pi)^d\frac{R^d}{k^d}\big[J_{d/2}(kR)\big]^2 \; .
\label{eq:Pphi_d_exact}
\end{equation}
From this expression, we expect oscillations of the power spectrum on scales $\Delta k \sim 1/R$, which is compatible with the early-time evolution of the power spectrum in figure~\ref{fig:PS-Evol}. For \(kR\gg1\), averaging over the rapid oscillations of the Bessel function gives
\begin{equation}
\big\langle \Delta_{\rm bubble}^{\rm step}(k)\big\rangle_{\rm osc}
=
v_\phi^2\,\frac{d}{\pi}\,\frac{V_R}{V}\,\frac{1}{kR}
+\mathcal{O}\!\big(k^{-2}\big) \; ,
\label{eq:Delta_d_asymp}
\end{equation}
where \(V_R=S_{d-1}R^d/d\) is the bubble volume. The thin-wall approximation therefore predicts a \(1/k\) tail for \(k\gg R^{-1}\), in agreement with the simulation results shown in figure~\ref{fig:PS-Evol}. 

\begin{figure}[t]
\centering
\begin{subfigure}[t]{0.495\linewidth}
\centering
\includegraphics[width=0.99\linewidth]{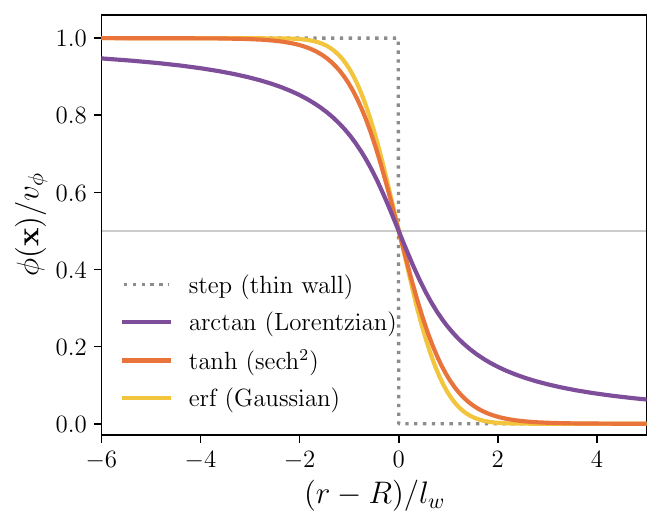}
\caption{Wall profiles in real space.}
\label{fig:wall_profiles_kernels}
\end{subfigure}
\begin{subfigure}[t]{0.495\linewidth}
\centering
\includegraphics[width=0.99\linewidth]{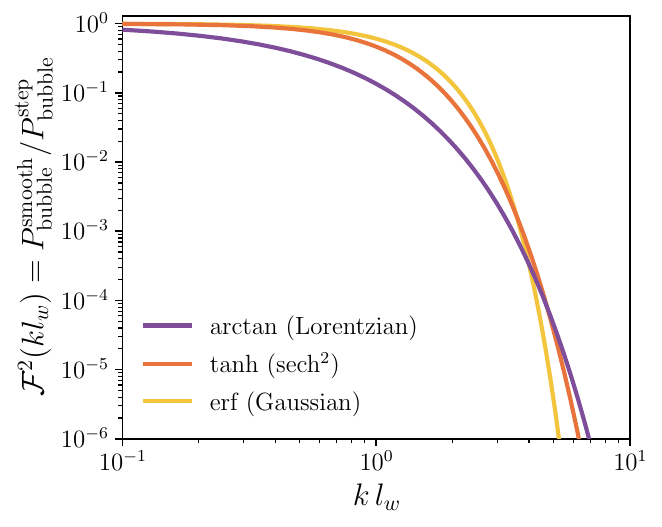}
\caption{Power-spectrum suppression.}
\label{fig:formfactor_suppression}
\end{subfigure}
\caption{The three finite-thickness wall models of eqs.~\eqref{eq:profiles}
and~\eqref{eq:formfactors}. \emph{Left:} the normalised field profile
$\phi(\mathbf x)/v_\phi$ versus $(r-R)/l_w$, from convolving the step profile
(dotted) with the arctan (Lorentzian), $\tanh$ (sech$^2$) and erf (Gaussian)
kernels. \emph{Right:} the resulting power-spectrum suppression
$\mathcal F^2(kl_w)=P_{\rm bubble}^{\rm smooth}/P_{\rm bubble}^{\rm step}$
versus $kl_w$. Modes with $k\ll l_w^{-1}$ are unchanged and those with
$k\gg l_w^{-1}$ suppressed; the arctan and $\tanh$ walls give an exponential
cutoff, the erf wall the steeper Gaussian tail favoured by the lattice data
(Figs.~\ref{fig:PS-Evol} and~\ref{fig:numericalPS_appendix}).}
\label{fig:wall_kernels}
\end{figure}

\paragraph{Finite wall thickness.}
We now consider three different smooth profiles centred
at $|\mathbf x|=R$ with finite width $l_w$ (see figure~\ref{fig:wall_kernels}):
\begin{equation}
\phi(\mathbf x) \,=\, \frac{v_\phi}{2}\times
\begin{cases}
\displaystyle 1+\frac{2}{\pi}\arctan\!\left(\dfrac{R-|\mathbf x|}{l_w}\right) \, ,
& \text{(arctan)} \, ,\\[12pt]
\displaystyle 1+\tanh\!\left(\dfrac{R-|\mathbf x|}{l_w}\right) \, ,
& \text{($\tanh$)} \, , \\[12pt]
\dfrac{v_\phi}{2}
\left[
1+\operatorname{erf}\!\left(\dfrac{R-|\mathbf x|}{l_w}\right)
\right] \, ,
& \text{(erf)} \; .
\end{cases}
\label{eq:profiles}
\end{equation}

\begin{figure}
\centering
\begin{subfigure}{0.75\linewidth}
\includegraphics[width=\linewidth]{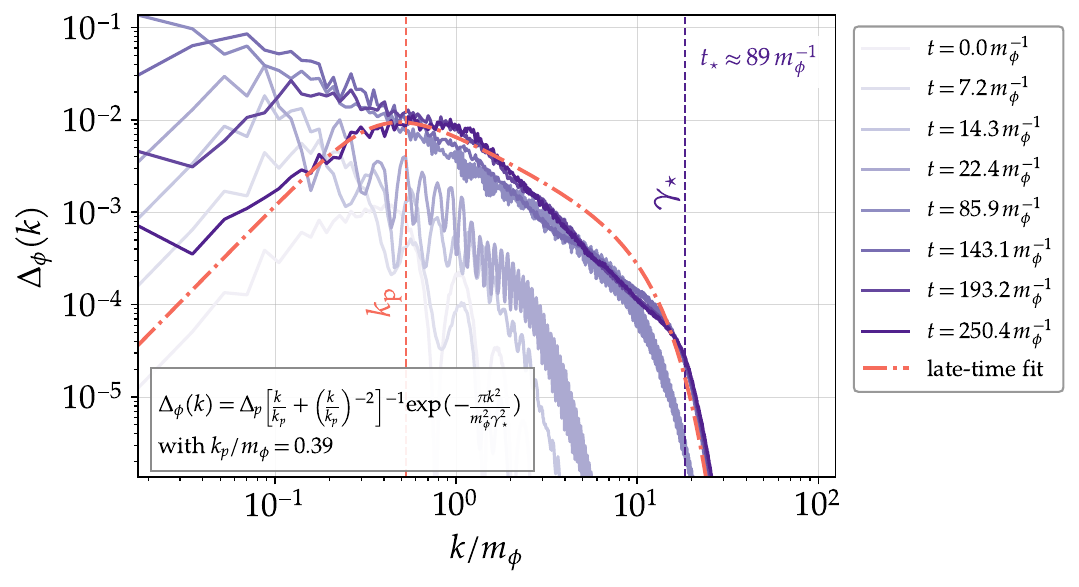}
\end{subfigure}
\begin{subfigure}{0.75\linewidth}
\includegraphics[width=\linewidth]{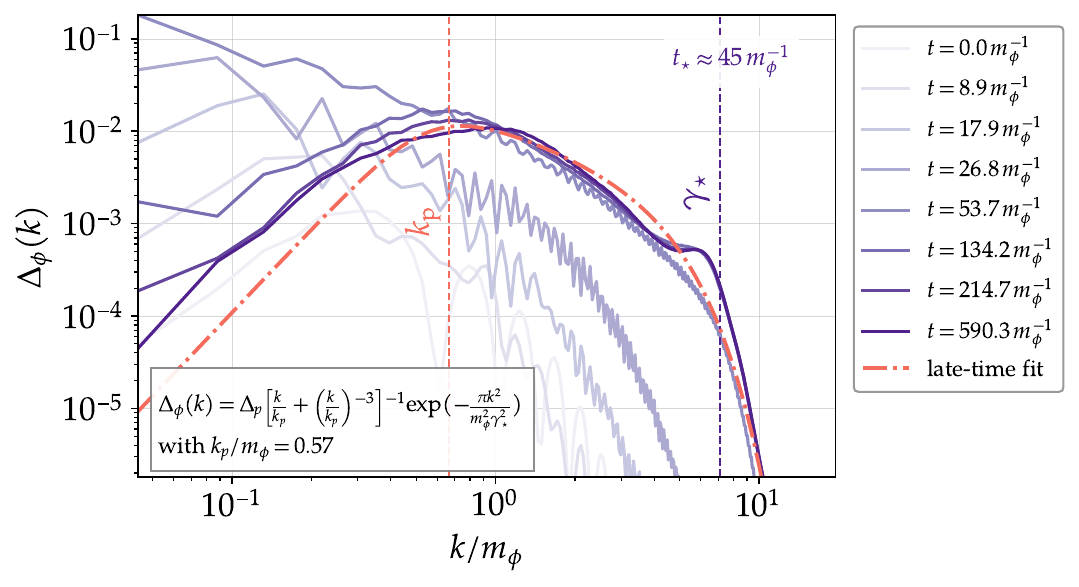}
\end{subfigure}
\caption{Evolution of the power spectra corresponding to the simulations
of section~\ref{sec:lattice}. Top: $2{+}1$ simulation with four bubbles,
$N=20\,000$, $L=80$, $\Delta t=0.002$ (cf.\
figure~\ref{fig:2Dsimulation}). Bottom: $3{+}1$ simulation
with four bubbles and $N=1200$, $L=24$, $\Delta t=0.006$ (cf.\ figure~\ref{fig:3D_simu_overview}). The small peak around $k/m_\phi \simeq \gamma_\star$ is an artefact of the UV cut-off: it disappears as
$\Delta x$ is reduced. In both panels, the dashed purple grid lines indicate the average Lorentz factor of the walls
$\gamma_\star$ at collision. The labels in the right top corner indicate the time of the first bubble collision. The best-fit analytical template for the late-time power spectra are plotted in dot-dashed lines. }
\label{fig:numericalPS_appendix}
\end{figure}

The $\tanh$ profile matches eq.~\eqref{eq:wall_profile} of the body of the paper after the identification $(R,v_\phi)\leftrightarrow(R_c,\phi_0)$, and is a typical ansatz for the bounce. However, all three functions are valid models for a bubble with finite wall thickness. These profiles can be written as a convolution of a step function smoothened with a normalised kernel 
\begin{equation}
    \frac{\phi(\textbf{x})}{v_\phi} = \Theta(R-|\textbf{x}|) \;*\;  g\left(\frac{r-R}{l_w}\right)\,,\qquad\int_{-\infty}^{+\infty} g(u)du = 1
\end{equation}
with $u = (r-R)/l_w$. The Fourier transform is then given by the product 
\begin{equation}
\tilde\phi_d^\mathrm{smooth}(k)
\,\simeq\, \mathcal F (kl_w) \,\tilde\phi_d^\mathrm{step}(k)\,,
\end{equation}
where the form factor $\mathcal F (kl_w)$ encapsulates the dependence on the wall thickness and is given by the Fourier transform of the kernel $g(u)$. Consequently, considering a smooth profile modifies our previous results for the power spectrum only by a multiplication   
\begin{equation}
    P_{\rm bubble}^{\rm smooth}= P_{\rm bubble}^{\rm step}(k) \, \mathcal{F}^2(kl_w)\,.
\end{equation}
The form factor only impacts high-k modes while the behaviour for $k\ll l_w^{-1}$ remains identical with that of the power spectrum of a step-like bubble. For the three profiles above, the kernels and corresponding form factors are given by
\begin{equation}
\begin{aligned}
\text{(arctan)}:\quad
& g(r) \,=\, \frac{1}{\pi(1+u^{2})} \, , &
& \mathcal F_\mathrm{arctan}(k l_w) \,=\, \exp(-k l_w) \, ,\\[4pt]
\text{($\tanh$)}:\quad
& g(u) \,=\, \tfrac{1}{2}\,\mathrm{sech}^{2}(u) \, , &
& \mathcal F_\mathrm{tanh}(k l_w)
\,=\, \frac{\pi k l_w/2}{\sinh(\pi k l_w/2)} \;,\\[4pt] 
\text{(erf)}:\quad
& g(u) \,=\, \frac{1}{\sqrt{\pi}}e^{-u^2} \, , &
& \mathcal F_\mathrm{erf}(k l_w)
\,=\, \exp(-\frac{(kl_w)^2}{4}) \; .
\end{aligned}
\label{eq:formfactors}
\end{equation}
Note that the form factor corresponding to a $\tanh$ profile behaves like $\sim (\pi k l_w)^{2}\,e^{-\pi k l_w}
$ for $kl_w\gg1$. Hence, all three profiles lead to a suppression of modes with $k\gg l_w^{-1}$, confirming that the wall width indeed sets the UV cut-off. The $\tanh$ and $\arctan$ profiles result in a simple exponential suppression while the erf profile gives rise to a Gaussian tail. We find that our lattice results in figs. \ref{fig:PS-Evol} and \ref{fig:numericalPS_appendix} are more compatible with a Gaussian decay (see figure~\ref{fig:wall_kernels}).

\begin{figure}[t]
    \centering
    \includegraphics[width=\linewidth]{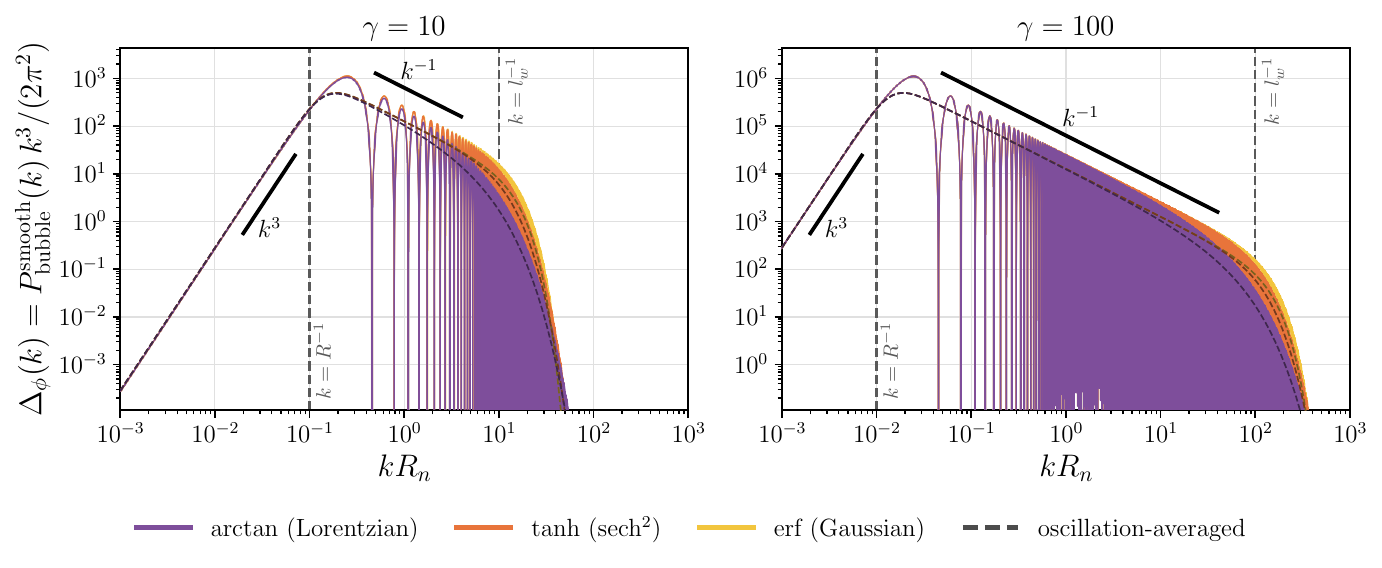}
        \caption{Dimensionless single-bubble power spectrum
    $\Delta_\phi(k)=k^3 P_{\rm bubble}^{\rm smooth}(k)/(2\pi^2)$ for the three
    wall kernels of eqs.~\eqref{eq:profiles} and~\eqref{eq:formfactors} (arctan,
    $\tanh$, erf), at $\gamma=10$ (left) and $\gamma=100$ (right). Solid: full
    spectrum; dashed: oscillation-averaged envelope. Black segments mark the
    $k^3$ and $k^{-1}$ slopes; vertical dashed lines the scales $k=R^{-1}$ and
    $k=l_w^{-1}$. The kernels coincide until the cut-off, where the erf
    (Gaussian) tail falls off fastest and the arctan (Lorentzian) slowest.}
    \label{fig:deltaphi_profiles_osc}
\end{figure}

\paragraph{Time evolution.}
Until now, our expressions did not include any dependence on the Lorentz factor $\gamma$. This parameter enters through the bubble radius and the contracted wall width
\begin{equation}
R\simeq \gamma R_n \, ,
\qquad
l_w=\frac{l_{w,0}}{\gamma} \; ,
\end{equation}
where \(R_n\) is the initial radius and \(l_{w,0}\sim m_\phi^{-1}\) the wall width in the wall rest frame. We conclude that the dimensionless power spectrum of an expanding bubble with an initial radius $R_n$, initial finite thickness $l_{w,0}$ and Lorentz factor $\gamma$ is given by
\begin{equation}
P_{\rm bubble}^{\rm smooth}(k,\gamma) \simeq\frac{v_\phi^2}{V}(2\pi)^d\frac{(\gamma R_n)^d}{k^d}\big[J_{d/2}(k\gamma R_n)\big]^2 \exp\left({- \frac{(kl_{w,0})^2}{2 \gamma^2}}\right) \; .
\label{eq:Pphi_d_exact_gamma}
\end{equation}
This analytically motivated spectrum is shown in figure~\ref{fig:deltaphi_profiles_osc} for two Lorentz factors ($\gamma=10$ and $100$), with the overall amplitude and effective radius adjusted to match; the predicted shift of the characteristic scales with increasing $\gamma$ is consistent with the simulation results. The choice of wall profile enters only through the UV form factor: as illustrated in the same figure, the arctan, $\tanh$ and erf walls yield identical spectra throughout the causality ($\propto k^3$) and relativistic ($\propto 1/k$) regimes and differ only beyond the cut-off $k\sim\gamma/l_{w,0}$ (i.e. $kR_n\sim\gamma$), where the Gaussian (erf) tail is the steepest.

\paragraph{Analytical template.}
The simulation results (section~\ref{subsec:numericalresults-PS}) show that after the collision the power spectrum stays roughly unchanged for modes $ k\geq m_\phi$, while the IR-part of the spectrum starts transitioning to an increasing power law as expected from causality. Combining the infrared causality scaling \(\Delta_\phi\propto k^3\) in three spatial dimensions with the intermediate \(1/k\) behaviour and the UV cut-off determined from the previous computations, we are led to the template
\begin{equation}
\Delta_\phi(k)\simeq \Delta_p\times
\begin{cases}
\left(\dfrac{k}{k_p}\right)^3 \, , & k\ll k_p \, ,\\[1.2ex]
\left(\dfrac{k}{k_p}\right)^{-1}\exp\!\left[-\dfrac{k^2}{k_\star^2}\right] \, , & k\gg k_p \; ,
\end{cases}
\label{eq:app_analytical}
\end{equation}
where \(\Delta_p\equiv \Delta_\phi(k_p)\), \(k_p\sim m_\phi\), and \(k_\star\sim l_w^{-1}\sim \gamma_\star m_\phi\). By fitting the numerical data we find $k_\star \simeq \gamma_\star m_\phi/\sqrt{\pi}$. This gives the broken power law template used in the main text (eq.~\eqref{eq:brokenPowerLaw}).

\paragraph{Normalisation from energy conservation.} The amplitude $\Delta_p $ can be estimated by requiring energy conservation. The total energy density of the scalar configuration can be expressed in terms of the dimensionless power spectrum: 
\begin{align}
    \langle \rho_\mathrm{tot} \rangle  &=\int_0^{\infty} \frac{dk}{k} \left( m_\phi^2 + k^2\right) \Delta_\phi(k) \\  
    & \simeq  \Delta_p m_\phi^2 \int \frac{dx}{x} \; \frac{x^2  + 1}{ 1 + x^4} \; \exp\!\left[-\pi x^2/x_\star^2\right] \\ & = \frac{\Delta_p  \gamma_\star m_\phi^2}{ 2}  + \mathcal{O}(1/\gamma_\star) \; .
\end{align}
In the second step, we set $k_p\simeq  m_\phi$, $k_\star = \gamma_\star m_\phi$ and used the smoothed broken power law of eq.~\eqref{eq:brokenPowerLaw}. The integral is rewritten as a function of $ x \equiv k/m_\phi$. Requiring that the total energy density is equal to the latent heat $\Delta V$ gives : 
\begin{equation}
    \Delta_p =\frac{ 2 \Delta V}{m_\phi^2 \gamma_\star}\,,
\end{equation}
which agrees with the values inferred from the fits in figures \ref{fig:PS-Evol} and \ref{fig:numericalPS_appendix}.

\section{Numerical results of the power spectrum}
\label{app:NumericalPS}

\paragraph{Additional lattice simulations.}
To complement the $3{+}1$ result discussed in
section~\ref{subsec:numericalresults-PS}
(figure~\ref{fig:PS-Evol}), figure~\ref{fig:numericalPS_appendix} shows
the time evolution of the dimensionless scalar power spectrum
$\Delta_\phi(k)$ for two additional runs: a $2{+}1$-dimensional
simulation with four bubbles, corresponding to
figure~\ref{fig:2Dsimulation} of the main text, and a $3{+}1$-dimensional
simulation corresponding to figure~\ref{fig:3D_simu_overview}. In both cases the dashed
vertical line marks $k=\gamma_\star m_\phi$, the wall width cut-off
inferred from the average bubble radius at collision, which is approximated as $R_\star \simeq L/2 N_b^{1/d}$.

\paragraph{Spectral evolution.}
After the first collision, the infrared part of the spectrum differs from the
single-bubble pre-collision spectrum discussed in appendix~\ref{app:analytical}, but in the later evolution
modes with \(k\lesssim m_\phi\) again relax towards the causal scaling
\(\Delta_\phi\propto k^d\).
The qualitative features anticipated by the analytic template
\eqref{eq:analytical_smooth} are clearly visible: a steep IR rise consistent
with the causality scaling $\Delta_\phi\propto k^{d}$, where $d$ is the number of spatial dimensions, a $\sim 1/k$
intermediate plateau between $k_p\simeq m_\phi$ and
$k_\star=\gamma_\star m_\phi$, and a sharp suppression beyond
$k_\star$. However, in the $2+1$ simulation the UV power law at later times seems steeper than $1/k$.  The small bump in the $3+1$ power spectrum just above $k_\star$ is a lattice artifact. It decreases with improved resolution (smaller grid spacing $\Delta x$), and does not invalidate the overall results since its amplitude is subdominant.

\section{Equation of state of a thermalised scalar}
\label{app:thermalisation}

This appendix derives, for a scalar that reaches thermal equilibrium after
collision, two quantities: its dimensionless power spectrum
$\Delta_\phi^{\rm th}$ and its EoS $\omega^{\rm th}$, the
latter quoted in section~\ref{subsec:returntoMD}. Both follow from the same
relation between the power spectrum and the energy density used throughout
section~\ref{subsec:PS}, now evaluated on a thermal occupation rather than on
the simulated post-collision spectrum. We first build the bridge from the
lattice spectrum to a kinetic phase-space density, then read off each result
in turn.

\paragraph{Phase space distribution.}
Both results rest on a single map between the field-theory spectrum measured
on the lattice and the kinetic phase-space density $f_\phi$. Working in $d$
spatial dimensions, with WKB frequency $\omega_k(a)=\sqrt{k^2/a^2+m_\phi^2}$
and dimensionless spectrum $\Delta_\phi=[S_{d-1}k^d/(2\pi)^d]\,P_\phi$,
$S_{d-1}=2\pi^{d/2}/\Gamma(d/2)$ [reducing to
eq.~\eqref{eq:power_spectrum_Delta_def} for $d=3$], we fix $f_\phi$ in
physical momentum $p=k/a$ by equating the field-theory energy density
$\rho_\phi=\int d\ln k\,\omega_k^2\,\Delta_\phi$
(eqs.~\eqref{eq:EK_of_PS}--\eqref{eq:EG_EV_of_PS}) to its kinetic-theory form
$\rho_\phi=a^{-d}\!\int\!(d^dk/(2\pi)^d)\,\omega_k\,f_\phi(k/a)$,
\begin{equation}
\label{eq:f_phi_P_phi}
f_\phi(k/a)=a^d\,\omega_k(a)\,P_\phi(k,a) \; .
\end{equation}
In equilibrium, the occupation is Bose--Einstein:
\begin{equation}
\label{eq:f_phi_th}
f_\phi^{\rm th}(p)=\frac{1}{e^{\omega_p/T}-1}\,,
\qquad
\omega_p=\sqrt{p^2+m_\phi^2}\,,
\end{equation}
and eq.~\eqref{eq:f_phi_P_phi} turns this single input into both results
below.

\paragraph{Thermal power spectrum.}
Inserting $P_\phi=f_\phi/(a^d\omega_k)$ from eq.~\eqref{eq:f_phi_P_phi} into
the definition of $\Delta_\phi$ and writing $k=ap$, the scale-factor powers
cancel and the dimensionless spectrum becomes a function of physical
momentum alone,
\begin{equation}
\label{eq:Delta_phi_th}
\boxed{\;
\Delta_\phi^{\rm th}(p)
=\frac{S_{d-1}}{(2\pi)^d}\,\frac{p^d}{\omega_p}\,f_\phi^{\rm th}(p)
=\frac{S_{d-1}}{(2\pi)^d}\,\frac{p^d}{\omega_p}\,
\frac{1}{e^{\omega_p/T}-1}\,,\;}
\end{equation}
which for $d=3$ ($S_2=4\pi$, so $S_2/(2\pi)^3=1/2\pi^2$) reduces to
\begin{equation}
\label{eq:Delta_phi_th_d3}
\Delta_\phi^{\rm th}(p)=\frac{1}{2\pi^2}\,\frac{p^3}{\omega_p}\,
\frac{1}{e^{\omega_p/T}-1}\,.
\end{equation}
This is the spectrum a thermalised scalar would display in place of the
free-streaming post-collision spectrum of
section~\ref{subsec:numericalresults-PS}. By construction it reproduces the
thermal energy density when weighted by $\omega_p^2$ as in
eq.~\eqref{eq:omega_EoS_PS}, and it furnishes the input for the equation of
state below.

\paragraph{Thermal equation of state.}
Feeding the thermal spectrum~\eqref{eq:Delta_phi_th} into the master
formula~\eqref{eq:omega_EoS_PS} turns the spectral integrals into the
standard kinetic-theory moments. The numerator becomes the gradient energy
and the denominator the kinetic energy $\langle\rho_\mathrm{kin}\rangle
=\langle\rho_\mathrm{grad}\rangle+\langle\rho_\mathrm{pot}\rangle$, with
\begin{equation}
\label{eq:Hubble_expansion_K_G_V_th_eq}
\langle\rho_\mathrm{grad}\rangle=\tfrac12\!\int\!\tfrac{d^dp}{(2\pi)^d}\,\tfrac{p^2}{\omega_p}\,f_\phi^{\rm th} \, ,
\qquad
\langle\rho_\mathrm{pot}\rangle=\tfrac12\!\int\!\tfrac{d^dp}{(2\pi)^d}\,\tfrac{m_\phi^2}{\omega_p}\,f_\phi^{\rm th} \; .
\end{equation}
The EoS collapses to the ideal-gas ratio
\begin{equation}
\label{eq:omega_th_ratio}
\omega^{\rm th}(T)=\frac1d\,
\frac{\int d^dp\,(p^2/\omega_p)\,f_\phi^{\rm th}}
{\int d^dp\,\omega_p\,f_\phi^{\rm th}}\,,
\end{equation}
the denominator being the sum of the gradient and potential integrals since
$\omega_p=p^2/\omega_p+m_\phi^2/\omega_p$. In the Maxwell--Boltzmann limit
$f_\phi^{\rm th}\simeq e^{-\omega_p/T}$ these moments are modified Bessel
functions,
\begin{equation}
\label{eq:Hubble_expansion_K_G_V_th_eq_MB}
\int\!\tfrac{d^dp}{(2\pi)^d}\,\tfrac{p^2}{\omega_p}\,e^{-\omega_p/T}
=2d\,c_d\,K_{\frac{d+1}{2}}(x) \, ,
\qquad
\int\!\tfrac{d^dp}{(2\pi)^d}\,\tfrac{m_\phi^2}{\omega_p}\,e^{-\omega_p/T}
=2x\,c_d\,K_{\frac{d-1}{2}}(x) \; ,
\end{equation}
with $x\equiv m_\phi/T$, $c_d=(T^2x/2\pi)^{(d+1)/2}$ and $K_\alpha$ the
modified Bessel functions of the second kind. Their ratio yields the thermal
EoS in $d$ dimensions,
\begin{equation}
\label{eq:omega_th_general}
\boxed{\;
\omega^{\rm th}(x)=\frac{K_{\frac{d+1}{2}}(x)}
{d\,K_{\frac{d+1}{2}}(x)+x\,K_{\frac{d-1}{2}}(x)}
\;\simeq\;
\begin{cases}
1/d, & x\ll1 \, ,\\[2pt]
1/x, & x\gg1 \; ,
\end{cases}\;}
\end{equation}
which reduces to eq.~\eqref{eq:omega_th_final} for $d=3$. As the universe expands and $x$ increases, it interpolates
from radiation ($x\ll1$) to matter ($x\gg1$). As discussed in section~\ref{subsec:returntoMD}, the initial value of $x$ can be obtained from energy conservation, assuming self-thermalisation into the scalar's own quanta.

\small
\providecommand{\href}[2]{#2}\begingroup\raggedright\endgroup

\end{document}